\documentclass[12pt]{article}
\usepackage{amsmath,amssymb,epsfig,amsfonts}
\usepackage{graphicx,subfigure}
\usepackage[usenames, dvipsnames]{color}
\usepackage[backref]{hyperref}
\usepackage{cite}
\usepackage{verbatim} %for comments
\usepackage{float}
\usepackage{floatflt}
\usepackage{lipsum}
\usepackage{rotating}
\usepackage{multirow}
\usepackage{cleveref}
\makeatletter

\DeclareMathOperator{\ord}{ord} 

\newcommand{\be}{\begin{equation}}
\newcommand{\ee}{\end{equation}}
\newcommand{\ba}{\begin{aligned}}
\newcommand{\ea}{\end{aligned}}

\newcommand{\bea}{\begin{eqnarray}}
\newcommand{\eea}{\end{eqnarray}}

\def\unit{{1\kern-.65ex {\rm l}}}
\def\1{{1\kern-.65ex {\rm l}}}

\def\fkD{{\mathfrak{D}}}

\def\fks{{\mathfrak{s}}}

\newcount\hour \newcount\minute
\hour=\time \divide \hour by 60
\minute=\time
\def\now{%
\ifnum \hour<13
  \ifnum \hour=0 \advance \hour by 12 \number\hour:\else \number\hour:\fi%
     \ifnum \minute<10 0\fi%
     \number\minute%
\ A.M.%
\else \advance \hour by -12 \number\hour:%
  \ifnum \minute<10 0\fi%
  \number\minute%
  \ P.M.%
\fi%
}

\makeatother

\begin{document}

% format
\baselineskip=18pt  % a la harvmac
\numberwithin{equation}{section}  % make eq labels (sec.num)
%\allowdisplaybreaks  % allow page breaks in displayed eqs

% print date, time and filename 
%\pagestyle{myheadings}
%\markright{{\tt \jobname.tex} -- \today{} \now}

%%%%%%%%%%%%%%%%%%%%%%%%%%%%%%%%%%%%%%%%%%%
%%%        TITLE BEGINS HERE
%%%%%%%%%%%%%%%%%%%%%%%%%%%%%%%%%%%%%%%%%%%

%% ========== title (note version) begins here ==========
%
%\vspace*{-1cm}
%\begin{center}
% {\Large\bf Title of the Document}
%\end{center}
%\vspace*{-.5cm}
%
%% ========== title (note version) ends here ==========

%% ========== title (paper version, a la harvmac) begins here ==========

\setcounter{tocdepth}{2} % Prevent subsubsections from being listed in the toc

\thispagestyle{empty}

% Report number
\vspace*{-2cm} 
\begin{flushright}
  {\tt KCL-MTH-14-21}\\
\end{flushright}

% title, authors, affiliation
\vspace*{0.8cm} 
\begin{center}

%{\LARGE Mordell-Weil meets Tate}
  
%{\LARGE $SU(5)$ Models from Standard Forms for Elliptic Fibrations with rank two Mordell-Weil
%Group}

%  {\LARGE $SU(5) \times U(1)^2$ Models from Tate's Algorithm}
%   {\LARGE Elliptic Fibrations for F-theory GUTs with two $U(1)$s
%     \\\smallskip from
%   Tate's Algorithm}

%   {\LARGE Generalized Tate Forms for F-theory GUTs \\\smallskip with two $U(1)$s}

   {\LARGE Tate's Algorithm for F-theory GUTs \\\smallskip with two $U(1)$s}

%   {\LARGE Tate for two $U(1)$s}

 \vspace*{1.5cm}
 Craig Lawrie\footnote{\tt {gmail:$\,$ craig.lawrie1729}} and 
 Damiano Sacco\footnote{\tt damiano.sacco@kcl.ac.uk} \\

\vspace*{0.5cm}
 {\it Department of Mathematics, King's College London \\
  The Strand, London WC2R 2LS, England }\\
%  {\tt {gmail:$\,$ craig.lawrie1729}}\\
%  {\tt {email:$\,$ damiano.sacco@kcl.ac.uk}}

\vspace*{0.8cm}
\end{center}
\vspace*{.5cm}

% abstract
\noindent
We present a systematic study of elliptic fibrations for F-theory realizations
of gauge theories with two $U(1)$ factors. In particular, we
determine a new class of ${SU(5) \times U(1)^2}$ fibrations, which can be used to
engineer Grand Unified Theories, with multiple,
differently charged, ${\bf 10}$ matter representations. To determine these
models we apply Tate's algorithm to elliptic fibrations with two 
$U(1)$ symmetries, which are realized in terms of a cubic in $\mathbb{P}^2$.
In the process, we find fibers which are not characterized solely
in terms of vanishing orders, but with some additional specialization, which
plays a key role in the construction of these novel $SU(5)$ models with multiple
${\bf 10}$ matter. We also determine a table of Tate-like forms for
Kodaira fibers with two $U(1)$s.

\newpage
%%%%%%%%%%%%%%%%%%%%%%%%%%%%%%%%%%%%%%%%%%%
%%%           TITLE ENDS HERE
%%%%%%%%%%%%%%%%%%%%%%%%%%%%%%%%%%%%%%%%%%%

%%%%%%%%%%%%%%%%%%%%%%%%%%%%%%%%%%%%%%%%%%%
%%%        MAIN TEXT BEGINS HERE
%%%%%%%%%%%%%%%%%%%%%%%%%%%%%%%%%%%%%%%%%%%

\tableofcontents
\newpage

%%%%%%%%%%%%%%%%%%%%%%%%%%%%%%%%%%%%%%
%%%%%%%%%%%%%%%%%%%%%%%%%%%%%%%%%%%%%%

\section{Introduction}

%The abelian theory should perhaps be considered the ur-example of a gauge
%field theory, first presented one hundred and fifty years ago, almost to the
%day, by Maxwell in his formulation of electrodynamics \cite{Maxwell01011865}.
% Abelian gauge factors play a crucial role 
%
%
%There are many reasons to consider gauge theories with abelian factors; for
%example one particularly relevant reason is that by imposing $U(1)$ charges on the matter
%fields of a Grand Unified Theory it is possible to suppress couplings that
%contribute to the unphysical decay of the proton. 
F-theory \cite{Vafa:1996xn,Morrison:1996pp, Morrison:1996na} on elliptically
fibered Calabi-Yau manifolds has proven to be a successful framework to
realize supersymmetric non-abelian gauge theories, in particular Grand Unified
Theories (GUTs) \cite{Donagi:2008ca,Beasley:2008dc,Beasley:2008kw}. 
Although GUTs are an appealing framework for supersymmetric model
building\footnote{See \cite{Weigand:2010wm, Heckman:2010bq,
Maharana:2012tu} for some nice reviews of GUT model building in F-theory.}, it is well known
that they
can suffer from fast proton decay, which, however, can be obviated by having
additional discrete or continuous symmetries. 
In this paper we  consider
F-theory compactifications that give rise to GUTs with two additional $U(1)$s,
which can potentially be used to suppress certain proton decay
operators\footnote{Discrete symmetries have been studied 
  in local and global F-theory model building in, e.g.
  \cite{BerasaluceGonzalez:2011wy, %Camara:2011jg, 
%    BerasaluceGonzalez:2012vb,
    BerasaluceGonzalez:2012zn,
  Mayrhofer:2014haa, Garcia-Etxebarria:2014qua, Mayrhofer:2014laa}.}.  In
F-theory abelian gauge factors have their genesis in geometric properties of
the compactification manifold, namely in the existence of additional rational
sections of the elliptic fibration. We carry out a systematic procedure to
constrain which such fibrations can give rise to gauge groups $G \times
U(1)^2$.

%theories with two
%abelian factors in addition to a non-abelian component. 

 %one $U(1)$ is often still too
%constraining to be physically viable.
It has been known for many years that abelian gauge symmetries in F-theory are
characterized by the Mordell-Weil group of the elliptically fibered Calabi-Yau
compactification space \cite{Morrison:1996pp, Morrison:1996na}, which is the
group formed by the rational sections of the fibration. In recent years
abelian gauge factors have been much studied in the context of four
dimensional GUTs arising from F-theory compactifications. In local F-theory
models $U(1)$s have a realization in terms of factored spectral covers as shown in
\cite{Hayashi:2008ba, Donagi:2009ra, Marsano:2009gv, Marsano:2009wr, Dudas:2009hu,
Hayashi:2010zp, Dudas:2010zb,  Dolan:2011iu}.
%\cite{Marsano:2009wr, Donagi:2009ra, Marsano:2009gv, Dudas:2009hu,
%Dudas:2010zb, Cvetic:2010rq, Dolan:2011iu, Marsano:2012yc, Hayashi:2010zp}.
Global models with one $U(1)$ were studied in \cite{Grimm:2010ez,
Grimm:2011tb, Braun:2011zm, Grimm:2011fx, Krause:2012yh, Morrison:2012ei,Cvetic:2012xn,
Mayrhofer:2012zy, Braun:2013yti, Kuntzler:2014ila},
%\cite{Kuntzler:2014ila, Grimm:2010ez, Braun:2011zm, Krause:2012yh,
%Grimm:2011fx, Morrison:2012ei, Cvetic:2012xn, Mayrhofer:2012zy,
%Braun:2013yti, Esole:2014dea} 
however phenomenologically one $U(1)$ factor is not sufficient to forbid all dangerous
couplings \cite{Krippendorf:2014xba}. It is then well motivated to consider elliptic fibrations with multiple $U(1)$ factors,
the constuction of which was initiated in 
%
%Models with multiple abelian factors have been constructed  in
\cite{Borchmann:2013jwa, Cvetic:2013nia, Braun:2013nqa, Cvetic:2013uta, 
Borchmann:2013hta,  Cvetic:2013jta, Cvetic:2013qsa, Klevers:2014bqa,
Braun:2014qka},
%\cite{Braun:2013nqa, Cvetic:2013uta, Borchmann:2013jwa, Cvetic:2013nia,
%Borchmann:2013hta, Cvetic:2013jta, Cvetic:2013qsa, Klevers:2014bqa}. 
with the realization of the $SU(5)\times U(1)^2$ models 
%appearing 
in these papers 
primarily based on constructions from 
%relying mainly on 
toric tops \cite{Candelas:1996su}.

It is natural to ask whether there is a systematic way to explore the full
range of possible low-energy theories with two additional abelian gauge
factors which have an F-theory realization. One approach to address this
question is to apply Tate's algorithm \cite{MR0393039, Bershadsky:1996nh, Katz:2011qp} to elliptic fibrations with two additional rational sections. This
is the approach that we take in this paper and indeed we show that there is a
large class of new elliptic fibrations with phenomenologically interesting
properties not seen from the top constructions.  While Tate's algorithm is a comprehensive method to obtain the
form of any elliptic fibration with two rational sections there is a
caveat that it is sometimes difficult in practice to proceed with the
algorithm without making simplifications at the cost of generality.

%Abelian
%gauge factors in a gauge theory with reductive gauge group have many applications in modern
%field theory. A pertinent example is that, by imposing $U(1)$ charges on the
%matter fields of a Grand Unified Theory it is possible to prevent the
%existence of uncharged couplings that contribute to the decay of the proton.
%String theory, as a UV completion of gauge theory, provides a compelling
%origin of $U(1)$ symmetries. More specifically, in gauge theories that arise
%as a compactification from F-theory \cite{Vafa:1996xn} the abelian gauge
%factors have their genesis in geometric properties of the compactification
%manifold. In this paper we carry out a systematic procedure to constrain
%which such manifolds can give rise to gauge theories with two abelian factors
%in addition to a non-abelian component.

The starting point for the application of Tate's algorithm in this context is
the realization of the elliptic fiber as a cubic in $\mathbb{P}^2$
\cite{Borchmann:2013jwa, Cvetic:2013nia, Borchmann:2013hta, Cvetic:2013jta}. Tate's algorithm involves the study
of the discriminant of this cubic equation,
%
%the elliptic fibration where the fibers are realized as a cubic in
%$\mathbb{P}^2$ \cite{Borchmann:2013hta, Cvetic:2013nia}.  
%
%
%
%
%
%Using Tate's algorithm we determine conditions that specify the elliptic fibrations with $U(1)$ factors in different ways and reproduce the.
%
%Tate's algorithm, in this context, as applied to the Weierstrass form
%\begin{equation}
%  y^2 = x^3 + f x + g \,,
%\end{equation}
%involves a study of the discriminant
%\begin{equation}
%  \Delta = 4f^3 + 27 g ^2 \,.
%\end{equation}
which captures the information about the singularities
of the fiber. The singular fibers of an elliptic surface were classified
by Kodaira \cite{MR0132556, MR0184257} and N{\'e}ron \cite{MR0179172}, and
they belong to an $ADE$-type classification; Tate's algorithm is a
systematic procedure to determine the type of singular fiber. The $ADE$ type of
the singular fiber determines the non-abelian part of the gauge symmetry.
%is then the non-abelian gauge symmetry of the gauge theory
%arising from the compactification\footnote{One can also realize the
%  non-simply-laced gauge symmetry as a quotient of the $ADE$ type singularity
%  \cite{Aspinwall:1998xj}.}. 
Tate's algorithm was applied to the Weierstrass form for an elliptic fibration
where there are generically no $U(1)$s in
\cite{Bershadsky:1996nh,Katz:2011qp}, and in \cite{Kuntzler:2014ila} to the quartic equation in
$\mathbb{P}^{(1,1,2)}$ which realizes a single $U(1)$ \cite{Morrison:2012ei}. The application of the algorithm to the cubic in
$\mathbb{P}^2$ will constrain the form of 
%
%  Instead of using the Weierstrass equation, which
%  was done for an elliptic fibration without the assumption of any additional
%  sections in \cite{Bershadsky:1996nh,Katz:2011qp}, or the quartic equation in
%  $\mathbb{P}^{(1,1,2)}$ used for one additional
%  section in \cite{Kuntzler:2014ila}, a cubic equation equivalent to the
%  specialized form of the Weierstrass equation that one gets when one demands
%  the existence of two additional rational sections was considered. The
%  algorithm then constrains the form of the 
the fibrations which realize a $G
  \times U(1)^2$ symmetry, for some non-abelian gauge group $G$,
  which
  are phenomenologically interesting for model building.

As a result of Tate's algorithm we find a collection of
elliptic fibrations which realize the gauge symmetry $SU(5)\times U(1)^2$
where the non-abelian symmetry is the minimal simple Lie group
containing the Standard Model gauge group. 
%Models with an $SU(5)$ symmetry
%have received much attention in F-theory, originating in part with
%\cite{Blumenhagen:2009yv, Marsano:2009wr, Grimm:2009yu, Knapp:2011ip}.
The fibrations found
encompass all of the $SU(5)$  models with two $U(1)$s in the literature which we are
aware of, and
includes previously unknown models which, in many cases, have exciting
phenomenological features, such as having multiple, differently charged, ${\bf
10}$ matter curves. We also determine fibrations that lead to $E_6$ and $SO(10)$ 
gauge groups with two $U(1)$s.

%The utility of elliptic fibrations with two $U(1)$s is not restricted to GUT model
%building, 

Our results are not restricted to F-theoretic GUT model building, and we hope
that they are also useful in other areas of F-theory, for example
%present results would also be useful in F-theory that is not
%GUT model building, for example 
in direct constructions of the Standard Model
\cite{Lin:2014qga, Grassi:2014zxa}, 
%as arising from multisections in genus one fibrations \cite{Morrison:2014era}, 
in the determination of the network of resolutions
of elliptic fibrations \cite{Hayashi:2013lra,
Hayashi:2014kca, Esole:2014bka, Braun:2014kla, Esole:2014hya}, or in the
recent relationship drawn between elliptic fibrations with $U(1)$s and genus
one fibrations with multisections \cite{Morrison:2014era, Braun:2014oya,
Anderson:2014yva}.

%Recently
%elliptic fibrations with $U(1)$s have been related to genus one fibrations
%with multisections \cite{Morrison:2014era}

%Outside of GUT model building, $U(1)$ factors have recently been used in the
%direct construction of the Standard Model from F-theory \cite{Lin:2014qga}. In
%\cite{Morrison:2014era} elliptic fibrations with $U(1)$ factors are related to genus one fibrations with
%multisections.
%The importance of abelian factors in determining the network of resolutions of
%elliptic fibrations has been studied in 
%$U(1)$ factors also contribute to the higher codimension resolution structure
%of elliptic fibrations as studied in 
%\cite{Hayashi:2013lra,
%Hayashi:2014kca, Esole:2014bka, Braun:2014kla, Esole:2014hya}.

%Recently $U(1)$ factors have been used in the direct construction of the
%Standard Model in \cite{Lin:2014qga}, and in \cite{Morrison:2014era} they have
%been studied in the context of genus one fibrations, where the existence of a
%section is not guaranteed.
%One can
%consider engineering the Standard Model directly where the $U(1)$ factors play
%the role of the $U(1)_Y$ as was recently done in \cite{Lin:2014qga}.  Recently
%genus one fibrations have been studied, where the existence of a section is
%not guaranteed \cite{Morrison:2014era}. 
%The presence of $U(1)$ factors also plays an important
%role in determining which distinct resolutions of the non-abelian singularity
%are possible \cite{Hayashi:2013lra, Hayashi:2014kca,
%Esole:2014bka, Braun:2014kla, Esole:2014hya}. 

In section \ref{sec:Overview} we present a summary where we highlight the
fibrations found in the application of Tate's algorithm to the cubic equation,
up to fibers realizing $SU(5)$. We also present a table of a particularly nice
kind of realizations for Kodaira fibers $I_n$ and $I_n^*$. In
section \ref{sec:Prelude} we recap the embedding of the elliptic fibration as
a cubic hypersurface in a $\mathbb{P}^2$ fibration and give details of the
resolution and intersection procedures. Section \ref{sec:TateAlg} contains
Tate's algorithm proper, up to the $I_5$, or $SU(5)$, singular
fibers. In section \ref{sec:Spectra} the $U(1)$ charges of the
various ${\bf 10}$ and ${\bf 5}$ matter curves that appear in the models from
the $SU(5)$ singular fibers are determined. 
In section \ref{sec:Exceptional}
Tate's algorithm is continued from where it was left off in section \ref{sec:TateAlg}
and we obtain fibrations that have a non-abelian component
corresponding to an exceptional Lie algebra.

%Finally in section \ref{sec:Exceptional}
%Tate's algorithm is continued from where it was left off in section \ref{sec:TateAlg}
%and realizations of elliptic fibrations that have a non-abelian component
%corresponding to an exceptional Lie algebra are obtained.

\section{Overview and Summary}\label{sec:Overview}
For the reader's
convenience, the key results are summarized in this section. For those
interested simply in the new $SU(5)$ models we refer to section
\ref{sec:Spectra}.

An elliptic fibration with two additional rational sections,
which gives rise to a gauge theory with two additional $U(1)$s, can be realized 
as a hypersurface in a $\mathbb{P}^2$ fibration, as in
\cite{Borchmann:2013jwa, Cvetic:2013nia, Borchmann:2013hta, Cvetic:2013jta}, given by the equation
\begin{equation} \label{eqn:P111cubic}
  \fks_1 w^3 + \fks_2 w^2 x + \fks_3 w x^2 + \fks_5 w^2 y + \fks_6 w x y + 
  \fks_7 x^2 y + \fks_8 w y^2 + \fks_9 x y ^2 = 0 \,,
\end{equation}
where $[w:x:y]$ are  projective
coordinates on the $\mathbb{P}^2$.
This fibration has three sections which have projective coordinates
\begin{equation}
  \Sigma_0: [0: 0: 1 ] \,, \quad \Sigma_2: [ 0: 1: 0 ] \,, \quad \Sigma_1:[ 0:
    \fks_9: -\fks_7 ] \,.
\end{equation}
The application of Tate's algorithm involves enhancing the singularity of this
elliptic fibration, where the particular enhancements are determined
by the discriminant. 
As the coefficients of the fibration are sections of holomorphic line bundles
over the base, one can look at an open neighbourhood around the singular locus
in the base with coordinate $z$ such that the singular locus is above $z = 0$,  
and consider the expansion in the coordinate $z$ of the $\fks_i$
\begin{equation}
  \fks_i = \sum_{j=0}^{\infty} s_{i,j} z^j \,.
\end{equation}
Often the pertinent information from the equation (\ref{eqn:P111cubic}) is
just the vanishing orders of the $\fks_i$ in $z$, which we will refer to
through
\begin{equation}
  n_i = \text{ord}_z(\fks_i) \,.
\end{equation}
A shorthand for the equation 
will be the tuple of positive integers $(n_1, n_2, n_3, n_5, n_6, n_7, n_8, n_9)$
representing the vanishing orders. It will not always be possible to express a
fibration just through a set of vanishing orders, but there will also be 
non-trivial relations among the coefficients of the equation. We will refer
to fibrations of this form as {\it non-canonical models}.  This will be the result of
solving in full generality the polynomials which appear in the discriminant as
a necessary condition for enhancing the singular fiber. In particular the fact
that the coefficients of our fibration belong to a unique factorization domain
\cite{MR0103906, Katz:2011qp} will be used. Schematically we will refer to these fibrations via the
shorthand notation
\begin{equation}
I_{nc^i}: \quad \left\{
  \begin{array}{c}
  (n_1, n_2, n_3, n_5, n_6, n_7, n_8, n_9) \cr
  [s_{1,n_1}, s_{2,n_2}, s_{3,n_3}, s_{5,n_5}, s_{6,n_6}, s_{7,n_7},
    s_{8,n_8}, s_{9,n_9}] \,
  \end{array} \right\} \,,
\end{equation}
where the term in square brackets denotes any specialization of the leading
non-vanishing coefficients in the expansion of the $\fks_i$, and the $I$
represents the Kodaira fiber type. Often, for ease
of reading, a dash will be inserted to indicate that a particular coefficient
is unspecialized. The exponent of the index $nc$ will signal how many
non-canonical enhancements of the discriminant were used in order to obtain
the singular fiber, that is, how many times solving a polynomial in the
discriminant did not require just setting some of the expansion coefficients to zero, but
also some additional cancellation.

There is a last piece of notation that needs to be explained before the
results can be presented. 
Since the elliptic fibration has three sections, it will be seen in section
\ref{sec:TateAlg}, where the algorithm is studied in detail, that the
discriminant will reflect the fact that the sections can intersect the
components of the resolved fiber in multiple different ways. Thus, 
a number of (non-)canonical forms for each Kodaira singular fiber will be
obtained
depending on which fiber component each of the sections intersects. To
represent this, denote by $I_{n}^{(012)}$ the case where all the three
sections intersect the same fiber component, and then introduce 
separation of the sections by means of the notation $I_n^{(0|^n 1 |^m 2)}$,
where the number of slashes will signal the distance between the fiber
components that the corresponding sections intersect. Consider the two
examples:
\begin{itemize} 
  \item $I_{4, nc^2}^{s(01|2)}$ will represent a Kodaira singular fiber $I_4$,
      obtained through two non-canonical enhancements of the discriminant. The
      sections $\Sigma_0$, $\Sigma_2$ will intersect one of the fiber
      components, while $\Sigma_1$ will sit on an adjacent fiber component
      (i.e. one which intersects the previous component). Depicting the
      $\mathbb{P}^1$ components of the singular fiber as lines, and the
      sections as nodes, the fiber $I_4^{s(01|2)}$ can be represented by the diagram 
      {\begin{center}\includegraphics[scale=0.2]{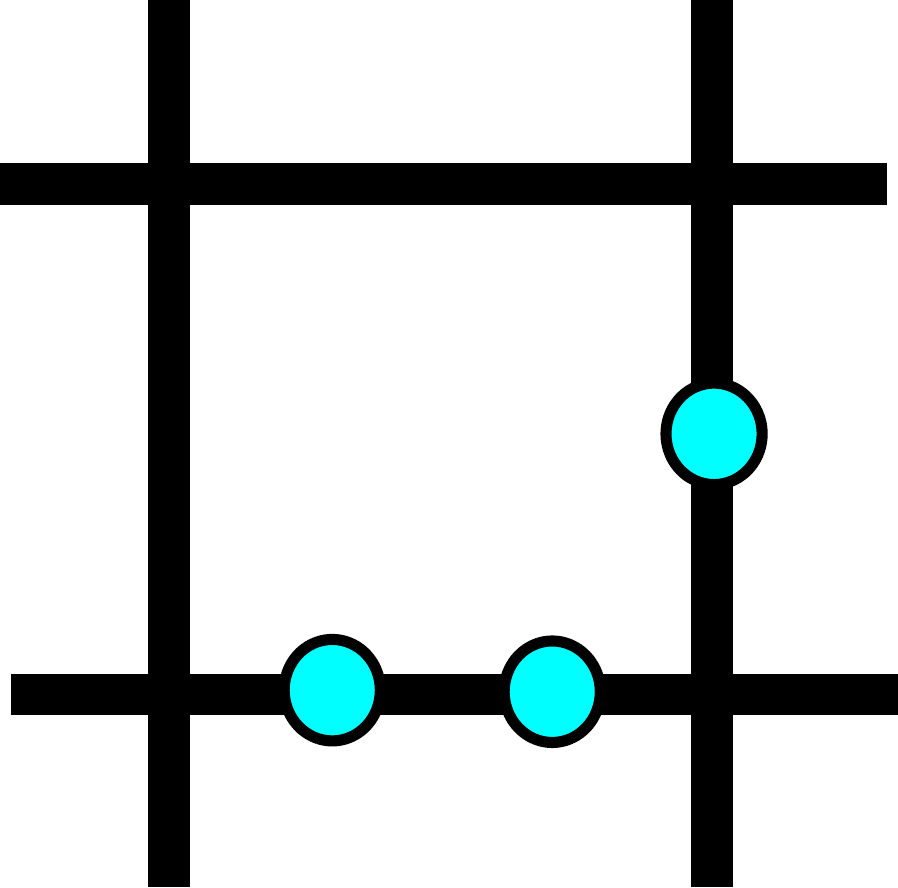}\end{center}}
  \item $I_{5, nc^3}^{s(01||2)}$ will represent an
        $I_5$ found upon imposing  non-canonical conditions on the
        coefficients of our equation three times, such that the fiber
        component intersected by $\Sigma_0$ and $\Sigma_1$ does not intersect
        the component that $\Sigma_2$ intersects. This
        $I_5^{s(01||2)}$ is
        represented pictorially as
      {\begin{center}\includegraphics[scale=0.2]{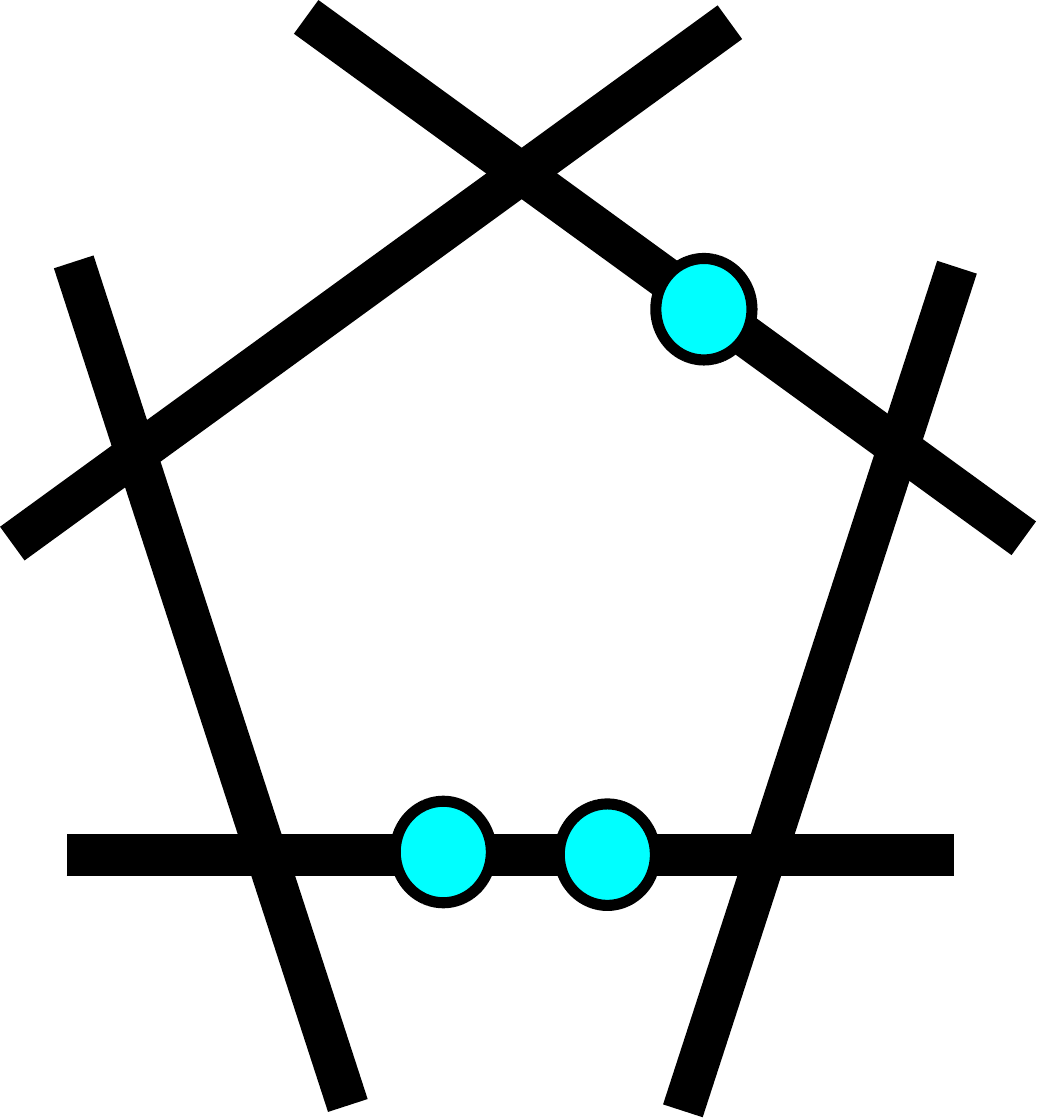}\end{center}}
\end{itemize} 
We refer to section \ref{sec:Exceptional} for more details about the 
notation for representing singular fibers corresponding to other types of
Kodaira singular fibers.

 All of the fibers found and determined are presented in the following summary
 tables, where the fibers are grouped first by the Kodaira type and then by
 the degree of canonicality:
\begin{itemize}
\item In table \ref{TableNonCanonicalord3} we list the singular fibers up to
  vanishing order $\ord_z(\Delta) = 3$. These include fibers of type
  $I_1$,$I_2$,$I_3$, $II$, and $III$.
\item In table \ref{TableNonCanonicalord4} we list the singular fibers at
  vanishing order $\ord_z(\Delta) = 4$. These include both type $I_4$ and type $IV$ Kodaira fibers.
\item In table \ref{TableNonCanonicalord5} we list the $I_5$ singular fibers.
\end{itemize}

For each of the $I_5$ singular fibers obtained through the algorithm 
the $U(1)$ charges are calculated and the results are presented in section
\ref{sec:Spectra}, along with the comparison with the $U(1)$ charges of the
known $SU(5)$ toric tops \cite{Candelas:1996su, Borchmann:2013hta, Mayrhofer:2012zy,
Borchmann:2013jwa, Braun:2013nqa}. 

Tate-like (that is, canonical) forms for generic Kodaira
singular fibers were also determined and they are presented in table
\ref{TableTateForms}. Appendix \ref{Forms for General Fibers} includes
explicit details of the resolutions of these forms.

\begin{table}[H]
  \centering{
  \begin{tabular}{|c|c|}\hline
    Singular Fiber & Vanishing Orders and Non-canonical Data \cr\hline
     $I_1^{(012)}$ & $(1,1,0,1,0,0,0,0)$  \cr\hline
  $I_2^{(012)}$ & $(2,1,0,1,0,0,0,0)$  \cr\hline
    $I_2^{(01|2)}$ & $(1,1,1,0,0,0,0,0)$  \cr\hline
    $I_{2,nc}^{(1|02)}$ & 
        $\begin{array}{c}
        (1,1,0,1,0,0,0,0) \cr 
         [-,-,\sigma_2\sigma_5,-,\sigma_2\sigma_4 +
      \sigma_3\sigma_5,\sigma_1\sigma_2,\sigma_3\sigma_4,\sigma_1\sigma_3]
       \end{array}$
       \cr
      \hline
    $II_{nc}^{(012)}$ & $\begin{array}{c}(1,1,0,1,0,0,0,0) \cr
      [-,-,\mu\sigma_1^2,-,2\mu\sigma_3\sigma_8,-,\mu\sigma_8^2,-]\end{array}$ \cr \hline
$I_3^{ns(012)}$ & $(3,2,0,2,0,0,0,0)$   \cr     \hline
    $I_3^{s(01|2)}$ & $(2,1,1,1,0,0,0,0)$   \cr\hline
  $I_3^{s(0|1|2)}$ & $    (1,1,1,1,0,0,1,0)$  \cr\hline
  
   $I_{3,nc}^{s(012)}$& $\begin{array}{c} (3,1,0,1,0,0,0,0)
    \cr [-,\sigma_1 \sigma_2,\sigma_2\sigma_5,\sigma_1\sigma_3, \sigma_2\sigma_4+\sigma_3 \sigma_5,-,\sigma_3 \sigma_4,-]\end{array}$   \cr\hline 
    $I_{3,nc}^{s(01|2)}$ & $\begin{array}{c} (2,1,1,0,0,0,0,0)
    \cr [-,\sigma_1 \sigma_2,\sigma_1\sigma_3,\sigma_2\sigma_5, \sigma_2\sigma_4+\sigma_3 \sigma_5,\sigma_3 \sigma_4,-,-]\end{array}$   \cr\hline    
    $I_{3,nc}^{s(02|1)}$ & $\begin{array}{c}(2,1,0,1,0,0,0,0)\cr
    [-,-,\sigma_2\sigma_5,-,\sigma_2\sigma_4 +
      \sigma_3\sigma_5,\sigma_1\sigma_2,\sigma_3\sigma_4,\sigma_1\sigma_3]\end{array}$ \cr\hline    
    
    $I_{3,nc}^{s(0|1|2)}$ & $\begin{array}{c}
    
    (1,1,1,0,0,0,0,0)\cr
    [-,-,-,\sigma_2\sigma_5,\sigma_2\sigma_4 +
      \sigma_3\sigma_5,\sigma_3\sigma_4,\sigma_1\sigma_2,\sigma_1\sigma_3]\end{array}$ \cr\hline
       
%    $I_{3,nc}^{s(012)}$ & $\begin{array}{c} (3,2,0,2,0,0,0,0)
%    \cr [-,-,\sigma_1\sigma_3,-,\sigma_1 \sigma_2+\sigma_3 \sigma_4,-,\sigma_2 \sigma_4,-]\end{array}$   \cr\hline    

  $III_{nc}^{(012)}$ & $\begin{array}{c}(2,1,0,1,0,0,0,0)\cr 
    [-,-,\mu\sigma_3^2,-,2\mu\sigma_3\sigma_8,-,\mu\sigma_8^2,-]\end{array}$ \cr\hline

    $III_{nc}^{(01|2)}$ & $\begin{array}{c}(1,1,1,0,0,0,0,0)\cr[-,-,-,\mu\sigma_5^2,2\mu\sigma_5\sigma_7,\mu\sigma_7^2,-,-]\end{array}$ \cr\hline

%$III_{nc^2}^{(1|02)}$ & $\begin{array}{c} (1,1,0,1,0,0,0,0)\cr
%    [-,-,\mu\xi_1^2,-,2\mu\xi_1\xi_4,\xi_1\xi_2,\mu\xi_4^2,\xi_2\xi_4]\end{array}$ \cr\hline
$III_{nc^2}^{(02|1)}$ & $\begin{array}{c} (1,1,0,1,0,0,0,0) \cr [-,-,\xi_2^2\xi_4,-,2\xi_2\xi_3\xi_4,\sigma_1\xi_2,\xi_3^2\xi_4,\sigma_1\xi_3]\end{array}$ \cr\hline
  \end{tabular}}
  \caption{Singular fibers where $\text{ord}_z(\Delta) \leq 3$.}
\label{TableNonCanonicalord3}
\end{table}

    \begin{table}[H]
  \centering
  \begin{tabular}{|c|c|}\hline
    Singular Fiber & Vanishing Orders and Non-canonical Data \cr\hline
    $I_4^{ns(012)}$ & $(4,2,0,2,0,0,0,0)$   \cr\hline
 $I_4^{s(01|2)}$ & $(3,2,1,1,0,0,0,0)$  \cr\hline
    $I_4^{ns(01||2)}$ & $(2,2,2,0,0,0,0,0)$   \cr\hline
   
$I_4^{s(0|1|2)}$ & $(2,1,1,1,0,0,1,0)$  \cr\hline
    
    %$I_4^{s(01||2)}$ & $(2,2,2,1,0,0,0,0)$  \cr\hline
    
    $I_{4,nc}^{s(012)}$ & $\begin{array}{c} (4,2,0,2,0,0,0,0)\cr
    [-,-, \sigma_1 \sigma_3,-,\sigma_1 \sigma_2 +\sigma_3 \sigma_4,-,\sigma_2 \sigma_4,-]\end{array}$   \cr\hline
     $I_{4,nc}^{s(01|2)}$ & $\begin{array}{c} (2,1,1,1,0,0,0,0)\cr [\sigma_3\sigma_4,\sigma_1\sigma_3,-,\sigma_2\sigma_4 +
      \sigma_3\sigma_5,\sigma_1\sigma_2,-,\sigma_2\sigma_5,- ]\end{array}$ \cr\hline
$I_{4,nc}^{s(02|1)}$ & $\begin{array}{c} (3,2,0,2,0,0,0,0)\cr 
    [-,-,\sigma_2 \sigma_5,-,\sigma_2 \sigma_4+\sigma_3 \sigma_5,\sigma_1 \sigma_2,\sigma_3\sigma_4,\sigma_1\sigma_3]\end{array}$ \cr\hline
$I_{4,nc}^{s(01||2)}$ & $\begin{array}{c} (2,2,2,0,0,0,0,0)\cr
    [-,-,-, \sigma_1 \sigma_3,\sigma_1 \sigma_2 +\sigma_3 \sigma_4,\sigma_2 \sigma_4,-,-]\end{array}$   \cr\hline 
   
    $I_{4,nc}^{s(01||2)}$ & $\begin{array}{c}(2,1,1,1,0,0,0,0)\cr
    [-,\sigma_1\sigma_2,\sigma_1\sigma_3,-,\sigma_2\sigma_4,\sigma_3\sigma_4,-,-]\end{array}$ \cr\hline
      
      $I_{4,nc}^{s(1|0|2)}$ & $\begin{array}{c} (1,1,1,1,0,0,1,0)\cr
    [\sigma_2\sigma_5,\sigma_2\sigma_4 + \sigma_3\sigma_5,
      \sigma_3\sigma_4,-,\sigma_1\sigma_2,\sigma_1\sigma_3,-,- ]\end{array}$ \cr\hline
    
  $I_{4,nc^2}^{s(02|1)}$ & $\begin{array}{c}(3,2,0,2,0,0,0,0)\cr
    [-,-,\sigma_3\xi_1\xi_2,-,\sigma_2\xi_1\xi_2+\sigma_3\xi_1\xi_3,\xi_2\xi_4,\xi_1\xi_3\sigma_2,\xi_3\xi_4]\end{array}$\cr \hline   
   
$I_{4,nc^2}^{s(0|1|2)}$ & $\begin{array}{c}(1,1,1,0,0,0,0,0)\cr
    [\xi_3\xi_4,\xi_2\xi_4 +
      \xi_3\xi_5,\xi_2\xi_5,\xi_3\sigma_5,\xi_3\sigma_4 +
      \xi_2\sigma_5,\xi_3\sigma_4,\sigma_1\xi_3,\sigma_1\xi_2]\end{array}$ 
     \cr\hline

 $I_{4,nc^2}^{s(1|0|2)}$ &
  $\begin{array}{c}
    (1,1,1,0,0,0,0,0) \cr
    [\xi_3\xi_4, \xi_2\xi_4 + \xi_3\xi_5, \xi_2\xi_5, \sigma_2\xi_1\xi_3,
      \sigma_2\xi_1\xi_3, \sigma_2\xi_1\xi_2 +
      \sigma_3\xi_1\xi_3,\sigma_3\xi_1\xi_2, \sigma_1\sigma_2,
      \sigma_1\sigma_3]
    \end{array}$
\cr\hline

%$I_{4,nc^2}^{s(1|02)}$ & $\begin{array}{c}(3,2,0,2,0,0,0,0)\cr
%    [-,-,\sigma_1\xi_1\xi_3,-,\sigma_1\xi_1\xi_2+\sigma_4\xi_1\xi_3,\xi_3\xi_4,\xi_1\xi_2\sigma_4,\xi_2\xi_4]\end{array}$\cr \hline

%$IV_{nc}^{ns(012)}$ & $\begin{array}{c} (2,2,0,2,0,0,0,0)\cr
%          [-,-,\mu \sigma_3^2,-,2\mu\sigma_3\sigma_8,-,\mu %\sigma_8^2,-]\end{array}$ \cr\hline
$IV^{s(01|2)}$ & $(2,1,1,1,1,0,0,0)$  \cr     \hline
        
      $IV^{s(0|1|2)}$ & $(1,1,1,1,1,0,1,0)$  \cr\hline
$IV_{nc^2}^{ns(012)}$ & $\begin{array}{c} (2,1,0,1,0,0,0,0)\cr
    [-,\xi_1 \xi_3,\mu \xi_3^2,\xi_1\xi_2,2\mu\xi_2\xi_3,-,\mu \xi_2^2,-]\end{array}$ \cr\hline
$IV_{nc^2}^{s(01|2)}$ & $\begin{array}{c} (1,1,1,0,0,0,0,0) \cr [\xi_2\xi_5,\xi_2\xi_4 + \xi_3\xi_5,\xi_3\xi_4,\mu\xi_2^2,2\mu\xi_2\xi_3,\mu\xi_3^2,-,-]\end{array}$ \cr\hline
$IV_{nc^2}^{s(02|1)}$ & $\begin{array}{c} (2,1,0,1,0,0,0,0)\cr
    [-,-, \xi_4 \xi_2^2,-,2\xi_2\xi_3\xi_4,\sigma_1  \xi_2, \xi_4\xi_3^2,\sigma_1  \xi_3]\end{array}$ \cr\hline

$IV_{nc^2}^{s(0|1|2)}$ & $\begin{array}{c} (1,1,1,0,0,0,0,0)\cr
    [-,-,-,\mu\xi_2^2,2\mu\xi_2\xi_3,\mu\xi_3^2,\xi_2\xi_4,\xi_3\xi_4]\end{array}$ \cr\hline
%$IV_{nc^2}^{s(0|1|2)}$ & $\begin{array}{c} (1,1,1,0,0,0,0,0)\cr
%    [-,-,-,\xi_2^2\xi_4,2\xi_2\xi_3\xi_4,\xi_3^2\xi_4,\sigma_1\xi_2,\sigma_1\xi_3]\end{array}$ \cr\hline

%$IV_{nc^2}^{s(1|02))}$ & $\begin{array}{c} (2,1,0,1,0,0,0,0)\cr
%    [-,-,\mu\xi_3^2,-,2\mu\xi_2\xi_3,\xi_1 \xi_3,\mu \xi_2^2, \xi_1 \xi_2]\end{array}$ \cr\hline

      $IV_{nc^3}^{s(012)}$ & $\begin{array}{c} (2,1,0,1,0,0,0,0)\cr
    [\delta_2\delta_4,\xi_3(\delta_1 \delta_2+\delta_3 \delta_4),\delta_1 \delta_3\xi_3^2,\xi_2(\delta_1\delta_2+\delta_3\delta_4),2\delta_1\delta_3\xi_2\xi_3,-,\delta_1\delta_3\xi_2^2,-]\end{array}$ \cr\hline
     \end{tabular}
     \caption{Singular fibers where $\text{ord}_z(\Delta) = 4$.}
\label{TableNonCanonicalord4}     
\end{table}

\begin{table}[H]{\small
  \begin{equation}\notag
    \begin{array}{|c|c|}\hline
      \text{Singular Fiber } & \text{Vanishing Orders and Non-canonical Data}
      \cr\hline
I_{5}^{ns(012)} &(5,3,0,3,0,0,0,0) \cr\hline
I_{5}^{s(01|2)} & (4,2,1,2,0,0,0,0) \cr\hline

   I_{5}^{s(01||2)} & (3,2,2,1,0,0,0,0) \cr\hline
             I_{5}^{s(0|1|2)} &(3,2,1,1,0,0,1,0) \cr\hline
I_{5}^{s(0|1||2)} &(2,2,2,1,0,0,1,0) \cr\hline
I_{5,nc}^{s(012)} & 
      \begin{array}{c} 
        (5,2,0,2,0,0,0,0) \cr
        [-,\sigma_1 \sigma_2,\sigma_2\sigma_5,\sigma_1\sigma_3, \sigma_2\sigma_4+\sigma_3 \sigma_5,-,\sigma_3 \sigma_4,-]
      \end{array} \cr\hline            
   
 I_{5,nc}^{s(01|2)} & 
      \begin{array}{c} 
        (3,2,1,1,0,0,0,0) \cr
        [\sigma_2\sigma_5,\sigma_2\sigma_4+\sigma_3\sigma_5,\sigma_3\sigma_4,\sigma_1\sigma_2,\sigma_1\sigma_3,-,-,-]
      \end{array} \cr\hline  
      I_{5,nc}^{s(02|1)} & 
      \begin{array}{c} 
        (4,2,0,2,0,0,0,0) \cr
        [-,-,\sigma_3\sigma_4,-,\sigma_2\sigma_4+\sigma_3\sigma_5,\sigma_1\sigma_3,\sigma_2\sigma_5,\sigma_1\sigma_2]
      \end{array} \cr\hline       
% I_{5,nc}^{s(012)} & 
%      \begin{array}{c} 
%      (5,3,0,3,0,0,0,0) \cr
%        [-,-,\sigma_1\sigma_3,-,\sigma_2\sigma_4 +
%          \sigma_3\sigma_5,-,\sigma_2\sigma_4,-]
%      \end{array} \cr\hline
              
      I_{5,nc}^{s(0|1|2)} & 
      \begin{array}{c} 
        (2,1,1,1,0,0,1,0) \cr
        [\sigma_1\sigma_3,\sigma_1\sigma_2,-,\sigma_3\sigma_4,\sigma_2\sigma_4,-,-,-]
      \end{array} \cr\hline
      I_{5,nc}^{s(1|0|2)} & 
      \begin{array}{c} 
        (3,2,1,1,0,0,0,0) \cr
        [-,-,-,-,\sigma_1\sigma_2,\sigma_1\sigma_3,\sigma_2\sigma_4,\sigma_3\sigma_4]
      \end{array} \cr\hline
      I_{5,nc}^{s(0|1||2)} & 
      \begin{array}{c} 
        (2,2,2,0,0,0,0,0) \cr
        [-,-,-,\sigma_2\sigma_5,\sigma_2\sigma_4 +
          \sigma_3\sigma_5,\sigma_3\sigma_4,\sigma_1\sigma_2,\sigma_1\sigma_3]
      \end{array} \cr\hline
     
      I_{5,nc}^{s(0|1||2)} & 
      \begin{array}{c} 
        (2,1,1,1,0,0,1,0) \cr
        [-,\sigma_1\sigma_2,\sigma_1\sigma_3,-,\sigma_2\sigma_4,\sigma_3\sigma_4,-,-]
      \end{array} \cr\hline

   I_{5,nc^2}^{s(02|1)} & 
      \begin{array}{c} 
        (4,2,0,2,0,0,0,0) \cr
        [-,-,\sigma_3\xi_2,-,\sigma_2\xi_2+\sigma_3\xi_3,\xi_2\xi_4,\sigma_2\xi_3,\xi_3\xi_4]
      \end{array} \cr\hline   
   
   I_{5,nc^2}^{s(01||2)} & 
      \begin{array}{c} 
        (2,1,1,1,0,0,0,0) \cr
        [\xi_3\xi_4,\sigma_2\xi_3,\sigma_3\xi_3,\xi_2\xi_4+\xi_3\xi_5,\sigma_2\xi_2,\sigma_3\xi_2,\xi_2\xi_5,-]
      \end{array} \cr\hline   
   
   I_{5,nc^2}^{s(01||2)} & 
          \begin{array}{c} 
            (2,2,2,0,0,0,0,0) \cr
            [\xi_3\xi_4,\xi_2 \xi_4+\xi_3 \xi_5,\xi_2\xi_5,\sigma_3\xi_3,\sigma_2\xi_3+\sigma_3\xi_2,\sigma_2 \xi_2,-,-]
          \end{array} \cr\hline
       I_{5,nc^2}^{s(1|0|2)} & 
      \begin{array}{c} 
        (2,1,1,1,0,0,0,0) \cr
        [\sigma_3\sigma_4,\sigma_3\xi_1\xi_3,-,\sigma_2\sigma_4+\sigma_3\xi_1\xi_2,\sigma_2\xi_1\xi_3,\xi_3\xi_4,\sigma_2\xi_1\xi_2,\xi_2\xi_4]
      \end{array} \cr\hline   
    I_{5,nc^2}^{s(0|1||2)} & 
      \begin{array}{c} 
        (2,1,1,1,0,0,0,0) \cr
        [-,\sigma_1\xi_3,\sigma_1\xi_2,-,\sigma_4\xi_3,\sigma_4\xi_2,\xi_3\xi_4,\xi_2\xi_4]
      \end{array} \cr\hline       

I_{5,nc^2}^{s(0|2||1)} & 
      \begin{array}{c} 
        (1,1,1,1,0,0,1,0) \cr
        [\xi_3\xi_4\xi_5\xi_6,\sigma_4\xi_5\xi_6+\sigma_3 \xi_3 \xi_4,\sigma_3\sigma_4,\xi_3\xi_5\xi_7+\xi_4\xi_6\xi_8,\xi_1\xi_3\xi_5\xi_6,\sigma_3\xi_1\xi_3,\xi_7\xi_8,\xi_1\xi_6\xi_8]
      \end{array} \cr\hline

%I_{5,nc^2}^{s(0|1||2)} & 
%      \begin{array}{c} 
%        (2,1,1,1,0,0,0,0) \cr
%        [-,\xi_3\xi_4,\xi_2\xi_4,-,\sigma_1\xi_3,\sigma_1\xi_2,\sigma_4\xi_3,\sigma_4\xi_2]
%      \end{array} \cr\hline    
%I_{5,nc^2}^{s(1|0|2)} & 
%      \begin{array}{c} 
%        (2,1,1,1,0,0,0,0) \cr
%        [\xi_7\xi_8,\xi_1\xi_6\xi_8,-,\xi_5\xi_7+\xi_4\xi_6\xi_8,\xi_1\xi_5\xi_6,\sigma_3\xi_1,\xi_4\xi_5\xi_6,\sigma_3\xi_4]
%      \end{array} \cr\hline 
%I_{5,nc^2}^{s(01||2)} & 
%      \begin{array}{c} 
%        (2,1,1,1,0,0,0,0) \cr
%        [\sigma_4\xi_3,\sigma_1\xi_3,\xi_3\xi_4,\sigma_4\xi_2+\sigma_5\xi_3,\sigma_1\xi_2,\xi_2\xi_4,\sigma_5\xi_2,-]
%      \end{array} \cr\hline

% I_{5,nc^2}^{s(1|02)} & 
%      \begin{array}{c} 
%        (4,2,0,2,0,0,0,0) \cr
%        [-,-,\sigma_1\xi_3,-,\sigma_1\xi_2+\sigma_4\xi_3,\xi_3\xi_4,\sigma_4\xi_2,\xi_2\xi_4]
%      \end{array} \cr\hline        

I_{5,nc^3}^{s(0|1||2)} & 
      \begin{array}{c} 
        (1,1,1,0,0,0,0,0) \cr
        [\xi_3\delta_3\delta_4,\delta_4(\delta_3\xi_2+\delta_2\xi_3),\xi_2 \delta_2\delta_4,\xi_3\delta_1\delta_3,\delta_1(\delta_2\xi_3+\delta_3 \xi_2),\delta_1\delta_2\xi_2,\sigma_1\xi_3,\sigma_1\xi_2]
      \end{array} \cr \hline
%I_{5,nc^3}^{s(0|1||2)} & 
%      \begin{array}{c} 
%        (1,1,1,0,0,0,0,0) \cr
%        [\xi_3\delta_2\delta_4,\delta_4(\delta_2\xi_2+\delta_3\xi_3),\xi_2 \delta_3\delta_4,\xi_1\xi_3\delta_1\delta_2,\delta_1\xi_1(\delta_2\xi_2+\delta_3 \xi_3),\delta_1\delta_3\xi_1\xi_2,\sigma_1\delta_1\delta_2,\sigma_1\delta_1\delta_3]
%      \end{array} \cr\hline       
\end{array}       
       \end{equation}}
    \caption{Singular fibers where $\text{ord}_z(\Delta) = 5$.}
       \label{TableNonCanonicalord5}

\end{table}

\begin{sidewaystable}
\centering
\begin{tabular}{|c|c|*{8}{c|}|c|c|}\hline
 &  & \multicolumn{8}{|c}{Vanishing Orders of the Coefficients}  &  &  \\ 
Fiber Type &Gauge Group &$\fks_1$&$\fks_2$&$\fks_3$&$\fks_5$&$\fks_6$&$\fks_7$&$\fks_8$&$\fks_9$&$\mathcal{O}(\Delta)$ & Restrictions
\\\hline \hline
      $I_{2k+1}^{s(0|^n1|^m2)}$ &$SU(2k+1)$ &$ 2k+1-(n+m)$&$ k-n$&$ m$&$ k+1-m$&$0$&$ 0$&$ n$&$0$ 
        &$2k+1$ & $  m + n \leq k $
      \\
$I_{2k+1}^{s(0|^n1|^m2)}$ &$SU(2k+1)$ &$ 2k+1-(n+m)$&$ m$&$ m$&$ n$&$0$&$ 0$&$ n$&$0$ 
        &$2k+1$ &$ m+n\leq \left \lfloor{ \frac{2}{3}(2k+1)}\right \rfloor$     
      \\
$I_{2k}^{s(0|^n1|^m2)}$ &$SU(2k)$ &$ 2k-(n+m)$&$ k-n$&$ m$&$ k-m$&$0$&$ 0$&$ n$&$0$ 
        & $2k$&$ m+n\leq k,\quad  m<k$     

\\
$I_{2k}^{s(0|^n1|^m2)}$ &$SU(2k)$ &$ 2k-(n+m)$&$ m$&$ m$&$ n$&$0$&$ 0$&$ n$&$0$ 
        & $2k$&$ m+n\leq \left \lfloor{\frac{4}{3}k}\right \rfloor$

      \\

$I_{2k}^{ns(01|^n2)}$ &$k$ &$ k$&$ k$&$ 0$&$ 0$&$0$&$ 0$&$ 0$&$0$ 
        & $2k$ &   
      \\

$I_{2k+1}^{ns(012)}$ &$-$ &$ 2k+1$&$ k+1$&$ 0$&$ k+1$&$0$&$ 0$&$ 0$&$0$ 
        &$2k+1$ &   
\\

$I_{2k}^{ns(012)}$ &$Sp(k)$ &$ 2k$&$ k$&$ 0$&$ k$&$0$&$ 0$&$ 0$&$0$ 
        &$2k$ &   
        \\

$I_{2k+1}^{*s(0|1||2)}$ &$SO(4k+10)$ &$ k+2$&$ k+2$&$ k+1$&$ 1$&$1$&$ 0$&$ 1$&$0$ 
        &     $2k+7$&
\\      
$I_{2k}^{*s(0|1||2)}$ &$SO(4k+8)$ &$ k+2$&$ k+1$&$ k+1$&$ 1$&$1$&$ 0$&$ 1$&$0$ 
        & $2k+6$    &
\\      
$I_{2k+1}^{*s(01||2)}$ &$SO(4k+10)$ &$ k+3$&$ k+2$&$ k+2$&$ 1$&$1$&$ 0$&$ 0$&$0$ 
        &    $2k+7$ &
\\

$I_{2k}^{*s(01||2)}$ &$SO(4k+8)$ &$ k+2$&$ k+2$&$ k+1$&$ 1$&$1$&$ 0$&$ 0$&$0$ 
        &     $2k+6$&
\\

$I_{2k+1}^{*ns(01|2)}$ &$SO(4k+10)$ &$ 2k+3$&$ k+2$&$ 1$&$ k+2$&$1$&$ 0$&$ 0$&$0$ 
        &     $2k+7$&
\\   
$I_{2k}^{*ns(01|2)}$ &$SO(4k+8)$ &$ 2k+2$&$ k+2$&$ 1$&$ k+1$&$1$&$ 0$&$ 0$&$0$ 
        &     $2k+6$&
\\   
$IV^{*(0|1|2)}$ &$E_6 $ & $ 2$&$ 2$&$ 1$&$ 2$&$1$&$ 0$&$ 1$&$0$ 
        &     $8$&
\\    
$IV^{*(01|2)}$ &$E_6 $ & $ 3$&$ 2$&$ 2$&$ 2$&$1$&$ 0$&$ 0$&$0$ 
        &     $8$&
\\    
$III^{*(01|2)}$ &$E_7 $ & $ 3$&$ 3$&$ 2$&$ 2$&$1$&$ 0$&$ 0$&$0$ 
        &     $9$&
\\    \hline
\end{tabular}

\caption{Tate-like Forms for Singular Fibers. We present Tate-like forms for
canonical fibers, listing the vanishing orders of the sections and of the
discriminant, the resulting gauge group, and if necessary the conditions of
validity for the forms. $I_n$ fibers have four Tate forms depending on section
separation and parity of $n$, while $I_m^*$ have six forms again depending on
where the sections intersect the fiber components and on parity of $m$.
Exceptional singular fibers that could be realized canonically are also
included.}

\label{TableTateForms}
\end{sidewaystable}

\newpage

\section{Setup}\label{sec:Prelude}

In this section the general setup for the discussion of singular
elliptic fibrations with a rank two Mordell-Weil group is provided. First it 
is explained in more detail that such a fibration can be embedded into a
$\mathbb{P}^2$ fibration via a cubic
hypersurface equation. This is done in section \ref{sec:Tate}. In section \ref{sec:sym} 
the symmetries of this cubic equation are detailed  and it is demonstrated how they
lead to a redundancy of singular fiber types. Some constraints are chosen,
listed at the head of section \ref{sec:sym}, to eliminate this redundancy.
All the properties of the
construction used in the resolution and study of the singular fibers
found are documented in section \ref{sec:Resolutions}.

\subsection{Embedding}\label{sec:Tate}

By the algebro-geometric construction in \cite{Morrison:2012ei,
Borchmann:2013jwa, Cvetic:2013nia, Borchmann:2013hta, Cvetic:2013jta}, an elliptic
fibration with rank two Mordell-Weil group can be embedded into a
$\mathbb{P}^2$ fibration by the hypersurface equation 
\begin{equation}
  \fks_1 w^3 + \fks_2 w^2 x + \fks_3 w x^2 + \fks_5 w^2 y + \fks_6 w x y + 
  \fks_7 x^2 y + \fks_8 w y^2 + \fks_9 x y ^2 = 0 \,,
\end{equation}
as seen in the previous section. Some explanation of this construction is
given in appendix
\ref{app:RR}. Here $[w: x: y]$ are the
projective coordinates of the fibration and the $\fks_i$ are elements of the
base coordinate ring, $R$. It can be seen that this has three marked points,
where $w, x$, and $y$ take values in the fraction field, $K$, associated to
$R$. Specifically the three marked points are
\begin{equation}
  [ 0: 0: 1 ] \,, \quad [ 0: 1: 0 ] \,, \quad [ 0: \fks_9: -\fks_7 ] \,,
\end{equation}
which we label as $\Sigma_0$, $\Sigma_2$, and $\Sigma_1$ respectively.

We will work in an open neighbourhood in the base, around the singular locus,
which has coordinate $z$ such that the singular
locus will occur at the origin of this open neighbourhood. In such a local
patch we can specify the $\fks_i$ as expansions in $z$,
\begin{equation}
  \fks_i = \sum_{j=0}^{\infty} s_{i,j} z^j \,.
\end{equation}
We also introduce the simplifying notation
\begin{equation}
  \fks_{i,k} = \sum_{j=k}^{\infty} s_{i,j} z^{j-k} \,.
\end{equation}
%Tate's algorithm requires to look at the discriminant of the fibration and, order by order in the expansion  in terms of the base coordinate $z$, to deform the fiber by imposing restricitions on the sections so as to enhance the order of the discriminant. Thus one is expected to reproduce the complete set of singular fibers classified by Kodaira.    
%Two aspects will nevertheless make the analysis richer and give rise to a variety of different singular fibers as explained in section \ref{sec:Overview}:
%\begin{itemize}
%\item We will try to solve every polynomial in the discriminant in full generality using the fact that the base coordinate ring is a unique factorization domain, allowing the singular fibers to be realized either {\it canonically} or {\it non-canonically}.
%\item Same Kodaira type fibers will be realized differing with respect to which components the three sections intersect.
%\end{itemize}

\subsection{Symmetries and Lops}\label{sec:sym}

In this section note is made of the symmetries inherent in the cubic equation
(\ref{eqn:P111cubic}), and a strategy is devised to remove the redundant
multiplicity of fiber types that occurs due to these symmetries. One finds that
the following sets of vanishing orders give rise to fibrations which have
codimension one singular fibers that are related by a relabelling of the
coefficients of (\ref{eqn:P111cubic}) 
\begin{align}
  (n_1,n_2,n_3,n_5,n_6,n_7,n_8,n_9) &\leftrightarrow
  (n_1,n_5,n_8,n_2,n_6,n_9,n_3,n_7) \cr
  (n_1+2,n_2+1,n_3,n_5+1,n_6,n_7,n_8,n_9) &\leftrightarrow 
  (n_1,n_2,n_3,n_5,n_6,n_7+1,n_8,n_9+1) \cr
  (n_1+1,n_2+1,n_3+1,n_5,n_6,n_7,n_8,n_9) &\leftrightarrow 
  (n_1,n_2,n_3,n_5,n_6,n_7,n_8+1,n_9+1), 
\end{align}
and any composition thereof. In the analysis of Tate's algorithm for the
quartic equation in $\mathbb{P}^{(1,1,2)}$ \cite{Kuntzler:2014ila} these kind of symmetries were
called lops. 
The first of these relations will be referred to as
the $\mathbb{Z}_2$ symmetry, and the second and third relations,
respectively, will be called lop
%\footnote{We use the
%  arboricultural terminology of 
%  \cite{Kuntzler:2014ila}.}
one and lop two.

These lop relations and the $\mathbb{Z}_2$ symmetry generate a family of
equivalences by applying them repeatedly and in different orders. To choose
an appropriate element of each equivalence class the procedure shall be as
follows:
\begin{itemize}
  \item Use the $\mathbb{Z}_2$ symmetry to fix $n_9 \geq n_7$.
%    $(n_1,n_2,n_3,n_5,n_6,n_7,n_8,n_9)$.
  \item Apply lop one to reduce $n_7$ to $0$.%: $(n_1 + 2n_7, n_2 + n_7, n_3, n_5 + n_7, n_6, 0, n_8, n_9 - n_7)$.
  \item Apply lop two to reduce the least valued of $n_8$ and $n_9-n_7$ to zero. 
  \item Apply the $\mathbb{Z}_2$ symmetry.
\end{itemize}

In this way one can often choose a representative of a particular
lop-equivalence class where $n_7 = n_9 = 0$. In the application of Tate's
algorithm enhancements which move a form out of this
lop-equivalence class will not be considered. In this way the redundancies inherent in
the cubic equation (\ref{eqn:P111cubic}) shall be removed. 
The remainder of this subsection shall be devoted to showing 
that these relations hold.

There is  a $\mathbb{Z}_2$ symmetry that comes from the interchange
\begin{equation}\label{eqn:Z2sym}
  (n_1,n_2,n_3,n_5,n_6,n_7,n_8,n_9) \leftrightarrow 
  (n_1,n_5,n_8,n_2,n_6,n_9,n_3,n_7) \,.
\end{equation}

One can see this by observing that the equations for each form,
\begin{equation}
  \fks_{1,n_1} w^3 + \fks_{2,n_2} w^2 x + \fks_{3,n_3} w x^2 + \fks_{5,n_5} w^2 y
  + \fks_{6,n_6} w x y + \fks_{7,n_7} x^2 y + \fks_{8,n_8} w y^2 
  + \fks_{9,n_9} x y ^2 = 0 \,,
\end{equation}
and
\begin{equation}
  \fks_{1,n_1} w^3 + \fks_{2,n_5} w^2 x + \fks_{3,n_8} w x^2 + \fks_{5,n_2} w^2 y
  + \fks_{6,n_6} w x y + \fks_{7,n_9} x^2 y + \fks_{8,n_3} w y^2 
  + \fks_{9,n_7} x y ^2 = 0 \,,
\end{equation}
have identical vanishing orders up to the redefinition $x \leftrightarrow y$.
This symmetry can be removed by only considering forms where, 
in order of preference,
\begin{align}\label{eqn:Z2prefs}
  &n_7 \geq n_9 \cr
  &n_3 \geq n_8 \cr
  &n_2 \geq n_5 \,.
\end{align}

Furthermore there are symmetries that can occur in the partially resolved
forms. One such, which was referred to as lop one above, is
an equivalence between the vanishing orders
\begin{equation}
  (n_1+2,n_2+1,n_3,n_5+1,n_6,n_7,n_8,n_9) \leftrightarrow 
  (n_1,n_2,n_3,n_5,n_6,n_7+1,n_8,n_9+1) \,.
\end{equation}
To see this consider first the geometry of the LHS after resolving the
singularity at the point $x = y = z_1 = 0$ by the blow up $(x, y, z_1;
\zeta_1)$\footnote{The notation of \cite{Lawrie:2012gg} is used to spectify 
blow ups throughout this paper.}. It
is clear that one can always do such a blow up as the $n_i$ are, by definition, non-negative. 
The partially resolved geometry is
\begin{align}
  \fks_{1,n_1+2} w^3 z_1^{n_1 + 2} \zeta_1^{n_1} 
  &+ \fks_{2,n_2+1} w^2 x z_1^{n_2 + 1} \zeta_1^{n_2}
  + \fks_{3,n_3} w x^2 z_1^{n_3} \zeta_1^{n_3} 
  + \fks_{5,n_5+1} w^2 y z_1^{n_5+1} \zeta_1^{n_5} \cr
  &+ \fks_{6,n_6} w x y z_1^{n_6} \zeta_1^{n_6} 
  + s_{7,n_7} x^2 y z_1^{n_7} \zeta_1^{n_7 + 1} 
  + \fks_{7, n_7+1} x^2 y z_1^{n_7+1} \zeta_1^{n_7+2} \cr
  &+ \fks_{8,n_8} w y^2 z_1^{n_8} 
  + s_{9,n_9} x y ^2 z_1^{n_9} \zeta_1^{n_9 + 1} 
  + \fks_{9,n_9+1} x y ^2 z_1^{n_9+1} \zeta_1^{n_9+2} =  0 \,,
\end{align}
with the Stanley-Reiser ideal
\begin{equation}
  \{ wxy, w\zeta_1, xyz_1\} \,.
\end{equation}
Similarly one can consider the RHS geometry after performing the small
resolution $(w, z_2; \zeta_2)$ to separate the reducible divisor $z_2$. The
geometry is
\begin{align}
  &s_{1,n_1} w^3 z_2^{n_1} \zeta_2^{n_1+2} 
  + s_{1,n_1+1} w^3 z_2^{n_1+1} \zeta_2^{n_1+3} 
  + \fks_{1,n_1+2} w^3 z_2^{n_1+2} \zeta_2^{n_1+4}  
  + s_{2,n_2} w^2 x z_2^{n_2} \zeta_2^{n_2+1} \cr
  &+ \fks_{2,n_2+1} w^2 x z_2^{n_2 + 1} \zeta_2^{n_2+2} 
  + \fks_{3,n_3} w x^2 z_2^{n_3} \zeta_2^{n_3} 
  + s_{5,n_5} w^2 y z_2^{n_5} \zeta_2^{n_5+1} 
  + \fks_{5,n_5+1} w^2 y z_2^{n_5+1} \zeta_2^{n_5+2} \cr
  &+ \fks_{6,n_6} w x y z_2^{n_6} \zeta_2^{n_6} 
  + \fks_{7, n_7+1} x^2 y z_2^{n_7+1} \zeta_2^{n_7} 
  + \fks_{8,n_8} w y^2 z_2^{n_8} 
  + \fks_{9,n_9+1} x y ^2 z_2^{n_9+1} \zeta_2^{n_9} =  0 \,,
\end{align}
with Stanley-Reiser ideal
\begin{equation}
  \{ wxy, wz_2, xy\zeta_2 \} \,.
\end{equation}
Under the identification $z_1 \leftrightarrow \zeta_2$ and
$\zeta_1 \leftrightarrow z_2$ it is observed that these equations and SR ideals
are equivalent. %It is important to note that $\fks_{i,j}$ is always a power
%series expansion in some variable $z$ starting with the constant term in $z$.
Any multiplicity arising from this redundancy in
(\ref{eqn:P111cubic}) can be removed by combining it with one of the earlier constraints
from the $\mathbb{Z}_2$ symmetry (\ref{eqn:Z2prefs}), $n_7 \geq n_9$, so as to
choose to consider only forms which have $n_9 = 0$.

There is another relation among the partially resolved geometries, which was
referred to as lop two, 
\begin{equation}
  (n_1+1,n_2+1,n_3+1,n_5,n_6,n_7,n_8,n_9) \leftrightarrow 
  (n_1,n_2,n_3,n_5,n_6,n_7,n_8+1,n_9+1) \,.
\end{equation}

Again this is seen by studying the partially resolved geometry explicitly. If
$(n_1+1,n_2+1,n_3+1,n_5,n_6,n_7,n_8,n_9)$ is resolved by the small resolution
$(y, z_1; \zeta_1)$ the blown up geometry is given by the equation
\begin{align}
  &s_{1,n_1+1} w^3 z_1^{n_1+1} \zeta_1^{n_1} 
  + \fks_{1,n_1+2} w^3 z_1^{n_1+2} \zeta_2^{n_1+1}  
  + s_{2,n_2+1} w^2 x z_1^{n_2+1} \zeta_1^{n_2} 
  + \fks_{2,n_2+2} w^2 x z_1^{n_2 + 2} \zeta_1^{n_2+1} \cr
  &+ s_{3,n_3+1} w x^2 z_1^{n_3+1} \zeta_1^{n_3} 
  + \fks_{3,n_3+2} w x^2 z_1^{n_3+2} \zeta_1^{n_3 + 1} 
  + \fks_{5,n_5} w^2 y z_1^{n_5} \zeta_1^{n_5} 
  + \fks_{6,n_6} w x y z_1^{n_6} \zeta_1^{n_6} \cr
  &+ \fks_{7,n_7} x^2 y z_1^{n_7} \zeta_1^{n_7} 
  + \fks_{8,n_8} w y^2 z_1^{n_8} \zeta_1^{n_8+1}
  + \fks_{9,n_9} x y^2 z_1^{n_9} \zeta_1^{n_9+1} = 0 \,,
\end{align}
with SR-ideal
\begin{equation}
  \{ wxy, yz_1, wx\zeta_1 \} \,.
\end{equation}
On the other side if $(n_1,n_2,n_3,n_5,n_6,n_7,n_8+1,n_9+1)$ is resolved by the resolution $(w,
x, z_2; \zeta_2)$ the geometry is then given as the vanishing of the
hypersurface polynomial
\begin{align}
  &\fks_{1,n_1} w^3 z_2^{n_1} \zeta_2^{n_1 + 1}  
  + \fks_{2,n_2} w^2 x z_2^{n_2} \zeta_2^{n_2 + 1} 
  + \fks_{3,n_3} w x^2 z_2^{n_3} \zeta_2^{n_3 + 1} 
  + \fks_{5,n_5} w^2 y z_2^{n_5} \zeta_2^{n_5} \cr
  &+ \fks_{6,n_6} w x y z_2^{n_6} \zeta_2^{n_6} 
  + \fks_{7,n_7} x^2 y z_2^{n_7} \zeta_2^{n_7} r
  + s_{8,n_8 + 1} w y^2 z_2^{n_8 + 1} \zeta_2^{n_8} 
  + \fks_{8,n_8 + 2} w y^2 z_2^{n_8 + 2} \zeta_2^{n_8 + 1} \cr
  &+ s_{9,n_9 + 1} x y^2 z_2^{n_9 + 1} \zeta_2^{n_9}
  + \fks_{9,n_9+2} x y^2 z_2^{n_9+2} \zeta_2^{n_9 + 1} = 0 \,,
\end{align}
with SR-ideal
\begin{equation}
  \{ wxy, wxz_2, \zeta_2y \} \,.
\end{equation} 
These two geometries describe the same partially resolved space, and can be
related by the interchange
\begin{equation}
  z_1 \leftrightarrow \zeta_2 \,, \quad \zeta_1 \leftrightarrow z_2 \,.
\end{equation}

\subsection{Resolutions, Intersections, and the Shioda
Map}\label{sec:Resolutions}

To determine the Kodaira type, including the distribution of the marked
points, of the codimension one singularity in the
fibration specified by (\ref{eqn:P111cubic}) one often explicitly constructs the
resolved geometry via a sequence of algebraic resolutions. In the context of
elliptic fibrations such resolutions have been constructed in
\cite{Grimm:2009yu, Krause:2011xj, Grimm:2011fx, Esole:2011sm, MS,
  Lawrie:2012gg, Braun:2013cb, Hayashi:2013lra, Braun:2014kla,
Esole:2014hya}. In this section we set up the framework to discuss the
resolved geometries and the intersection computations, for example of $U(1)$ charges of
matter curves, that are carried out as part of the analysis of the singular
fibers found. In particular details are given about the embedding of the
fibration as a hypersurface in an ambient fivefold, the details of how 
 the intersection numbers between curves and fibral divisors are computed, and on
the construction of the $U(1)$ charge generators.

%As this is a codimension one singularity the resolution is
%unique [CITE], and thus the fiber type is independent of the particular 
%sequence of resolutions we use.

Consider the ambient fivefold $X_5 =
\mathbb{P}^2(\mathcal{O}\oplus\mathcal{O}(\alpha)\oplus\mathcal{O}(\beta))$
which is the projectivization of line bundles over a base space
$B_3$. The elliptically fibered Calabi-Yau fourfold will be realized as the
hypersurface in this $X_5$ cut out by the cubic equation (\ref{eqn:P111cubic}).
%Consider the equation (\ref{eqn:P111cubic}) as a hypersurface in the
%fibration $X_5 =
%\mathbb{P}^2(\mathcal{O}\oplus\mathcal{O}(\alpha)\oplus\mathcal{O}(\beta))$. 
%\cl{I have no idea.} 
The terms in the
homogeneous polynomial are then sections of the following line bundles
%\begin{floatingtable}[r]{
\begin{equation}
  \begin{array}{c|c}
    \text{Section} & \text{Bundle} \cr\hline
      w & \mathcal{O}(\sigma) \cr
      x & \mathcal{O}(\sigma + \alpha) \cr
      y & \mathcal{O}(\sigma + \beta) \cr
      z & \mathcal{O}(S_G) \cr
      \fks_{1,j} & \mathcal{O}(c_1 + \alpha + \beta - jS_G) \cr
      \fks_{2,j} & \mathcal{O}(c_1 + \beta - jS_G) \cr
      \fks_{3,j} & \mathcal{O}(c_1 - \alpha + \beta - jS_G) \cr
      \fks_{5,j} & \mathcal{O}(c_1 + \alpha - jS_G) \cr
      \fks_{6,j} & \mathcal{O}(c_1 - jS_G) \cr
      \fks_{7,j} & \mathcal{O}(c_1 - \alpha - jS_G) \cr
      \fks_{8,j} & \mathcal{O}(c_1 + \alpha - \beta - jS_G) \cr
      \fks_{9,j} & \mathcal{O}(c_1 - \beta - jS_G) \cr
  \end{array}
\end{equation}
%\end{floatingtable}

Here $c_1$ is a shorthand notation for $\pi^*c_1(B_3)$.
In practice, the first step in any explicit determination of a singular fiber 
is to blow up the $\mathbb{P}^2$ fibration to
a $dP_2$ fibration by the substitution $w \rightarrow l_1 l_2 w$, $x \rightarrow l_1 x$,
and $y \rightarrow l_2 y$ and taking the proper transform, as was also the
procedure in \cite{Borchmann:2013jwa, Cvetic:2013nia, Borchmann:2013hta,
Cvetic:2013jta}.

The geometry is then specified by the equation 
\begin{equation}
  \fks_1 l_1^2 l_2^2 w^3 
  + \fks_2 l_1^2 l_2 w^2 x 
  + \fks_3 l_1^2 w x^2 
  + \fks_5 l_1 l_2^2 w^2 y 
  + \fks_6 l_1 l_2 w x y 
  + \fks_7 l_1 x^2 y 
  + \fks_8 l_2^2 w y^2 
  + \fks_9 l_2 x y^2 = 0 \,,
\end{equation}
in $dP_2$. After these blow ups the fiber coordinates in this equation are
sections of the line bundles
\begin{equation}
  \begin{array}{c|c}
    \text{Section} & \text{Bundle} \cr\hline
      w & \mathcal{O}(\sigma - F_1 - F_2) \cr
      x & \mathcal{O}(\sigma + \alpha - F_1) \cr
      y & \mathcal{O}(\sigma + \beta - F_2) \cr
      l_1 & \mathcal{O}(F_1) \cr
      l_2 & \mathcal{O}(F_2) \cr
  \end{array}
\end{equation}
As can be seen from the blow ups which mapped $\mathbb{P}^2$ to $dP_2$ the
marked point $[0:0:1]$ has been mapped to the exceptional divisor $l_1$,
similarly for $[0:1:0]$ and $l_2$. As such the marked points
$\Sigma_0$, $\Sigma_1$, and $\Sigma_2$ have been related to the divisors $l_1$, $w$, and $l_2$
respectively.

As the marked points form sections they are restricted to 
intersect, in codimension one, only a single multiplicity one component of the
singular fiber \cite{MR1078016}. 
%In codimension two the sections can however wrap
%entire fiber components, which has been noted and studied in [REF]

%\cl{Words on section multi.} Each of the three sections intersects exactly one component of the resolved,
%codimension one fiber, and that component must be of multiplicity one \cite{MR1078016}.

%, ie, $f^{-1}(\pi^{-1}(b \in B) \cap \{ w = 0
%\}) = [ 0, \fks_9(b), - \fks_7(b) ]$, where $f$ is of course the birational map
%corresponding to the blow up.

The $dP_2$ intersection ring is not freely generated due to the projective
relations which hold in $dP_2$. These relations are, using standard projective
coordinate notation,
\begin{equation}
  [w l_1 l_2: x l_1: y l_2] \,, \quad [ w: x ] \,, \quad [ w: y ] \,.
\end{equation}
These correspond to the relations in the intersection ring
\begin{align}
  \sigma \cdot (\sigma + \alpha) \cdot (\sigma + \beta) &= 0 \cr
  (\sigma - F_1 - F_2) \cdot (\sigma - F_1) &= 0 \cr
  (\sigma - F_1 - F_2) \cdot (\sigma - F_2) &= 0 \,.
\end{align}
The strategy, as it was in \cite{MS, Lawrie:2012gg}, will be to choose a basis
of the intersection ring and repeatedly apply these relations, including any
that come from exceptional divisor classes introduced in the resolution. In
this way the intersection numbers between curves and fibral
divisors can be computed. In this paper the resolutions and intersections were carried out 
using the Mathematica package {\tt Smooth} \cite{Smooth}.

Given an elliptic fibration with multiple rational sections there remains the
construction of the generators of the $U(1)$ symmetries, that is the
generators of the Mordell-Weil group. The Mordell-Weil group is a finitely
generated abelian group 
\cite{MR1312368}
\begin{equation}
  \mathbb{Z} \oplus \cdots \oplus \mathbb{Z} \oplus \mathcal{G} \,,
\end{equation}
where $\mathcal{G}$ is some finite torsion group\footnote{We shall not
  concern ourselves with $\mathcal{G}$ in this paper, but some investigations
  are \cite{Mayrhofer:2014opa, Aspinwall:1998xj}.}. There is a map, known as the
Shioda map, which constructs from rational sections the generators of the
Mordell-Weil group. This map is discussed in detail in \cite{Park:2011ji,
Morrison:2012ei, MR1030197}.

The Shioda map associates to each rational section, $\sigma_i$, a divisor
$s(\sigma_i)$ such that
%The Shioda map associates to each rational section, $\sigma_i$, a homology class in
%$H^{1,1}(Y_4)$, $s(\sigma_i)$, with the following properties
\begin{align}\label{eqn:Shioda}
  s(\sigma_i) \cdot F_{j} &= 0 \cr
  s(\sigma_i) \cdot B &= 0 \,,
\end{align}
where $F_{j}$ are the exceptional curves and $B$ is the dual to the
class of the base $B_3$. Reduction on the $F_{j}$ gives rise to gauge bosons
which should be uncharged under the abelian gauge symmetry. This is ensured by
the conditions (\ref{eqn:Shioda}).

%, which is an exceptional
%$\mathbb}{P}^1$ fibered over $z = 0$. Reduction on these exceptional
%$\mathbb{P}^1$s gives rise to the gauge bosons, which should be uncharged. 

%which correspond to the gauge
%bosons of the gauge group associated to the singular fiber type, and so 
%should be uncharged if we are to treat $s(\sigma_i)$ as a
%generator of a $U(1)$ symmetry. The $\pi^*D_H$ represents all horizontal divisors,
%that is, those divisors in the Calabi-Yau which are pullbacks from divisors on
%$B_3$.

The charge of a particular matter curve $C$ with respect to the $U(1)$
generator associated to the rational section $\sigma_i$ is given by the
intersection number $s(\sigma_i) \cdot C$. The constraints (\ref{eqn:Shioda})
determine the $U(1)$ charges from $s(\sigma_i)$ up to an overall scale. We
shall always consider the zero-section to be the rational section associated
with the introduction of the $l_1$ in the blow up to $dP_2$.

As was alluded to in section \ref{sec:Tate} it is not always the case that 
a fibration that arises from the algorithm can be specified purely in terms of the
vanishing orders of the coefficients. Sometimes it is necessary to also
include some specialization of the coefficients in the $z$-expansion of the
coefficients of the equation. Consider a discriminant of the form
\begin{equation}
  \Delta = (AB - CD)z^n + \mathcal{O}(z^{n+1}) \,.
\end{equation}
An enhancement that would enhance this singularity would be where $AB - CD =
0$. The solution of this polynomial cannot in general be specified in terms of
the vanishing order of $A$, $B$, $C$, and $D$. In appendix \ref{app:poly}  we collect
the solutions to several polynomials of this form which come up repeatedly in
the application of Tate's algorithm to (\ref{eqn:P111cubic}). The
solution to this particular polynomial is 
\begin{align}
  A &= \sigma_1\sigma_2 \cr
  B &= \sigma_3\sigma_4 \cr
  C &= \sigma_1\sigma_3 \cr
  D &= \sigma_2\sigma_4 \,,
\end{align}
where the pairs $(\sigma_1, \sigma_4)$ and $(\sigma_2, \sigma_3)$ are coprime.
It is not generally possible
to perform some shift of the coordinates in (\ref{eqn:P111cubic}) to return
this solution to an expression involving just vanishing orders. This is
notably different from Tate's algorithm as carried out on the Weierstrass
equation in \cite{Katz:2011qp}; there the equation includes monic terms
unaccompanied by any coefficient, which often allows one to shift the
variables to absorb these non-canonical like solutions into higher
vanishing orders of the model. 
%\cl{If we assume that we can remove any base
%  section where the codimension two enhancement at that locus is non-minimal
%  things might be different and we may be able to get rid of some models. We
%should maybe write something about this.}

\section{Tate's Algorithm}\label{sec:TateAlg}
In this section we will proceed through the algorithm \cite{Bershadsky:1996nh,
Katz:2011qp},
considering the discriminant of the elliptic fibration order by order in the
expansion in terms of the base coordinate $z$. By enhancing the fiber of our
elliptic fibration, we will see under which conditions on the sections
$\fks_i$  the order of the discriminant will enhance and then study the
resulting singular fibers. This will be done systematically up to singular
$I_5$ fibers for phenomenological reasons and in section \ref{sec:Exceptional}
we will provide details for some of the exceptional singular fibers.  In a step-by-step application of Tate's algorithm to the
elliptic fibration (\ref{eqn:P111cubic}) we find the various different types
of Kodaira singular fibers decorated with the information of which sections
intersect which components. The discriminant reflects the different ways in which the sections can
intersect the multiplicity one fiber components (as explained in section
\ref{sec:Resolutions}), thus giving rise to an increased number of singular
fibers over fibrations with fewer rational sections. The analysis will be carried out in parallel both for canonical models
(determined only by the vanishing orders of the sections) and for
non-canonical models (which require additional specialization arising from
solving polynomials in the discriminant.)

\subsection{Starting Points}

In the following we will assume that the fibration develops a singularity
along the locus $z=0$ in the base. A singularity can be characterized by one
of the following two criteria:
\begin{itemize}
\item The leading order of the discriminant as a series expansion in $z$ must vanish.
\item The derivatives of $\fkD|_{z=0}$ in an affine patch must vanish along
  the $z = 0$ locus, where $\fkD$ is the equation for the fibration.
\end{itemize}  
Since the leading order of the discriminant is a complicated and 
unenlightening expression, we will not present it here and instead study the derivatives
of the equation of the fibration. This will turn out to be
significantly simpler and we will see that the discriminant will enhance upon
substitution of the conditions found by the derivative analysis. On the other
hand, throughout our study of higher order singularities we will look only at
the discriminant ignoring the derivative approach.

Let us then study the equation for the elliptic fibration in the affine patch
with coordinates $(x,y)$, that is, where we can scale such that $w=1$. Along
the locus $z=0$ we assume that the fiber becomes singular at the point
$(x_0,y_0)$ and require the derivatives to vanish
\begin{equation}\label{partials}
\begin{aligned}
\fkD|_{z=0}=&s_{1,0} + s_{2,0} x_0 + s_{3,0} x_0^2 + s_{5,0} y_0 + s_{6,0} x_0 y_0 + s_{7,0} x_0^2 y_0 + 
 s_{8,0} y_0^2 + s_{9,0} x_0 y_0^2=0\\
\partial_x\fkD|_{z=0}=& s_{2.0} + 2 s_{3,0} x_0 + s_{6,0} y_0 + 2 s_{7,0} x_0 y_0 + s_{9,0} y_0^2=0\\
\partial_y\fkD|_{z=0}=& s_{5,0} + s_{6,0} x_0 + s_{7,0} x_0^2 + 2 s_{8,0} y_0
+ 2 s_{9,0} x_0 y_0=0 \,.
\end{aligned}
\end{equation}
We can solve for $s_{2,0}$ and $s_{5,0}$ from the last two equations
\begin{equation}
\begin{aligned}
s_{2,0} =& -2 s_{3,0} x_0 - y_0 (s_{6,0} + 2 s_{7,0} x_0 + s_{9,0} y_0)\\
s_{5,0} =& -x_0 (s_{6,0} + s_{7,0} x_0) - 2 (s_{8,0} + s_{9,0} x_0) y_0 \,.
\end{aligned}
\end{equation}
Upon substitution in the first equation we can solve for $s_{1,0}$
\begin{equation}
s_{1,0}=s_{3,0} x_0^2 + y_0 (s_{8,0} y_0 + x_0 (s_{6,0} + 2 s_{7,0} x_0 + 2
s_{9,0} y_0)) \,.
\end{equation}
When $s_{1,0},s_{2,0}$ and $s_{5,0}$ satisfy the above requirements the discriminant indeed enhances to first order. We can bring the equation of the fibration in a canonical form, depending only on the vanishing orders of the coefficients, by performing the following coordinate shift
\begin{equation}
\left( \begin{array}{c} x \\ y \end{array} \right) \rightarrow \left(
  \begin{array}{c} x - x_0  w \\ y  -y_0  w \end{array} \right) \,.
\end{equation}
We see that the singularity now sits at the origin of the affine patch and has
generic coefficients in addition to $\{ s_{1,0}=s_{2,0}=s_{5,0}=0 \}$. This is
an $I_1$ singular fiber, which is the only fiber at vanishing order
$\ord_z(\Delta) = 1$ in Kodaira's classification. That this is indeed an $I_1$ fiber
can also be seen by performing a linear approximation around the singular
point and noting that we obtain two distinct tangent lines, which shows that
this is indeed an ordinary double point. Since there is only one fiber
component, all the three sections will intersect it, and we will denote the
singular fiber
\begin{equation}
I_1^{(012)}: \quad (1,1,0,1,0,0,0,0) \,.
\end{equation}

This does not exhaust the possible ways to solve the three equations in
(\ref{partials}). Indeed, we can look at the affine subspace $y=0$ and see
that we can find additional solutions. Note that we will not consider here
the case $x=0$ as this is related by the $\mathbb{Z}_2$ symmetry discussed in
section \ref{sec:sym}. The partial
derivatives now read
\begin{equation}
\begin{aligned}
\fkD|_{z=y=0}=&s_{1,0} + s_{2,0} x_0 + s_{3,0} x_0^2=0 \\
\partial_x\fkD|_{z=y=0}=& s_{2,0} + 2 s_{3,0} x_0=0 \\
\partial_y\fkD|_{z=y=0}=& s_{5,0} + s_{6,0} x_0 + s_{7,0} x_0^2=0 \,.
\end{aligned}
\end{equation}
We see that if we require $\{s_{1,0}=s_{2,0}=s_{3,0}=0\}$ the three equations are satisfied for the two solutions of the quadratic equation $\{s_{5,0} + s_{6,0} x_0 + s_{7,0} x_0^2=0 \}$, which are the two singular points of an $I_2$ Kodaira fiber as the discriminant enhances to vanishing order $\Delta(z^2)$.  
Indeed, looking at the equation of the fiber, we see that this splits in two components
\begin{equation}
\begin{aligned}
D_1:& \qquad z=y=0 \cr
D_2:& \qquad z =  s_{5,0} w^2 +  s_{6,0} w x + s_{7,0} x^2 +  (
s_{8,0} w + s_{9,0} x) y=0 \,.
\end{aligned}
\end{equation}
The two components indeed intersect in two different points, thus showing that
this is an $I_2$ singular fiber. One of the sections intersects one component,
while the two remaining sections intersect the other, so we will denote this fiber as
\begin{equation}
I_2^{(01|2)}:\quad (1,1,1,0,0,0,0,0) \,.
\end{equation}
These two fibers represent the starting points for the analysis to be carried
out in the remainder of this section. Given the equation for the fibration, we
can ask whether $z$ divides any of the coefficients $\fks_i$. Then we can
conclude, inside our preferred lop-equivalence class, the following:
\begin{itemize}
\item If $z \nmid \fks_1$ and $z \nmid \fks_2$ then the fiber over the locus $\{z=0\}$ is smooth.
\item If $z \mid \fks_1,z \mid \fks_2$ and $z \mid \fks_3$ then we can carry on the analysis as in the next section and check whether the singularity is simply $I_2^{(01|2)}$ or some other enhanced kind.
\item  If $z \mid \fks_1,z \mid \fks_2$ and $z \mid \fks_5$ we will instead start our analysis from an $I_1^{(012)}$ singular fiber. It is important to notice that in this part of the algorithm we will not let $z \mid \fks_3$ as this case is covered in the previous branch.
\end{itemize}
 
\subsection{Enhancements from $\ord_z(\Delta) = 1$}

From the previous section we have found exactly one starting point for the
algorithm which has a discriminant linear in $z$: $(1,1,0,1,0,0,0,0)$. In this
section we shall study the various ways that this $I_1$ singular fiber can
enhance. The discriminant of the $(1,1,0,1,0,0,0,0)$ fibration is 
\begin{equation}\label{eqn:discord1}
  \Delta = s_{1,1}s_{3,0}s_{8,0}(s_{6,0}^2 - 4s_{3,0}s_{8,0})(s_{7,0}^2s_{8,0}
  - s_{6,0}s_{7,0}s_{9,0} + s_{3,0}s_{9,0}^2)z + \mathcal{O}(z^2) \,,
\end{equation}
up to numerical factors. The discriminant factors into five distinct terms
which will enhance the discriminant, and thus the singular fiber, when they
vanish. As this set of vanishing orders is specifying a fibration where $z
\nmid \fks_3$ then we cannot consider the situation where $s_{3,0} = 0$.
Equivalently, because of the $\mathbb{Z}_2$ symmetry explained in
section \ref{sec:sym}, we cannot consider $s_{8,0}$ vanishing. 

\begin{figure}
  \centering
    \includegraphics[height=3cm]{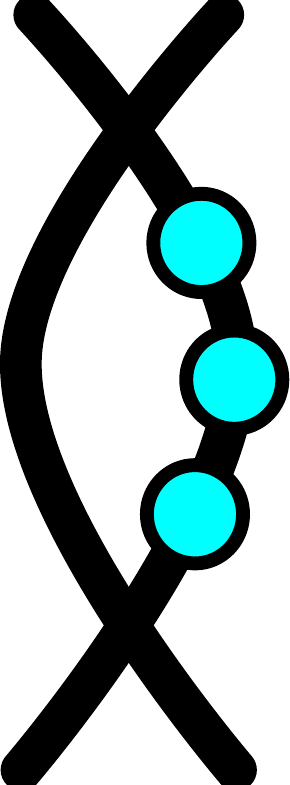} \qquad\qquad
    \includegraphics[height=3cm]{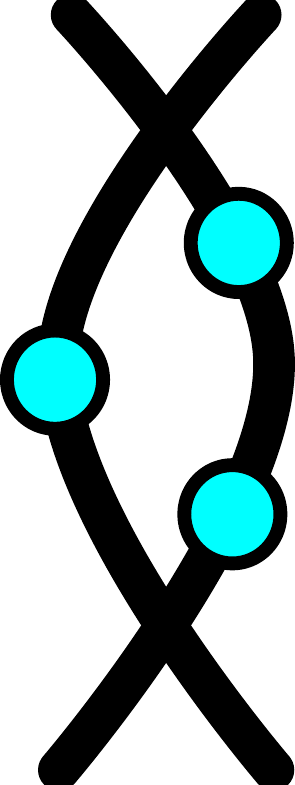} \qquad\qquad
    \includegraphics[height=3cm]{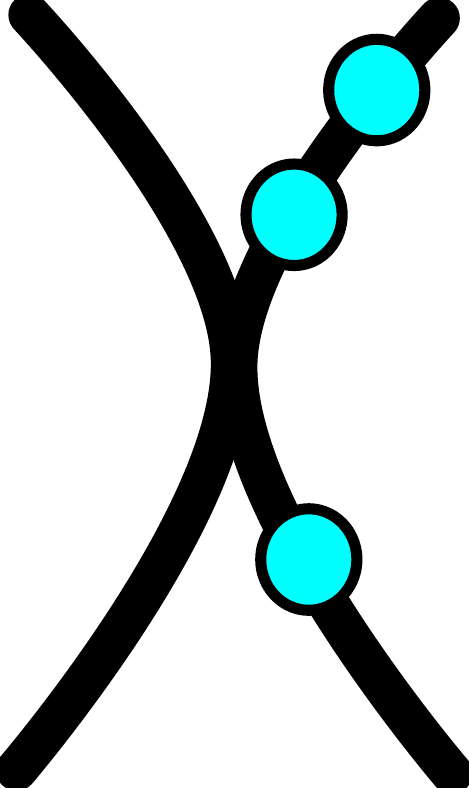} \qquad\qquad
    \includegraphics[height=3cm]{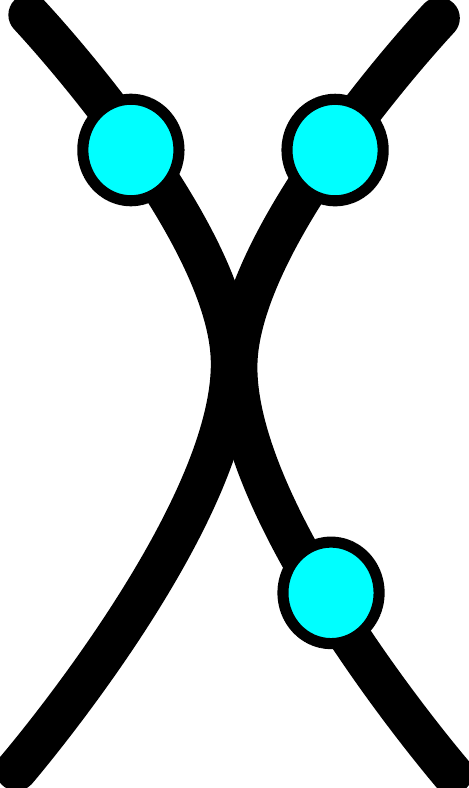}
    \caption{The type $I_2$ and type $III$ singular fibers with the possible
    locations of the three marked points denoted by the blue nodes.
    Respectively these are $I_2^{(ijk)}$, $I_2^{(i|jk)}$, $II^{(ijk)}$ and
    $II^{(i|jk)}$ fibers.}
  \label{I2Fibs}
\end{figure}

First let us consider the simple case where $s_{1,1} = 0$, which is equivalent
to stating that $z^2 \mid \fks_1$. Then $z^2 \mid \Delta$ and the singular fiber type,
determined by resolving the singularity explicitly as explained in section
\ref{sec:Resolutions}, is $I_2$. The three rational sections all intersect one
of the two components of the $I_2$ fiber
\begin{equation}
  I_2^{(012)} \, : \, (2,1,0,1,0,0,0,0) \,,
\end{equation}
listed in  table \ref{TableNonCanonicalord3}. 

The discriminant can also be enhanced in order by allowing $z$ to divide
either of the two polynomials in (\ref{eqn:discord1}). Let us first consider
the situation where $s_{6,0}^2 - 4s_{3,0}s_{8,0}$ vanishes. The solution to
this equation over this unique factorization domain is given in
appendix \ref{app:poly} and states that
\begin{align}\label{solutionpoly:II_nc^012}
  s_{6,0} &= \mu \sigma_3 \sigma_8 \cr
  s_{3,0} &= \mu \sigma_3^2 \cr
  s_{8,0} &= \mu \sigma_8^2 \,.
\end{align}
The discriminant then enhances so that $z^2 \mid \Delta$. To determine the
type of singular fiber here let us consider the equation of the single
component of the $I_1$ fiber which is being enhanced
\begin{equation}
  (s_{3,0}x^2 + s_{6,0}xy + s_{8,0}y^2) + xy(s_{7,0}x + s_{9,0}y) = 0 \,. 
\end{equation}
If $s_{6,0}^2 - 4s_{3,0}s_{8,0} = 0$ then the quadratic part of the equation
factors into a square which does not divide the cubic terms; this is exactly
the form of the equation for a cusp, which is a type $II$ fiber. Therefore we
have observed the fiber
\begin{equation}
  II_{nc}^{(012)}: \quad \left\{
  \begin{array}{c}
   (1,1,0,1,0,0,0,0) \cr
    [-,-,\mu\sigma_3^2,-,2\mu\sigma_3\sigma_8,-,\mu\sigma_8^2,-] 
\end{array} \right\} \,.
\end{equation}
from  table \ref{TableNonCanonicalord3}.

Finally we can consider the singular fiber that occurs when the second
polynomial in $\Delta$ vanishes: $s_{7,0}^2s_{8,0} - s_{6,0}s_{7,0}s_{9,0} +
s_{3,0}s_{9,0}^2 = 0$. Appendix \ref{app:poly} lists four generic
solutions of this polynomial, three canonical and one non-canonical, which are:
\begin{gather}
  s_{7,0} = s_{9,0} = 0 \cr
  s_{7,0} = s_{3,0} = 0 \cr
  s_{8,0} = s_{9,0} = 0 \cr
  s_{7,0} = \sigma_1\sigma_2 \,,\quad s_{9,0} = \sigma_1\sigma_3 \,,\quad
  s_{8,0} = \sigma_3\sigma_4 \,,\quad s_{3,0} = \sigma_2\sigma_5 \,,\quad
  s_{6,0} = \sigma_2\sigma_4 + \sigma_3\sigma_5 \,.
\end{gather}
Any of the three canonical solutions will remove us from our preferred
lop-equivalence class and so we do not consider them
as they will give rise to a redundancy of singular fiber types. The only
solution to consider therefore is the non-canonical one. The fiber found at
this locus is another $I_2$ fiber, which can be written as
\begin{equation}
I_{2,nc}: \quad \left\{
  \begin{array}{c}
  (1,1,0,1,0,0,0,0) \cr
  [-,-,\sigma_2\sigma_5,-,\sigma_2\sigma_4 +
        \sigma_3\sigma_5,\sigma_1\sigma_2,\sigma_3\sigma_4,\sigma_1\sigma_3]
\end{array} \right\} \,.
\end{equation}
Table \ref{TableNonCanonicalord3} is then complete up to second order, once we also include the $I_2^{(01|2)}$
which was found in the previous section as one of the alternate starting
points in the $z \mid \fks_3$ branch.

\subsection{Enhancements from $\ord_z(\Delta) = 2$}\label{sec:Delta=O(z^2)}

We will now consider the enhancement of the four previously found fibrations
which have a discriminant with vanishing order two in $z$. In this section we shall include the
details only of those enhancements that have some non-standard behaviour.

\begin{figure}
  \centering
    \includegraphics[height=3cm]{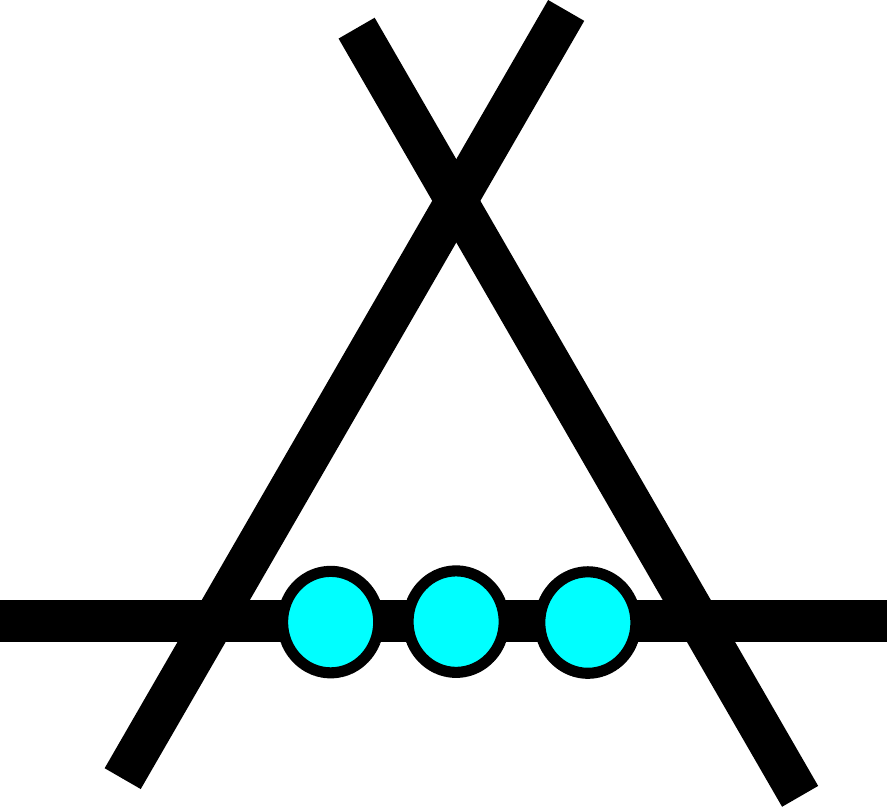} \qquad\qquad
    \includegraphics[height=3cm]{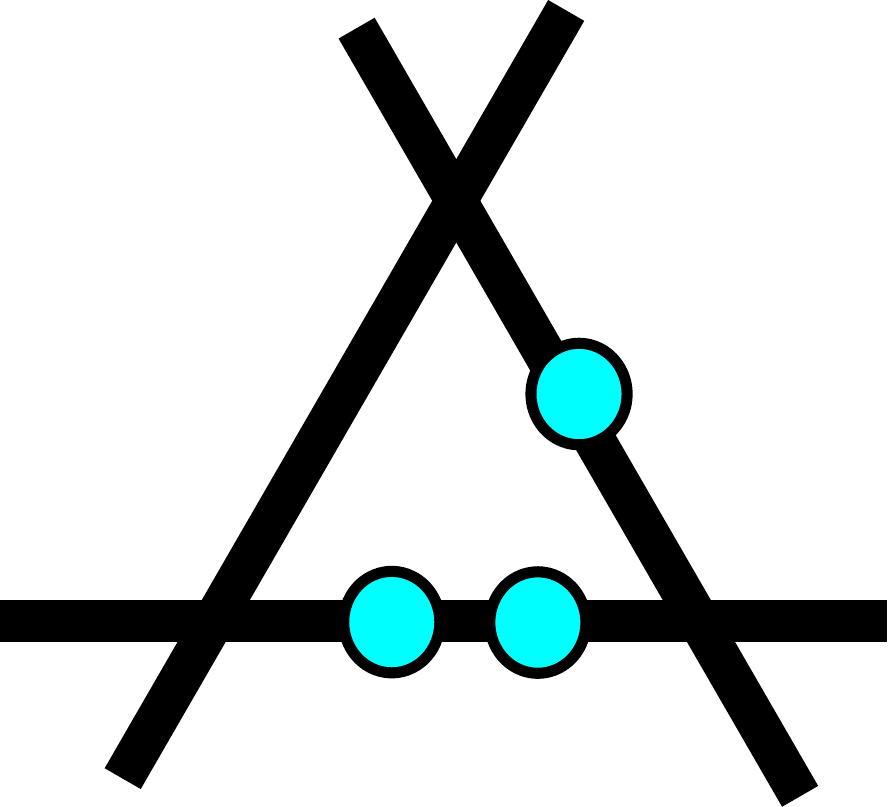} \qquad\qquad
    \includegraphics[height=3cm]{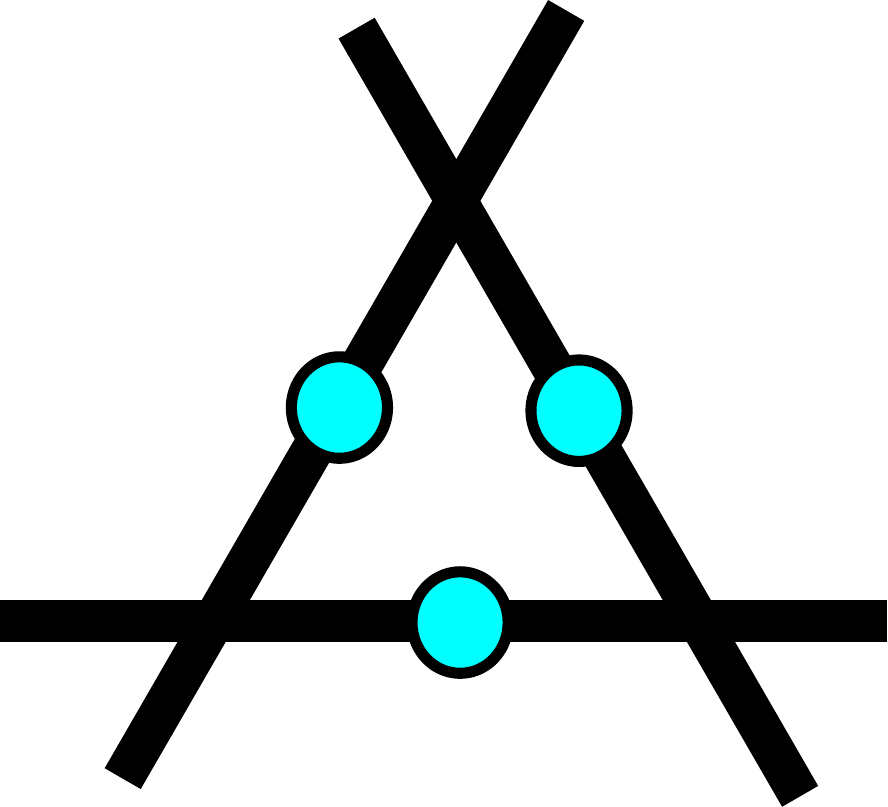}
    \caption{The type $I_3$ singular fibers with the locations of the 
    three marked points denoted by the blue nodes. Respectively these are $I_3^{(ijk)}$, $I_3^{(ij|k)}$ and $I_3^{(i|j|k)}$ fibers.}
  \label{I3Fibs}
\end{figure}

The fibrations $(2,1,0,1,0,0,0,0)$ and $(1,1,1,0,0,0,0,0)$ can contain,
respectively, in their discriminants polynomials with five and seven terms.
These are not polynomials that are discussed in appendix \ref{app:poly} as
their solutions are not known in full generality. In lieu of a complete
solution we consider non-generic but canonical type solutions which allow us
to obtain singular fibers of a particular type which would be unobtainable
without determining a full, generic solution to these polynomials. The
subbranches which follow from enhancements where it has been necessary to
consider a non-generic solution will also therefore be non-generic, however
all remaining branches are determined in full generality.

\subsubsection{Polynomial enhancement in the $z \nmid \fks_3$ branch}

The discriminant of the equation for the $(2,1,0,1,0,0,0,0)$ singular fiber
contains the polynomial
\begin{equation}
  P = s_{8,0} s_{2,1}^2-s_{5,1} s_{6,0}
              s_{2,1}+s_{1,2}  s_{6,0}^2+s_{3,0} \left(s_{5,1}^2-4 s_{1,2}
                          s_{8,0}\right) \,.
\end{equation}
As the most general solution for this five-term polynomial is not known we propose here two specific  solutions. The first is a canonical solution obtained by setting $s_{1,2}=s_{2,1}=s_{5,1}=0$. As a consequence $z^3\mid \Delta$ and we find an $I_3^{ns(012)}$ singular fiber
\begin{equation}
  I_{3}^{ns(012)} \, : \, (3,2,0,2,0,0,0,0) \,.
\end{equation}
Recalling the split/non-split monodromy distinction in Tate's algorithm, we see
only two components in this singular fiber. One of the fiber curves decomposes
when the component of the discriminant, $s_{6,0}^2-4s_{3,0}s_{8,0}$ has the
form of a perfect, non-zero square. 

%This condition is satisfied, as explained
%in appendix \ref{app:poly}, when we have:
%\begin{equation}
%s_{3,0}=\sigma_1 \sigma_3, \qquad s_{6,0}=\sigma_1 \sigma_2+\sigma_3\sigma_4,\qquad s_{8,0}=\sigma_2 \sigma_4
%\end{equation}
%Upon substitution we then find the split singular $I_3^{s(012)}$:
%\begin{equation}
%  I_{3,nc}^{s(012)} \,:\, (3,2,0,2,0,0,0,0) \,,\,
%  [-,-,\sigma_1\sigma_3,-,\sigma_1\sigma_2 +
%    \sigma_3\sigma_4,-,\sigma_2\sigma_4,-] \,.
%\end{equation}
%Since this is not the most general solution to the five-term polynomial, it will lead to particular $I_5$ models whose spectra will be discussed in section \ref{sec:Spectra}.

The second non-general solution to the five-term polynomial we consider here
is found by canonically setting $s_{1,2}=0$, and then the five term polynomial reduces to
\begin{equation}
P|_{(s_{1,2}=0)}=s_{2,1}^2s_{8,0}-s_{2,1}s_{6,0}s_{5,1} +s_{5,1}^2s_{3,0} \,.
\end{equation}
We notice that we cannot set $s_{3,0}$ to zero  because we are in the $z
\nmid\fks_3$ part of the algorithm (and by $\mathbb{Z}_2$ symmetry  we cannot
set to zero $s_{8,0}$ either). Moreover we just considered the canonical
solution given by setting $s_{2,1}=s_{5,1}=0$. We are then left with imposing
the non-canonical solution given in appendix \ref{app:poly}
\begin{equation}
 s_{2,1} = \sigma_1\sigma_2 \,,\quad s_{5,1} = \sigma_1\sigma_3 \,,\quad
  s_{8,0} = \sigma_3\sigma_4 \,,\quad s_{3,0} = \sigma_2\sigma_5 \,,\quad
  s_{6,0} = \sigma_2\sigma_4 + \sigma_3\sigma_5 \,.
\end{equation}
The resulting singular fiber is then an $I_{3,nc}^{s(012)}$
\begin{equation}
  I_{3,nc}^{s(012)}: \quad \left\{
  \begin{array}{c}
    (3,1,0,1,0,0,0,0) \cr
    [-,\sigma_1\sigma_2,\sigma_2\sigma_5,\sigma_1\sigma_3,\sigma_2\sigma_4 +
          \sigma_3\sigma_5,-,\sigma_3\sigma_4,-] 
\end{array} \right\} \,.
\end{equation}

\subsubsection{Polynomial enhancement in the $z \mid \fks_3$ branch}

The other relevant details we will provide concern enhancements from the singular $I_2^{(01|2)}$ which has vanishing orders $(1,1,1,0,0,0,0,0)$. The discriminant contains a seven-term polynomial
\begin{equation}
\begin{aligned}
    P &= s_{3,1}^2s_{5,0}^2 + s_{7,0}(s_{2,1}^2s_{5,0} -
            s_{1,1}s_{2,1}s_{6,0} + s_{1,1}^2s_{7,0}) + \cr
                    &\qquad +s_{3,1}(-s_{2,1}s_{5,0}s_{6,0} + s_{1,1}(s_{6,0}^2
                    - 2s_{5,0}s_{7,0}))
                            \,.
  \end{aligned}
\end{equation}
Since a generic solution is not known for this polynomial, we again take advantage of a simple canonical solution given by $s_{1,1}=s_{2,1}=s_{3,1}=0$. We see that $z^4\mid \Delta$ and we observe a singular $I_4^{ns(01||2)}$
\begin{equation}
I_{4}^{ns(01||2)} \,:\, (2,2,2,0,0,0,0,0) \,.
\end{equation}
As in the previous case, we notice that the component of the discriminant $s_{6,0}^2-4 s_{5,0}s_{7,0}$ provides the condition for the split/non-split distinction. If this quantity is a perfect, non-zero square, then applying the solution given in appendix \ref{app:poly} we have a split $I_{4,nc}^{s(01||2)}$
\begin{equation}
  I_{4,nc}^{s(01||2)}: \quad \left\{
  \begin{array}{c}
    (2,2,2,0,0,0,0,0) \cr
    [-,-,-,\sigma_1\sigma_3,\sigma_1\sigma_2 +
          \sigma_3\sigma_4,\sigma_2\sigma_4,-,-] 
\end{array} \right\} \,.
\end{equation}

As in the previous subsection, we notice that if we only require $s_{1,1}=0$ the seven term polynomial reduces to the usual three-term one
\begin{equation}
  P|_{(s_{1,2}=0)}=s_{5,0}(s_{3,1}^2s_{5,0}-s_{3,1}s_{6,0}s_{2,1} +s_{2,1}^2s_{7,0}) \,.
\end{equation}
The solution involving setting $s_{5,0}$ to zero in addition to $s_{1,2}$
would give the fibration defined by the vanishing orders $(2,1,1,1,0,0,0,0)$
which is an $I_3^{s(01|2)}$ fiber
\begin{equation}
  I_3^{s(01|2)} \,:\, (2,1,1,1,0,0,0,0) \,.
\end{equation}
We can also apply the non-canonical solution of appendix \ref{app:poly} to the
three-term component
\begin{equation} 
 s_{2,1} = \sigma_1\sigma_2 \,,\quad s_{3,1} = \sigma_1\sigma_3 \,,\quad
  s_{7,0} = \sigma_3\sigma_4 \,,\quad s_{5,0} = \sigma_2\sigma_5 \,,\quad
  s_{6,0} = \sigma_2\sigma_4 + \sigma_3\sigma_5 \,.
\end{equation}
Upon substitution we find an $I_{3,nc}^{s(01|2)}$ singular fiber
\begin{equation}
  I_{3,nc}^{s(01|2)}: \quad \left\{
  \begin{array}{c}
    (2,1,1,0,0,0,0,0) \cr
    [-,\sigma_1\sigma_2,\sigma_1\sigma_3,\sigma_2\sigma_5,\sigma_2\sigma_4 +
          \sigma_3\sigma_5,\sigma_3\sigma_4,-,-]
\end{array} \right\} \,.
\end{equation}

%\subsubsection{A remark about coprimality}
%
%The final comment we want to make in this section is related to repeatedly
%non-canonical models and coprimality conditions that can arise between
%sections. We will take as an example of this feature the non-canonical type
%$II_{nc}^{(012)}$ fiber. The discriminant contains a factor
%\begin{equation}\label{TwoTermforCoprimality}
%\Delta \supset P= (s_{7,0}\sigma_8-\sigma_3s_{9,0}) \,.
%\end{equation}
%Recalling that the $\sigma_i$ were introduced in (\ref{solutionpoly:II_nc^012}), we refer to appendix \ref{app:poly} for an explanation of the coprimality between $\sigma_3$ and $\sigma_8$, meaning that they do not share any irreducible components. This turns out to be relevant because the non-canonical solution to (\ref{TwoTermforCoprimality}) reads
%\begin{equation}
%s_{7,0}=\xi_1\xi_2,\qquad \sigma_8=\xi_3 \xi_4,\qquad
%\sigma_3=\xi_1\xi_3,\qquad s_{9,0}=\xi_2\xi_4 \,.
%\end{equation}
%But, as we just remarked, coprimality between $\sigma_3$ and $\sigma_8$
%implies that we must have $\xi_3 = 1$ as otherwise it would represent a non-trivial common factor. Then we find a type $III_{nc^2}^{(1|02)}$ singular fiber
%\begin{equation}
%  III_{nc^2}^{(1|02)} \,:\, (1,1,0,1,0,0,0,0) \,,\,
%  [-,-,\mu\xi_1^2,-,2\mu\xi_1\xi_4,\xi_1\xi_2,\mu\xi_4^2,\xi_2\xi_4] \,.
%\end{equation}

\subsection{Enhancements from $\ord_z(\Delta) = 3$}

We now proceed to consider enhancements of the
discriminant starting from the fibers with $\ord_z(\Delta) = 3$, listed in
table
\ref{TableNonCanonicalord3}, and we report here the cases that
deserve mention due to some peculiarity. In particular we will consider
distinctions between split and non-split singular fibers and an instance where
we will need to consider the structure of the algorithm in order
not to reproduce singular fibers already obtained.

\begin{figure}
  \centering
    \includegraphics[height=3cm]{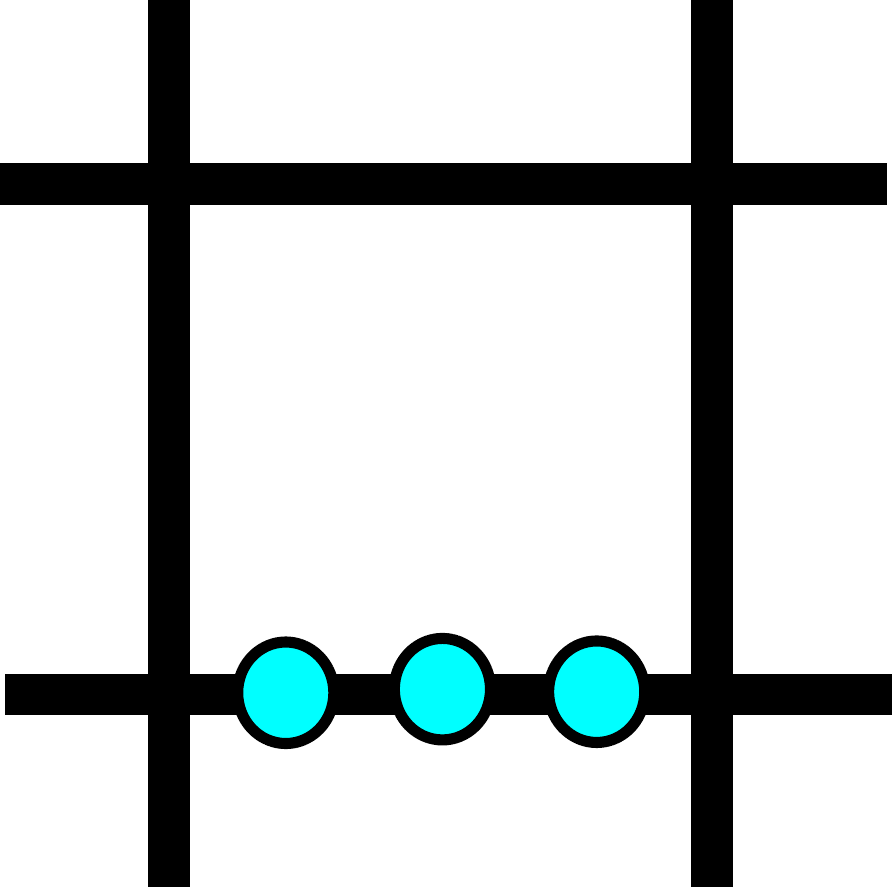} \qquad
    \includegraphics[height=3cm]{I412s3.pdf} \qquad
    \includegraphics[height=3cm]{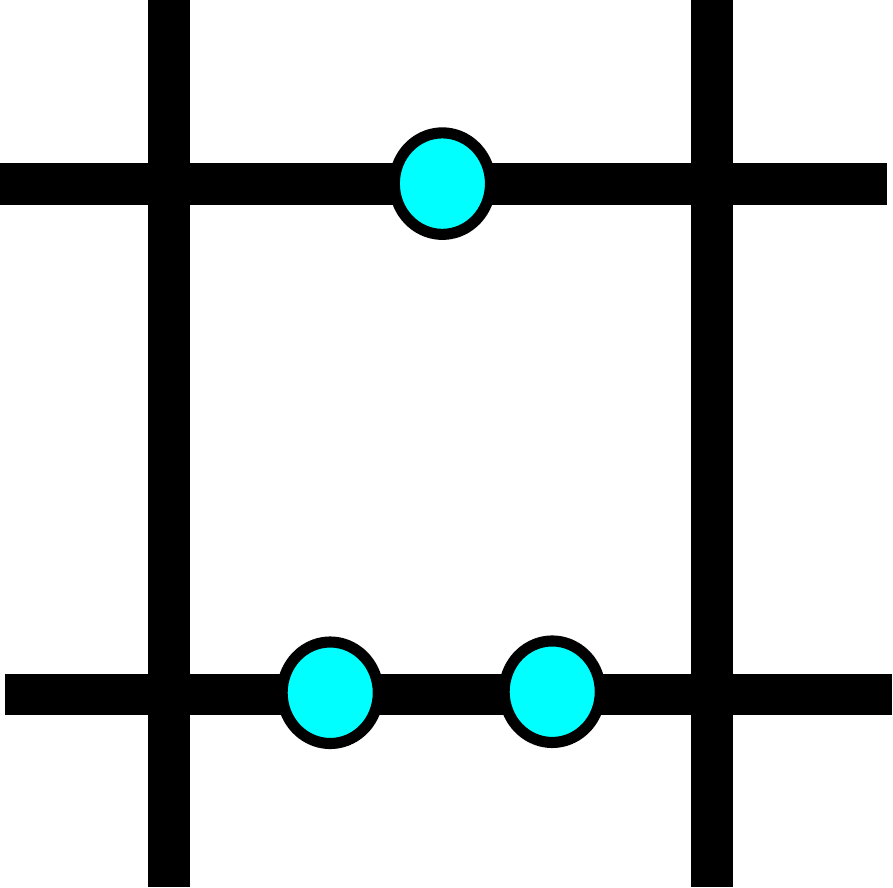} \qquad
    \includegraphics[height=3cm]{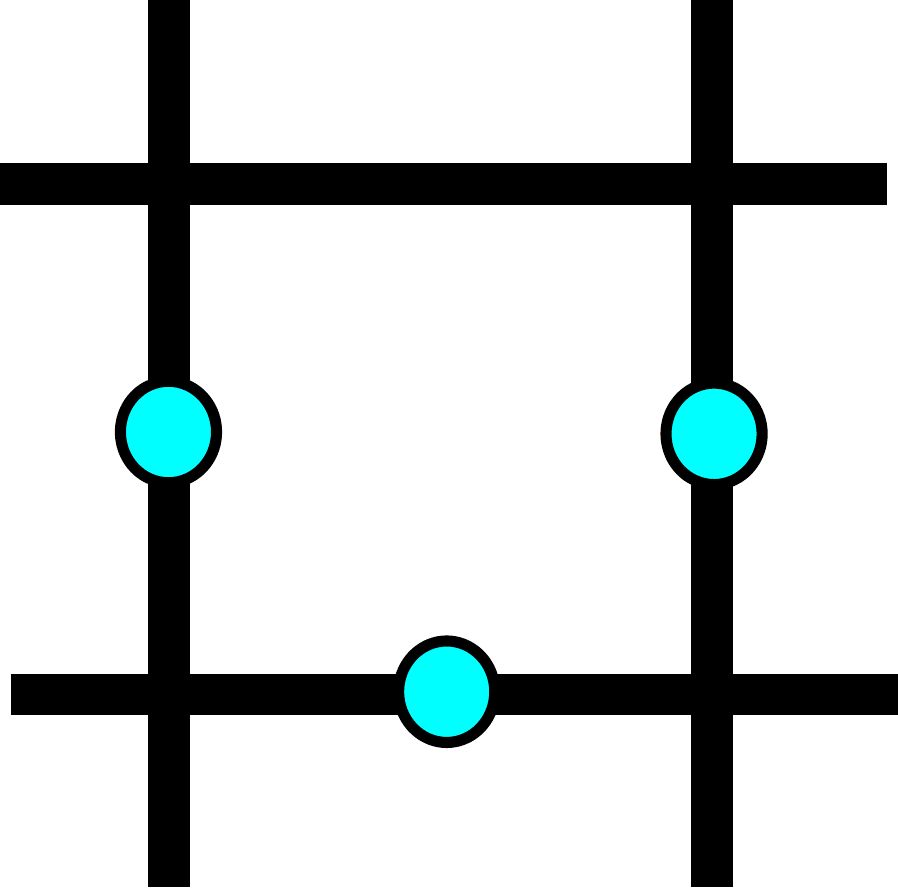} 
    \caption{The $I_4$ singular fibers and the decorations detailing where the
    rational sections can intersect. The fibers shown are $I_4^{(ijk)}$, $I_4^{(ij|k)}$, $I_4^{(ij||k)}$ and
    $I_4^{(i|j|k)}$ fibers.}
  \label{I4Fibs}
\end{figure}

\subsubsection{Split/non-split distinction}

We recall that in the previous section we found an $I_3^{ns(012)}$ singular
fiber and we now determine the enhancements of this fiber. The discriminant takes the form
\begin{equation}
  \Delta = s_{1,3}s_{3,0}s_{8,0}(s_{6,0}^2 - 4s_{3,0}s_{8,0})
  (s_{7,0}^2s_{8,0} - s_{6,0}s_{7,0}s_{9,0} + s_{3,0}s_{9,0}^2)z^3 
  + \mathcal{O}(z^4) \,.
\end{equation}
The simple enhancement $s_{1,3}=0$ will produce an $I_4^{ns(012)}$ singular fiber
\begin{equation}
I_{4}^{ns(012)} \,:\, (4,2,0,2,0,0,0,0) \,.
\end{equation}
As already observed, the discriminant component $s_{6,0}^2 - 4s_{3,0}s_{8,0}$
indicates that when this quantity is a perfect, non-zero square, we obtain the
split version of the singular fiber. Applying the solution in appendix \ref{app:poly} we then find the singular $I_{4,nc}^{s(012)}$
\begin{equation}
  I_{4,nc}^{s(012)}: \quad \left\{
  \begin{array}{c}
     (4,2,0,2,0,0,0,0) \cr
     [-,-,\sigma_1\sigma_3,-,\sigma_1\sigma_2 +
           \sigma_3\sigma_4,-,\sigma_2\sigma_4,-]
\end{array} \right\} \,.
\end{equation}

Another instance where the split/non-split distinction arises is in the case
of type $IV$  fibers. Consider the singular $III_{nc}^{(012)}$ listed in table
\ref{TableNonCanonicalord3}. This has discriminant
\begin{equation}
\Delta = \mu^6
\sigma_3\sigma_8(s_{5,1}\sigma_3-s_{2,1}\sigma_8)(s_{9,0}\sigma_3-s_{7,0}\sigma_8)z^3+\mathcal{O}(z^4)
\,.
\end{equation}
We remark that this was obtained in the algorithm by an application of the non-canonical solution to $s_{6,0}^2-4s_{3,0} s_{8,0}=0$ and therefore $\sigma_3$ and $\sigma_8$ are coprime. Enhancing the discriminant by solving non-canonically the first of the two-term polynomials requires setting
\begin{equation}
s_{5,1}=\xi_1\xi_2,\qquad\sigma_3=\xi_3,\qquad s_{2,1}=\xi_1\xi_3,\qquad
\sigma_8=\xi_2 \,.
\end{equation}
Where coprimality of $(\sigma_3,\sigma_8)$ was used in order to set $\xi_4 =
1$. The singular fiber corresponding to this enhancement is a type $IV_{nc^2}^{ns(012)}$
\begin{equation}
  IV_{nc^2}^{ns(012)}: \quad \left\{
  \begin{array}{c}
     (2,1,0,1,0,0,0,0) \cr
     [-,\xi_1\xi_3,\mu\xi_3^2,\xi_1\xi_2,2\mu\xi_2\xi_3,-,\mu\xi_2^2,-]
\end{array} \right\} \,.
\end{equation}
Then the discriminant indicates the quantity that needs to be a perfect square
in order for the fiber to become a split type $IV_{nc^3}^{s(012)}$. This is
$\xi_1^2-4 \mu s_{1,2}$, and applying the solution in appendix \ref{app:poly} we find
\begin{equation}
  IV_{nc^3}^{s(012)}: \quad \left\{
  \begin{array}{c}
     (2,1,0,1,0,0,0,0) \cr
      [\delta_2\delta_4,\xi_3(\delta_1 \delta_2+\delta_3 \delta_4),\delta_1
        \delta_3\xi_3^2,\xi_2(\delta_1\delta_2+\delta_3\delta_4),2\delta_1\delta_3\xi_2\xi_3,-,\delta_1\delta_3\xi_2^2,-]
\end{array} \right\} \,.
\end{equation}

\begin{figure}
  \centering
    \includegraphics[height=3cm]{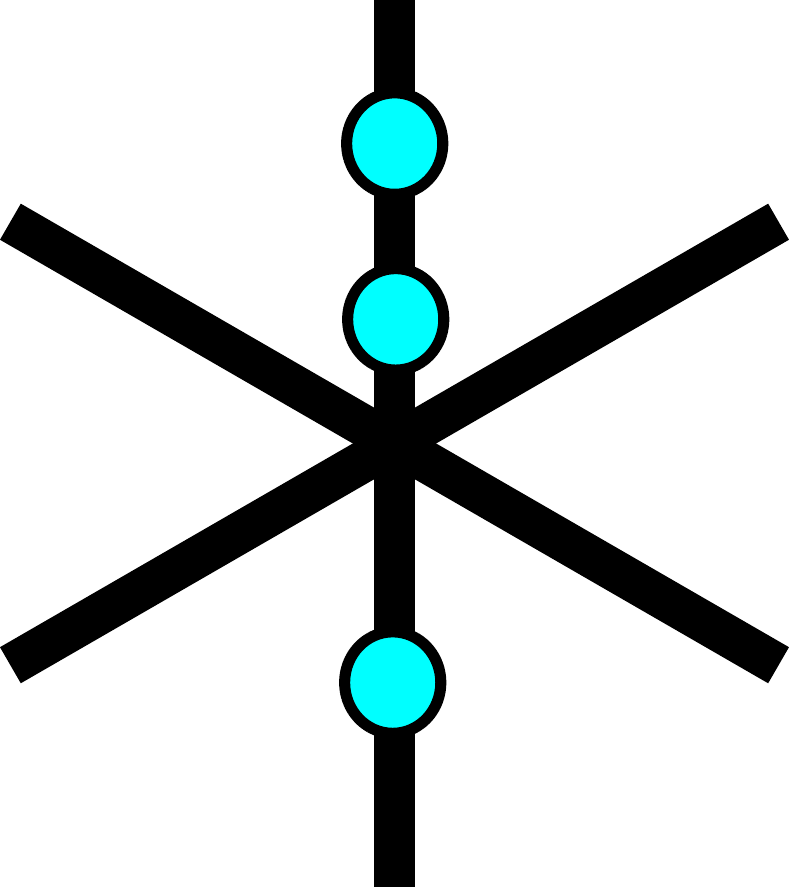} \qquad\qquad
    \includegraphics[height=3cm]{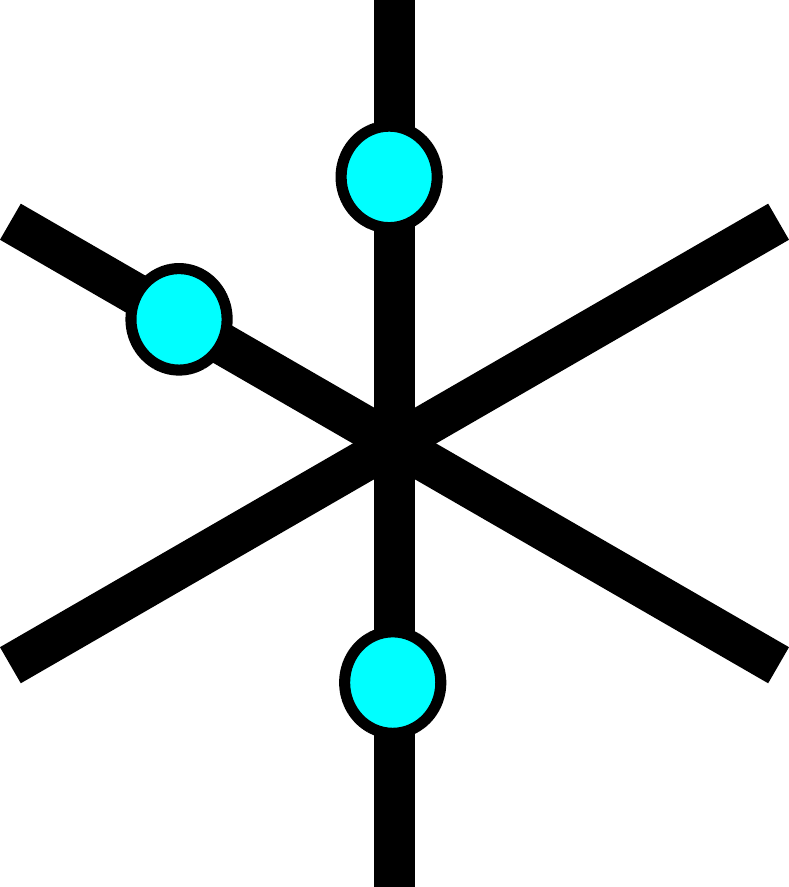} \qquad\qquad
    \includegraphics[height=3cm]{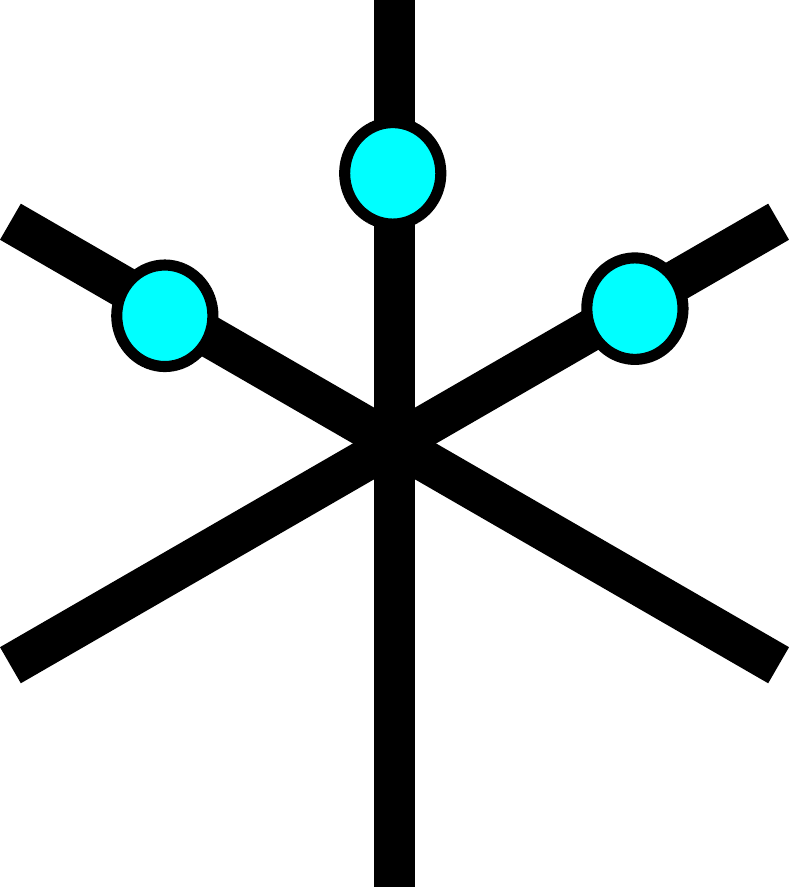} 
    \caption{The $IV$ fibers. We denote by the blue nodes the components of
    the fiber which are intersected by the sections. In the order, the fiber shown are $IV^{(ijk)}$, $IV^{(i|jk)}$ and $IV^{(i|j|k)}$ fibers.}
  \label{IVFibs}
\end{figure}

\subsubsection{Commutative enhancement structure of the algorithm}

We consider enhancements from the $III_{nc^2}^{(1|02)}$ fiber type. This was
found by applying twice the solutions in appendix \ref{app:poly}. Schematically
\begin{equation}
I_1^{(012)}\longrightarrow I_{2,nc}^{(1|02)}\longrightarrow
III_{nc^2}^{(1|02)} \,.
\end{equation} 
Noting that in the last step a coprimality condition had to be imposed, the discriminant of this singular fiber takes the form
\begin{equation}
\Delta = (s_{1,1}^3\xi_2^6\xi_3^6\xi_4^6 ) z^3 +\mathcal{O}(z^4) \,.
\end{equation}
We see that requiring the vanishing of any of the $\xi_i$ would imply setting to zero
two among $s_{7,0},s_{9,0},s_{3,0},s_{8,0}$, but we are not allowing the
vanishing of any of the those
sections to remain in our lop equivalence class or because we are in the $z
\nmid \fks_{3}$ branch of 
the algorithm. Moreover, we have considered the case $s_{1,1}=0$ in another part of the algorithm (specifically in going $I_1^{(012)}\rightarrow I_2^{(012)}$).
We can therefore conclude that all the enhancements would just reproduce
singular fibers found in other parts of the algorithm. The order in which the
enhancements are carried out is of no importance, but it is crucial,
in particular with non-canonical fibers, to keep track of which enhancements
would reproduce fiber types already obtained.  

\subsection{Enhancements from $\ord_z(\Delta) = 4$}
In this section we will proceed with the algorithm by again mentioning only
enhancements which require comment. In particular we will deal with the structure of obstructions to
full generality due to the complexity of polynomials in the discriminant, we will
encounter the distinction between split and semi-split fibers for $I_0^*$ and we will provide
details for one of the $I_{5,nc^3}$, obtained by solving non-canonically
polynomials in the discriminant three times.

\begin{figure}
  \centering
    \includegraphics[height=3cm]{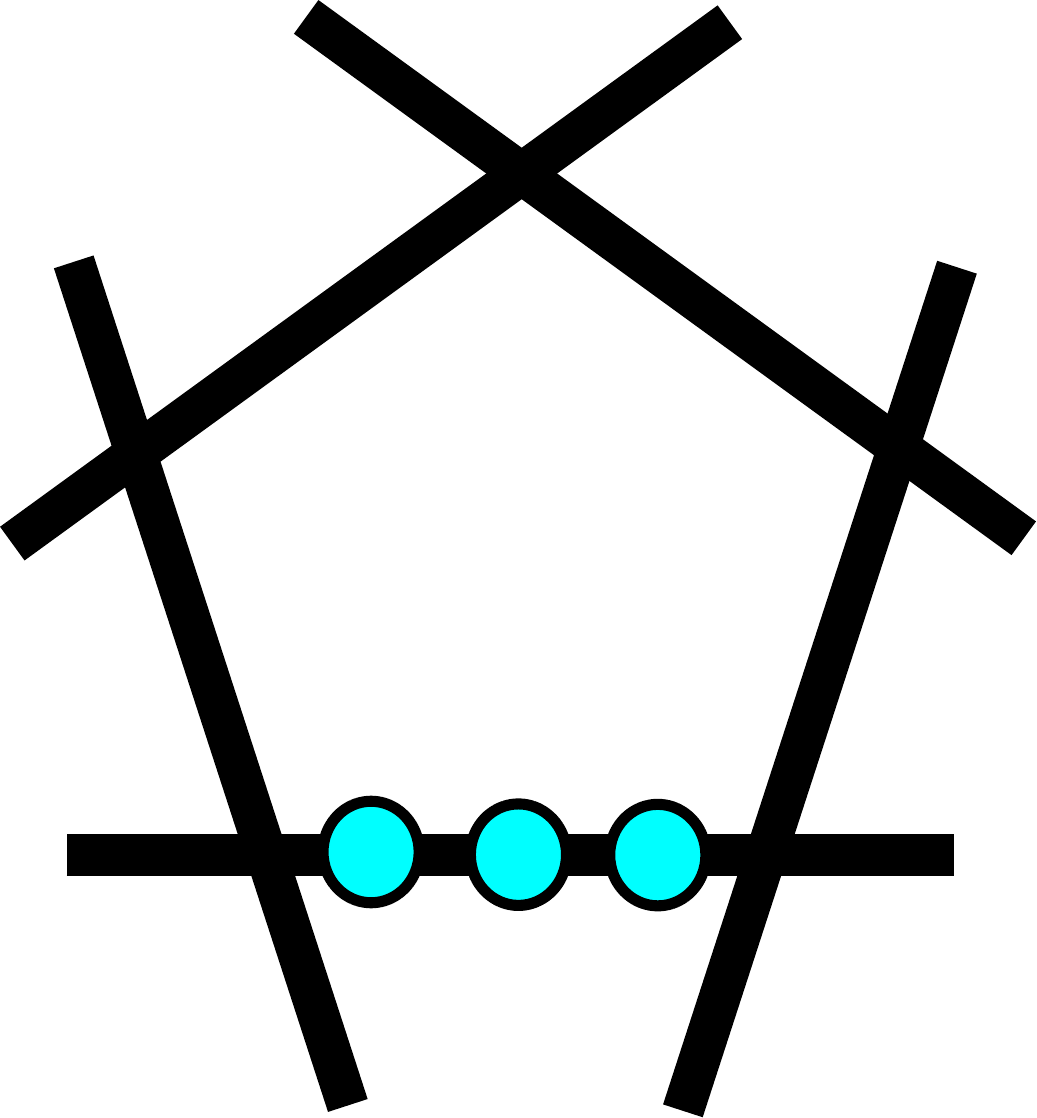} \qquad\qquad\quad
    \includegraphics[height=3cm]{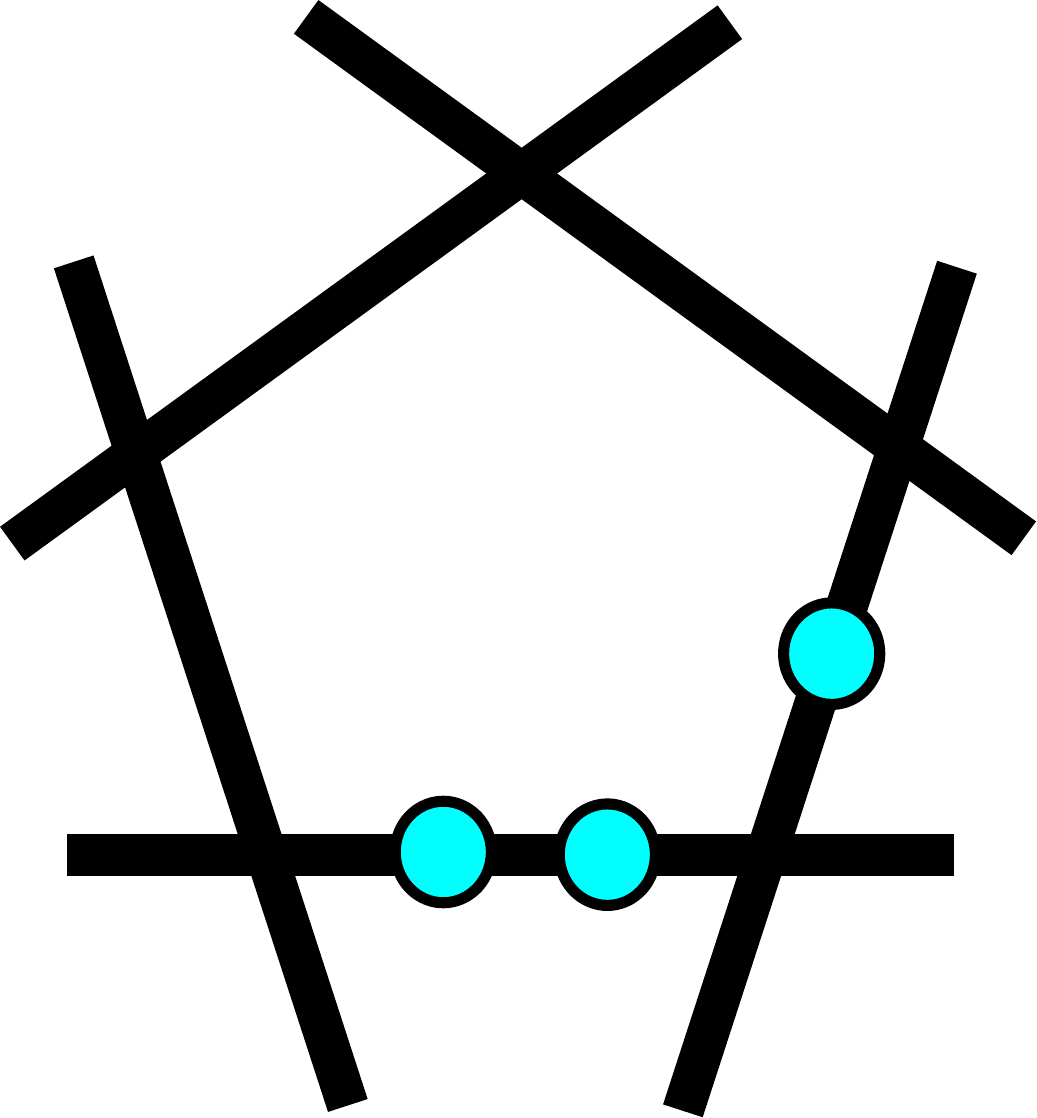} \qquad\qquad\quad
    \includegraphics[height=3cm]{I512ss3.pdf} \qquad\qquad\quad
    \includegraphics[height=3cm]{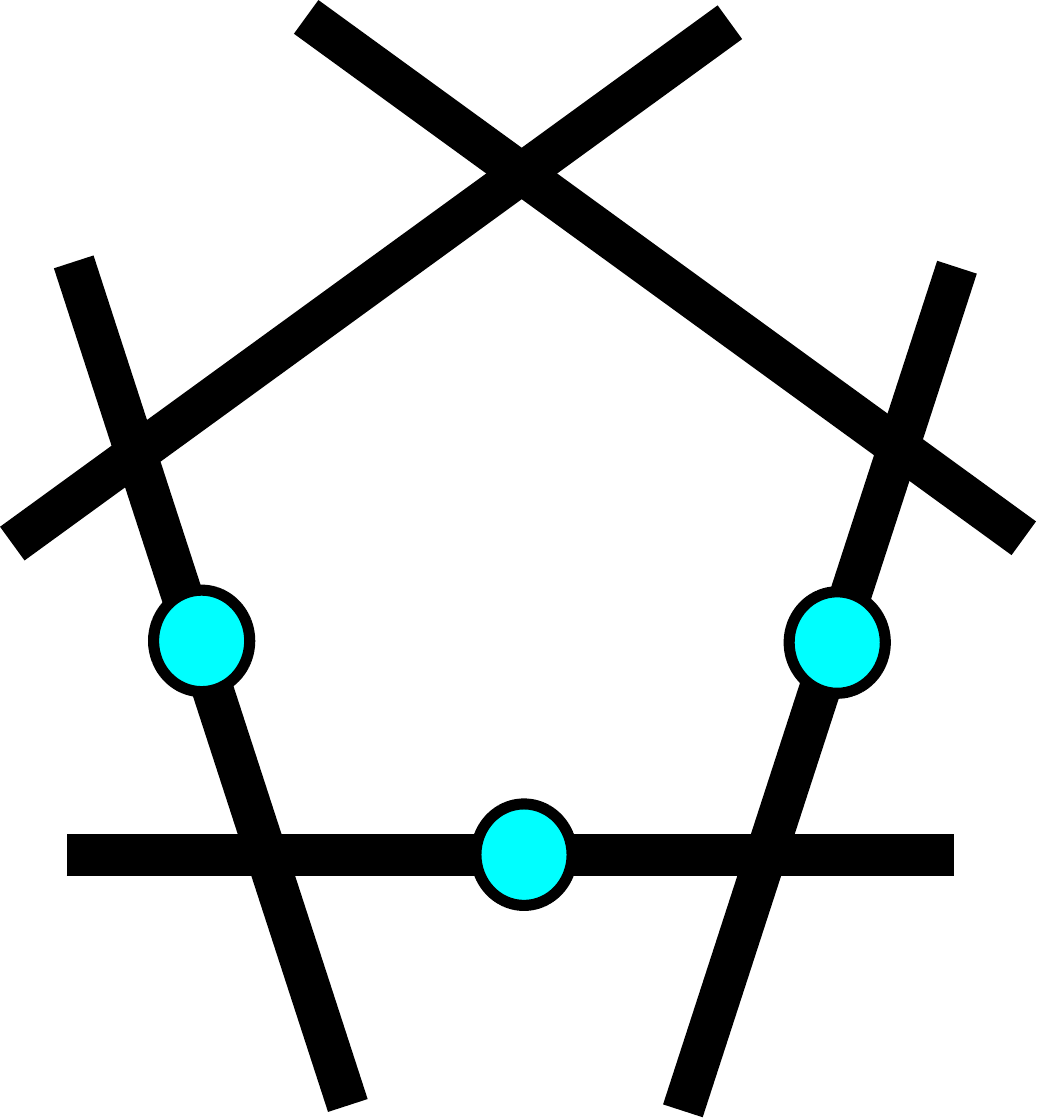} \qquad\qquad\quad
    \includegraphics[height=3cm]{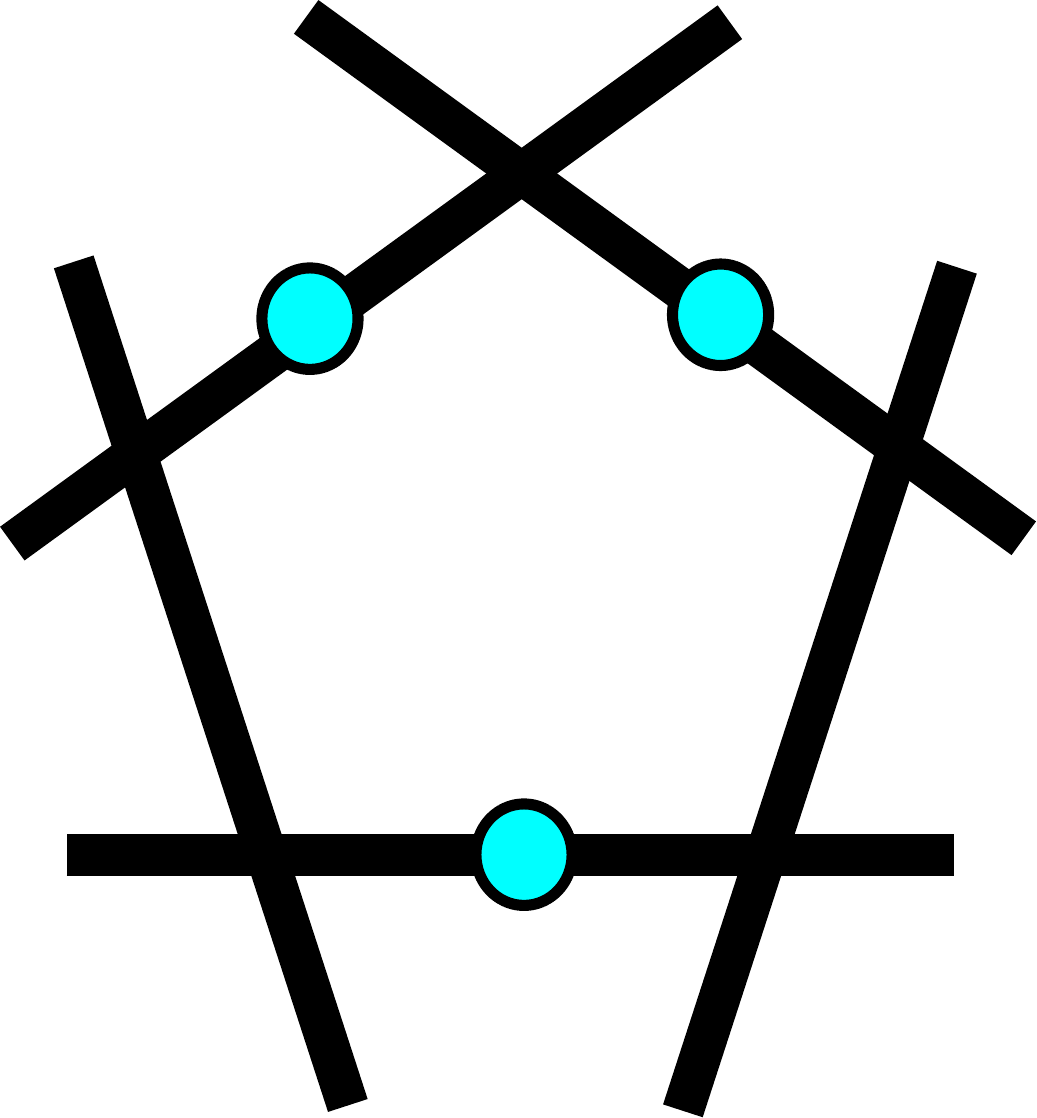} 
    \caption{The $I_5$ singular fibers. The possible intersections of the
    sections with the singular fibers are denoted by the positions of the blue
  nodes. The fibers shown in the first row are $I_5^{(ijk)}$, $I_5^{(ij|k)}$ and $I_5^{(ij||k)}$, whereas the fibers shown in the second row are, respectively, $I_5^{(i|j|k)}$ and $I_5^{(i|j||k)}$.}
  \label{I5Fibs}
\end{figure}

\subsubsection{Obstruction from polynomial enhancement}
At vanishing order of the discriminant $\ord_z(\Delta) = 4$ we find again the
two obstructions to full generality encountered at $\ord_z(\Delta) = 2$,
i.e. the same five-term and  seven-term polynomials. These come up
respectively in the discriminant of the singular fibers $I_4^{ns(012)}$ and
$I_4^{ns(01||2)}$, and will in fact be present at every even vanishing order
in the discriminants of $I_{2n}^{ns(012)}$ and $I_{2n}^{ns(01|^n2)}$. We
therefore review the singular fibers that we obtain from the enhancements.
More details can be found in section \ref{sec:Delta=O(z^2)}.

The discriminant of the singular fiber $I_4^{ns(012)}:(4,2,0,2,0,0,0,0)$ contains a component
\begin{equation}
\Delta\supset P = s_{8,0} s_{2,2}^2-s_{5,2} s_{6,0}
             s_{2,2}+s_{1,4}  s_{6,0}^2+s_{3,0} \left(s_{5,2}^2-4 s_{1,4}
                        s_{8,0}\right) \,.
\end{equation}
As in section \ref{sec:Delta=O(z^2)}  we consider two specific solutions.
The first one consists of setting $s_{1,4}=s_{2,2}=s_{5,2}=0$. This gives the singular fiber $I_5^{ns(012)}$
\begin{equation}
  I_{5}^{ns(012)} \,:\, (5,3,0,3,0,0,0,0) \,.
\end{equation}
Upon imposing the perfect square condition $s_{6,0}^2-4 s_{3,0}s_{8,0}=p^2$ we
find the singular fiber $I_{5}^{s(012)}$.
Alternatively, we set $s_{1,4}=0$ and we solve non-canonically, as in appendix
\ref{app:poly}, the resulting three-term polynomial polynomial
\begin{equation}
P|_{(s_{1,4}=0)}=s_{2,2}^2s_{8,0}-s_{2,2}s_{6,0}s_{5,2} +s_{5,2}^2s_{3,0} \,.
\end{equation}
This gives the non-canonical singular fiber $I_{5,nc}^{s(012)}$
\begin{equation}
  I_{5,nc}^{s(012)}: \quad \left\{
  \begin{array}{c}
     (5,2,0,2,0,0,0,0) \cr
     [-,\sigma_1\sigma_2,\sigma_2\sigma_5,\sigma_1\sigma_3,\sigma_2\sigma_4 +
           \sigma_3\sigma_5,-,\sigma_3\sigma_4,-]
\end{array} \right\} \,.
\end{equation}

The second obstruction that we encounter is, again, the seven-term polynomial in
the discriminant of the singular $I_4^{ns(01||2)}$
\begin{equation}
\begin{aligned}
\Delta \supset     P &= s_{3,2}^2s_{5,0}^2 + s_{7,0}(s_{2,2}^2s_{5,0} -
            s_{1,2}s_{2,2}s_{6,0} + s_{1,2}^2s_{7,0}) + \cr
                    &\qquad +s_{3,2}(-s_{2,2}s_{5,0}s_{6,0} + s_{1,2}(s_{6,0}^2
                    - 2s_{5,0}s_{7,0}))          \,.    
                           \
\end{aligned}                           
\end{equation}
The canonical solution that we consider requires $s_{1,2}=s_{2,2}=s_{3,2}=0$.
This gives a singular $I_6^{ns(01|||2)}$
\begin{equation}
  I_{6}^{ns(01|||2)} \,:\, (3,3,3,0,0,0,0,0) \,.
\end{equation}
The split version $I_6^{s(01|||2)}$ is found upon imposing that $s_{6,0}^2-4
s_{5,0} s_{7,0}$ is a perfect, non-zero square. 
We can also consider the solution where $s_{1,2} = 0$ and the three-term
polynomial component of the resulting polynomial is solved non-canonically.
This enhancement now produces an $I_{5,nc}^{s(01||2)}$, but this is
just a non-generic specialization of one of the $I_{5,nc^2}^{s(01||2)}$ fibers
also found in the algorithm and so we do not consider it further.

\subsubsection{Split/semi-split Distinction}

The split/semi-split distinction arises for singular fibers of Kodaira type
$I_0^*$. The example we provide concerns the possible enhancement of the
canonical type $IV^{s(01|2)}$, which was found schematically by
\begin{equation}
I_2^{(01|2)}\longrightarrow I_3^{s(01|2)}\longrightarrow IV^{s(01|2)} \,.
\end{equation}
The discriminant takes a rather simple form
\begin{equation}
\Delta = s_{2,1} s_{7,0} s_{8,0} z^4+\mathcal{O}(z^5) \,.
\end{equation}
The enhancement we will consider here is when $s_{2,1}=0$. As a consequence
$z^2 \mid \fks_2$ and $z^6 \mid \Delta$. This way we have found the semi-split
$I_0^{*ss(01|2)}$
\begin{equation}
 I_{0}^{*ss(01|2)} \,:\, (2,2,1,1,1,0,0,0) \,.\,
\end{equation}
In order for one of the curves of the $I_0^{*ss(01|2)}$ to split into two
separate non-intersecting components, we need to satisfy a perfect square
condition for the quantity $s_{5,1}^2-4 s_{1,2}s_{8,0}$. Following appendix \ref{app:poly} we find the split $I_{0,nc}^{*s(01|2)}$
\begin{equation}
  I_{0,nc}^{*s(01|2)}: \quad \left\{
  \begin{array}{c}
     (2,2,1,1,1,0,0,0) \cr
     [\sigma_1 \sigma_3,-,-,\sigma_1\sigma_2+\sigma_3\sigma_4,-,-,\sigma_2
       \sigma_4,-]
\end{array} \right\} \,.
\end{equation}

\subsubsection{A thrice non-canonical $I_5$}

In this section we provide details for an $I_{5,nc^3}^{s(0|1||2)}$. This singular fiber is observed in the algorithm by schematically enhancing
\begin{equation}
I_2^{(01|2)}\longrightarrow I_{3,nc}^{s(0|1|2)}\longrightarrow
I_{4,nc^2}^{s(0|1|2)}\longrightarrow I_{5,nc^3}^{s(0|1||2)} \,.
\end{equation}
All the three arrows represent non-canonical enhancements. In particular enhancing from $I_2^{(01|2)}$ to  $I_{3,nc}^{s(0|1|2)}$ requires solving a three-term polynomial present in the discriminant. This is $\Delta\supset (s_{8,0}^2s_{7,0}-s_{8,0}s_{6,0}s_{9,0}+s_{9,0}^2s_{5,0})$, which is solved by requiring
\begin{equation}
s_{8,0}=\sigma_1\sigma_2,\quad s_{9,0}=\sigma_1\sigma_3, \quad
s_{7,0}=\sigma_3 \sigma_4, \quad s_{5,0}=\sigma_2 \sigma_5, \quad
s_{6,0}=\sigma_2\sigma_4+\sigma_3 \sigma_5 \,.
\end{equation} 
Note that this solution implies that $(\sigma_2,\sigma_3)$ are coprime. This
gives an $I_{3,nc}^{s(0|1|2)}$. Looking at the discriminant of this singular fiber we see that one of the components is  $\Delta\supset (\sigma_3^2s_{1,1}-\sigma_2\sigma_3s_{2,1}+\sigma_2^2s_{3,1})$. We apply again the same solution to this three-term polynomial
\begin{equation}
\sigma_3=\xi_2,\quad \sigma_2=\xi_3, \quad s_{1,1}=\xi_3 \xi_4, \quad
s_{3,1}=\xi_2 \xi_5, \quad s_{2,1}=\xi_2\xi_4+\xi_3 \xi_5 \,.
\end{equation}
Where we used that $(\sigma_2,\sigma_3)$ are coprime to set $\xi_1 = 1$.
We have now enhanced the singular fiber to an $I_{4,nc^2}^{s(0|1|2)}$. To obtain
the thrice non-canonical $I_5$ we now consider the two-term polynomial
contained in the discriminant at fourth order: $\Delta\supset (\sigma_4
\xi_4-\sigma_5 \xi_5)$. Applying the non-canonical solution in appendix
\ref{app:poly}
\begin{equation}
\sigma_4=\delta_1 \delta_2,\quad \xi_4=\delta_3 \delta_4,\quad
\sigma_5=\delta_1 \delta_3,\quad \xi_5=\delta_2 \delta_4 \,.
\end{equation}
We have now reached the singular fiber $I_{5,nc^3}^{s(0|1||2)}$
\begin{equation}
  I_{5,nc^3}^{*s(0|2||1)}: \quad \left\{
  \begin{array}{c}
     (1,1,1,0,0,0,0,0) \cr
     [\xi_3\delta_3\delta_4,\delta_4(\delta_3\xi_2+\delta_2\xi_3),\xi_2
       \delta_2\delta_4,\xi_3\delta_1\delta_3,\delta_1(\delta_2\xi_3+\delta_3
       \xi_2),\delta_1\delta_2\xi_2,\sigma_1\xi_3,\sigma_1\xi_2]
\end{array} \right\} \,.
\end{equation}

\section{$U(1)$ Charges of $SU(5)$ Fibers}\label{sec:Spectra}

In section \ref{sec:TateAlg} a variety of different, canonical and
non-canonical, $I_5$ type singular fibers were found, and are listed in table 
\ref{TableNonCanonicalord5}. As elliptic fibrations with $SU(5)$ singular
fibers are phenomenologically interesting in this section the
$U(1)$ charges of the matter loci are determined for the $I_5$ fibers
obtained,
which lie in the chosen
lop-equivalence class. The $U(1)$ charges are calculated from the intersection
number of the matter curve with the Shioda mapped rational section, as
explained in section \ref{sec:Resolutions}. For the canonical $I_5$ singular
fibers we find, as expected, the same results that were found from the study of
toric tops. Details of the relationship between the canonical models
and the $SU(5)$ top models and their charges as found in
\cite{Borchmann:2013hta} are given.  
In the algorithm a number of non-canonical models which, as far as the authors are
aware have not been seen before, were found, some of which can realize two or 
three distinctly charged ${\bf 10}$ matter curves, potentially a desirable
feature, also some models realize as many as seven differently charged ${\bf 5}$
matter curves, which are of some interest in light of the phenomenological
study in  \cite{Krippendorf:2014xba}.

\subsection{Canonical $I_5$ Models}

The $U(1)$ charges of the canonical models are found in table \ref{tbl:CU1}.
Models with these particular $U(1)$ charges are well-studied in the
literature. In this subsection we provide a short comparison to the known
toric constructions from tops \cite{Candelas:1996su} , which were constructed with two extra sections
in \cite{Borchmann:2013hta, Mayrhofer:2012zy, Borchmann:2013jwa,
Braun:2013nqa}.

The toric tops as extracted from \cite{Borchmann:2013hta} are also given by
vanishing orders of the coefficients of the cubic polynomial (\ref{eqn:P111cubic}) 
 and are related to what we called canonical models. 
In
order to see this we need to perform a series of lop transformations to bring them to the
equivalence class of singular fibers considered in this paper. 
Section \ref{sec:sym} contains the details of the lop transformations.  All the tops were found
as part of the algorithm and exhaust the canonical models. The $U(1)$ charges
of the matter content matched the results found here identically for what was called tops 1 and 2, 
whereas for tops 3 and 4 a different linear combination of the $U(1)$
charges was used. The details of this linear combination are given in terms of
our choice of $U(1)$ generators. 

\begin{table}[t]
  \centering\footnotesize
  \begin{equation*}
    \begin{array}{|c|c|c|c|}
      \hline
      \text{Fiber} & \text{Model} & \text{Matter Locus} & \text{Matter}
      \cr\hline

      I_5^{s(0|1||2)} & 
      \begin{array}{c}
        (2,2,2,1,0,0,1,0) \cr
      \end{array} &
      \begin{array}{c}
        s_{1,2} \cr
        s_{6,0} \cr
        s_{7,0} \cr
        s_{9,0} \cr
        s_{6,0}s_{8,1} - s_{5,1}s_{9,0} \cr
        s_{3,2}s_{6,0}^2 - s_{2,2}s_{6,0}s_{7,0} + s_{1,2}s_{7,0}^2 \cr
      \end{array} &
      \begin{array}{c}
        {\bf 5}_{3,1} \oplus {\bf \overline{5}}_{-3,-1} \cr
        {\bf 10}_{1,2} \oplus {\bf \overline{10}}_{-1,-2} \cr
        {\bf 5}_{3,-4} \oplus {\bf \overline{5}}_{-3,4} \cr
        {\bf 5}_{3,6} \oplus {\bf \overline{5}}_{-3,-6} \cr
        {\bf 5}_{-2,1} \oplus {\bf \overline{5}}_{2,-1} \cr
        {\bf 5}_{-2,-4} \oplus {\bf \overline{5}}_{2,4} \cr
      \end{array} \cr\hline
      
      I_5^{s(01||2)} & 
      \begin{array}{c}
        (3,2,2,1,0,0,0,0) \cr
      \end{array} &
      \begin{array}{c}
        s_{6,0} \cr
        s_{7,0} \cr
        s_{8,0} \cr
        s_{1,3}s_{6,0} - s_{2,2}s_{5,1} \cr
        s_{3,2}s_{6,0} - s_{2,2}s_{7,0} \cr
        s_{6,0}s_{9,0} - s_{7,0}s_{8,0} \cr
      \end{array} &
      \begin{array}{c}
        {\bf 10}_{-1,0} \oplus {\bf \overline{10}}_{1,0} \cr
        {\bf 5}_{2,-1} \oplus {\bf \overline{5}}_{-2,1} \cr
        {\bf 5}_{-3,-1} \oplus {\bf \overline{5}}_{3,1} \cr
        {\bf 5}_{2,0} \oplus {\bf \overline{5}}_{-2,0} \cr
        {\bf 5}_{-3,0} \oplus {\bf \overline{5}}_{3,0} \cr
        {\bf 5}_{2,1} \oplus {\bf \overline{5}}_{-2,-1} \cr
      \end{array} \cr\hline
      
      I_5^{s(0|1|2)} & 
      \begin{array}{c}
        (3,2,1,1,0,0,1,0) \cr
      \end{array} &
      \begin{array}{c}
        s_{3,1} \cr
        s_{6,0} \cr
        s_{7,0} \cr
        s_{9,0} \cr
        s_{5,1}s_{9,0} - s_{6,0}s_{8,1} \cr
        s_{3,1}s_{5,1}^2 - s_{2,2}s_{5,1}s_{6,0} + s_{1,3}s_{6,0}^2 \cr
      \end{array} &
      \begin{array}{c}
        {\bf 5}_{-3,1} \oplus {\bf \overline{5}}_{3,-1} \cr
        {\bf 10}_{-1,2} \oplus {\bf \overline{10}}_{1,-2} \cr
        {\bf 5}_{2,-4} \oplus {\bf \overline{5}}_{-2,4} \cr
        {\bf 5}_{2,6} \oplus {\bf \overline{5}}_{-2,-6} \cr
        {\bf 5}_{2,1} \oplus {\bf \overline{5}}_{-2,-1} \cr
        {\bf 5}_{-3,-4} \oplus {\bf \overline{5}}_{3,4} \cr
      \end{array} \cr\hline
      
      I_5^{s(01|2)} & 
      \begin{array}{c}
        (4,2,1,2,0,0,0,0) \cr
      \end{array} &
      \begin{array}{c}
        s_{3,1} \cr
        s_{6,0} \cr
        s_{7,0} \cr
        s_{8,0} \cr
        s_{6,0}s_{9,0} - s_{7,0}s_{8,0} \cr
        s_{1,4}s_{6,0}^2 - s_{2,2}s_{5,2}s_{6,0} + s_{2,2}^2s_{8,0} \cr
      \end{array} &
      \begin{array}{c}
        {\bf 5}_{4,0} \oplus {\bf \overline{5}}_{-4,0} \cr
        {\bf 10}_{-2,0} \oplus {\bf \overline{10}}_{2,0} \cr
        {\bf 5}_{-1,1} \oplus {\bf \overline{5}}_{1,-1} \cr
        {\bf 5}_{4,1} \oplus {\bf \overline{5}}_{-4,-1} \cr
        {\bf 5}_{-1,-1} \oplus {\bf \overline{5}}_{1,1} \cr
        {\bf 5}_{-1,0} \oplus {\bf \overline{5}}_{1,0} \cr
      \end{array} \cr\hline
      
    \end{array}
  \end{equation*}
  \caption{$U(1)$ charges of the canonical $I_5$ models from table
    \ref{TableNonCanonicalord5}.}
    \label{tbl:CU1}
\end{table}

In table \ref{tbl:toprel} the tops are listed with the numbering and vanishing orders
as in appendix A of \cite{Borchmann:2013hta}(polygon 5), the lop equivalent
models as found in Tate's algorithm and details for the linear combination of
the $U(1)$ charges for top 3 and top 4.

\begin{table}[H]
  \centering %\footnotesize
  \begin{tabular}{|c|c|c|c|c|}
    \hline
  	Top  & Fiber & Vanishing Orders & Lop-equivalent model & $U(1)$ Linear Combination \cr \hline
    Top 1 & $I_5^{s(0|1|2)}$ &$(2,2,2,0,0,1,0,0)$ & $(3,2,1,1,0,0,1,0)$ & -  \cr\hline 
    Top 2 & $I_5^{s(0|1||2)}$ & $(1,2,3,0,0,1,0,0)$ &$(2,2,2,1,0,0,1,0)$&-\\\hline 
    Top 3 & $I_5^{s(01||2)}$ & $(1,1,2,0,0,2,0,0)$ & $(3,2,2,1,0,0,0,0)$ & 
      $\begin{array}{c}
        u_1 = - \text{w}_1 \cr 
        u_2 = \frac{1}{5}(\text{w}_2 - \text{w}_1)
      \end{array}$ \\\hline 
    Top 4 & $I_5^{s(01|2)}$ & $(1,1,1,0,0,2,0,1)$ &$(4,2,1,2,0,0,0,0)$& 
      $\begin{array}{c}
        u_1 = - \text{w}_1 \cr 
        u_2 = \frac{1}{5}(\text{w}_2 - \text{w}_1)
      \end{array}$ \\\hline 
  \end{tabular}
  \caption{The lop-equivalent models of the four tops from
    \cite{Borchmann:2013hta}. The linear combination of the $U(1)$ charges
    gives the charges found in table \ref{tbl:CU1}, $u_1$ and $u_2$, in
  terms of the $U(1)$ charges of the top model, w$_1$ and w$_2$. The reason the charges of tops
  3 and 4 differ is because the lop translation involves the $\mathbb{Z}_2$
symmetry, which exchanges two of the marked points.}
\label{tbl:toprel}
\end{table}

\subsection{Non-canonical $I_5$ Models}

Listed in \cref{tbl:NC1,tbl:NC2,tbl:NC3} are the $U(1)$ charges
of the, respectively once, twice, and thrice, non-canonical $I_5$ models found
in the algorithm. The $U(1)$ charge generators are given by the Shioda map, as
described in section \ref{sec:Resolutions}, where the zero-section of the
fibration corresponds to the divisor $l_1 = 0$ after the $\mathbb{P}^2$
fibration 
ambient space has been blown up into $dP_2$. As opposed to the canonical
models the majority of the models tabulated in this section were previously
unknown. Some of these models appear to have interesting properties for
phenomenology, such as the above noted multiple differently charged ${\bf 10}$
and 
%the profusion of 
${\bf 5}$ curves. 

While Tate's algorithm provides a generic procedure there are some caveats
that were introduced in the application of it studied in this paper. There are situations
where we were not able to solve for the enhancement locus in the discriminant
to a reasonable degree of generality. In these cases we have sometimes, as
discussed in section \ref{sec:TateAlg}, used a less generic solution where it
was obtainable in such a way that it did not lead to obvious irregularities
with the model. In cases where no such solution was obtained we have left that
particular subbranch of the Tate tree unexplored. 

Throughout the application of Tate's algorithm the fibrations have remained
inside the chosen lop-equivalence class and so each
each model in these tables then
represents an entire lop-orbit of fibrations. The $\mathbb{Z}_2$ symmetry
which acts inside this orbit interchanges two of the three marked points of
the fibration, which correspond, in the $dP_2$ hypersurface, to the exchange
of $l_1$ and $l_2$. As the $U(1)$ charges are computed from a Shioda map where
the zero-section is taken to be $l_1 = 0$ the $U(1)$ charges are rewritten as
a linear combination under this symmetry in an identical manner to the linear
combinations occurring in the tops in table \ref{tbl:toprel}.

One may point out the surprising paucity of non-minimal matter loci in these
models with highly specialised coefficients. In the fibrations which are at
least twice non-canonical there can occur polynomial enhancement loci where
some of the terms in the solutions (as given in appendix \ref{app:poly}) are
fixed by a coprimality condition coming from a previously solved polynomial.
Were these terms not fixed to unity by the algorithm then they would
contribute non-minimal loci to the fibrations.

In \cite{Borchmann:2013hta, Mayrhofer:2012zy} there were listed tops
corresponding to an $SU(4)$ non-abelian singularity with two additional
rational sections, and it was noted that one expects multiple {\bf 10} matter
curves where these tops are specialized with some non-generic coefficients of
the defining polynomial, and such a model, which realizes multiple ${\bf 10}$
curves, was constructed there from the $SU(4)$ tops.
Included in table \ref{tbl:su4top} are the relations (via the lops) between
these $SU(4)$ tops and the $SU(4)$ canonical models which underlie the once
non-canonical $SU(5)$ models obtained in the algorithm. 
\begin{table}[H]
\centering
\begin{tabular}{|c|c|c|c|}
  \hline
	Top Model & Fiber & Vanishing Orders & Lop-equivalent Model  \\ \hline
  Top 1 & $I_4^{(0|1|2)}$ & $(1,1,2,0,0,1,0,0)$ & $(2,1,1,1,0,0,1,0)$ \\\hline 
  Top 2 & $ I_4^{(01||2)}$ & $(0,1,2,0,0,2,0,0)$ & $(2,2,2,1,0,0,0,0)$ \\\hline 
  Top 3 & $I_4^{(01|2)}$ & $(1,1,1,0,0,1,0,1)$ & $(3,2,1,1,0,0,0,0)$ \\\hline 
  Top 4 & $I_4^{(01||2)}$ & $(0,1,2,0,0,1,0,1)$ & $(2,2,2,1,0,0,0,0)$ \\\hline 
  Top 5 & $I_4^{(01|2)}$ & $(0,0,1,0,0,2,0,1)$ & $(3,2,1,1,0,0,0,0)$ \\\hline 
\end{tabular}
\caption{The $SU(4)$ tops associated to polygon 5 in appendix B of \cite{Borchmann:2013hta}
are related to the canonical $I_4$ models listed in table
\ref{TableNonCanonicalord4} by lop-equivalence.}
\label{tbl:su4top}
\end{table}
Note that for $SU(4)$ top 4 is lop equivalent to top 2 and top 5 is lop
equivalent top 3, and their $U(1)$ charges, as listed in appendix B of 
\cite{Borchmann:2013hta}, can be written as a linear combination of the
lop-equivalent model.

\begin{table}[H]
  \centering\footnotesize
  \begin{equation*}
    \begin{array}{|c|c|c|c|}
      \hline
      \text{Fiber} & \text{Model} & \text{Matter Locus} & \text{Matter}
      \cr\hline

      I_5^{s(0|1||2)} & 
      \begin{array}{c}
        (2,2,2,0,0,0,0,0) \cr
        [-, -, -, \sigma_2\sigma_5, \sigma_2\sigma_4 + \sigma_3\sigma_5,
          \sigma_3\sigma_4, \sigma_1\sigma_2, \sigma_1\sigma_3]
      \end{array} &
      \begin{array}{c}
        \sigma_1 \cr
        \sigma_3 \cr
        \sigma_4 \cr
        \sigma_2\sigma_4 - \sigma_3\sigma_5 \cr
        (\ref{NL1}) \cr
        (\ref{NL2}) \cr
        (\ref{NL3}) \cr
      \end{array} &
      \begin{array}{c}
        {\bf 5}_{-3,-6} \oplus {\bf \overline{5}}_{3,6} \cr
        {\bf 5}_{2,-6} \oplus {\bf \overline{5}}_{-2,6} \cr
        {\bf 5}_{-3,4} \oplus {\bf \overline{5}}_{3,-4} \cr
        {\bf 10}_{-1,-2} \oplus {\bf \overline{10}}_{1,2} \cr
        {\bf 5}_{-3,-1} \oplus {\bf \overline{5}}_{3,1} \cr
        {\bf 5}_{2,-1} \oplus {\bf \overline{5}}_{-2,1} \cr
        {\bf 5}_{2,4} \oplus {\bf \overline{5}}_{-2,-4} \cr
      \end{array} \cr\hline

      I_5^{s(0|1|2)} & 
      \begin{array}{c}
        (2,1,1,1,0,0,1,0) \cr
        [\sigma_1\sigma_3, \sigma_1\sigma_2, -, \sigma_3\sigma_4,
          \sigma_2\sigma_4, -, -, -]
      \end{array} &
      \begin{array}{c}
        \sigma_2 \cr
        \sigma_4 \cr
        s_{7,0} \cr
        s_{9,0} \cr
        \sigma_4s_{3,1} - \sigma_1s_{7,0} \cr
        \sigma_2s_{8,1} - \sigma_3s_{9,0} \cr
        (\ref{NL4}) \cr
      \end{array} &
      \begin{array}{c}
        {\bf 10}_{1,-2} \oplus {\bf \overline{10}}_{-1,2} \cr
        {\bf 10}_{1,3} \oplus {\bf \overline{10}}_{-1,-3} \cr
        {\bf 5}_{-2,4} \oplus {\bf \overline{5}}_{2,-4} \cr
        {\bf 5}_{-2,-6} \oplus {\bf \overline{5}}_{2,6} \cr
        {\bf 5}_{3,-1} \oplus {\bf \overline{5}}_{-3,1} \cr
        {\bf 5}_{3,4} \oplus {\bf \overline{5}}_{-3,-4} \cr
        {\bf 5}_{-2,-1} \oplus {\bf \overline{5}}_{2,1} \cr
      \end{array} \cr\hline
      
      I_5^{s(0|1||2)} & 
      \begin{array}{c}
        (2,1,1,1,0,0,1,0) \cr
        [-, \sigma_1\sigma_2, \sigma_1\sigma_3, -, \sigma_2\sigma_4,
          \sigma_3\sigma_4, -, -]
      \end{array} &
      \begin{array}{c}
        \sigma_2 \cr
        \sigma_3 \cr
        \sigma_4 \cr
        s_{9,0} \cr
        \sigma_4s_{1,2} - \sigma_1s_{5,1} \cr
        \sigma_2\sigma_4s_{8,1} - s_{5,1}s_{9,0} \cr
        (\ref{NL5}) \cr
      \end{array} &
      \begin{array}{c}
        {\bf 10}_{-1,-2} \oplus {\bf \overline{10}}_{1,2} \cr
        {\bf 5}_{-3,4} \oplus {\bf \overline{5}}_{3,-4} \cr
        {\bf 10}_{-1,3} \oplus {\bf \overline{10}}_{1,-3} \cr
        {\bf 5}_{-3,-6} \oplus {\bf \overline{5}}_{3,6} \cr
        {\bf 5}_{-3,-1} \oplus {\bf \overline{5}}_{3,1} \cr
        {\bf 5}_{2,4} \oplus {\bf \overline{5}}_{-2,-4} \cr
        {\bf 5}_{2,-1} \oplus {\bf \overline{5}}_{-2,1} \cr
      \end{array} \cr\hline
      
      I_5^{s(1|0|2)} & 
      \begin{array}{c}
        (3,2,1,1,0,0,0,0) \cr
        [-,-,-,-,\sigma_1\sigma_2,\sigma_1\sigma_3,\sigma_2\sigma_4,\sigma_3\sigma_4]
      \end{array} &
      \begin{array}{c}
        \sigma_1 \cr
        \sigma_2 \cr
        \sigma_3 \cr
        \sigma_4 \cr
        s_{3,1} \cr
        (\ref{NL6}) \cr
        (\ref{NL7}) \cr
      \end{array} &
      \begin{array}{c}
        {\bf 10}_{-2,2} \oplus {\bf \overline{10}}_{2,-2} \cr
        {\bf 10}_{3,2} \oplus {\bf \overline{10}}_{-3,-2} \cr
        {\bf 5}_{-1,6} \oplus {\bf \overline{5}}_{1,-6} \cr
        {\bf 5}_{4,6} \oplus {\bf \overline{5}}_{-4,-6} \cr
        {\bf 5}_{4,1} \oplus {\bf \overline{5}}_{-4,-1} \cr
        {\bf 5}_{-1,1} \oplus {\bf \overline{5}}_{1,-1} \cr
        {\bf 5}_{-1,-4} \oplus {\bf \overline{5}}_{1,4} \cr
      \end{array} \cr\hline
      
      I_5^{s(01|2)} & 
      \begin{array}{c}
        (3,2,1,1,0,0,0,0) \cr
        [\sigma_2\sigma_5,\sigma_2\sigma_4+\sigma_3\sigma_5,\sigma_3\sigma_4,\sigma_1\sigma_2,\sigma_1\sigma_3,-,-,-]
      \end{array} &
      \begin{array}{c}
        \sigma_1 \cr
        \sigma_3 \cr
        \sigma_4 \cr
        s_{7,0} \cr
        s_{8,0} \cr
        \sigma_1\sigma_3s_{9,0} - s_{7,0}s_{8,0} \cr
        (\ref{NL8}) \cr
      \end{array} &
      \begin{array}{c}
        {\bf 10}_{2,0} \oplus {\bf \overline{10}}_{-2,0} \cr
        {\bf 10}_{-3,0} \oplus {\bf \overline{10}}_{3,0} \cr
        {\bf 5}_{-4,0} \oplus {\bf \overline{5}}_{4,0} \cr
        {\bf 5}_{1,-1} \oplus {\bf \overline{5}}_{-1,1} \cr
        {\bf 5}_{-4,-1} \oplus {\bf \overline{5}}_{4,1} \cr
        {\bf 5}_{1,1} \oplus {\bf \overline{5}}_{-1,-1} \cr
        {\bf 5}_{1,0} \oplus {\bf \overline{5}}_{-1,0} \cr
      \end{array} \cr\hline
      
      I_5^{s(1|02)} & 
      \begin{array}{c}
        (4,2,0,2,0,0,0,0) \cr
        [-, -, \sigma_3\sigma_4, -, \sigma_2\sigma_4 + \sigma_3\sigma_5,
          \sigma_1\sigma_3, \sigma_2\sigma_5, \sigma_1\sigma_2]
      \end{array} &
      \begin{array}{c}
        \sigma_1 \cr
        \sigma_2 \cr
        \sigma_3 \cr
        \sigma_4 \cr
        \sigma_5 \cr
        \sigma_2\sigma_4 - \sigma_3\sigma_5 \cr
        (\ref{NL9}) \cr
        (\ref{NL10}) \cr
      \end{array} &
      \begin{array}{c}
        {\bf 5}_{0,6} \oplus {\bf \overline{5}}_{0,-6} \cr
        {\bf 5}_{1,6} \oplus {\bf \overline{5}}_{-1,-6} \cr
        {\bf 5}_{-1,1} \oplus {\bf \overline{5}}_{1,-1} \cr
        {\bf 5}_{1,1} \oplus {\bf \overline{5}}_{-1,-1} \cr
        {\bf 5}_{-1,-4} \oplus {\bf \overline{5}}_{1,4} \cr
        {\bf 10}_{0,2} \oplus {\bf \overline{10}}_{0,-2} \cr
        {\bf 5}_{0,-4} \oplus {\bf \overline{5}}_{0,4} \cr
        {\bf 5}_{0,1} \oplus {\bf \overline{5}}_{0,-1} \cr
      \end{array} \cr\hline
      
      I_5^{s(012)} & 
      \begin{array}{c}
        (5,2,0,2,0,0,0,0) \cr
        [-, \sigma_1\sigma_2, \sigma_2\sigma_5, \sigma_1\sigma_3,
          \sigma_2\sigma_4 + \sigma_3\sigma_5, -, \sigma_3\sigma_4, -]
      \end{array} &
      \begin{array}{c}
        \sigma_2 \cr
        \sigma_3 \cr
        \sigma_4 \cr
        \sigma_5 \cr
        \sigma_2\sigma_4 - \sigma_3\sigma_5 \cr
        \sigma_3s_{7,0} - \sigma_2s_{9,0} \cr
        \sigma_4s_{7,0} - \sigma_5s_{9,0} \cr
        (\ref{NL11}) \cr
      \end{array} &
      \begin{array}{c}
        {\bf 5}_{-1,0} \oplus {\bf \overline{5}}_{1,0} \cr
        {\bf 5}_{1,1} \oplus {\bf \overline{5}}_{-1,-1} \cr
        {\bf 5}_{-1,-1} \oplus {\bf \overline{5}}_{1,1} \cr
        {\bf 5}_{1,0} \oplus {\bf \overline{5}}_{-1,0} \cr
        {\bf 10}_{0,0} \oplus {\bf \overline{10}}_{0,0} \cr
        {\bf 5}_{0,-1} \oplus {\bf \overline{5}}_{0,1} \cr
        {\bf 5}_{0,1} \oplus {\bf \overline{5}}_{0,-1} \cr
        {\bf 5}_{0,0} \oplus {\bf \overline{5}}_{0,0} \cr
      \end{array} \cr\hline
      
    \end{array}
  \end{equation*}
  \caption{$U(1)$ charges of the once non-canonical $I_5$ models from table \ref{TableNonCanonicalord5}.}
  \label{tbl:NC1}
  % Goddammit the caption doesn't fit on the page.
\end{table}

\begin{table}[H]
  \centering\footnotesize
  \begin{equation*}
    \begin{array}{|c|c|c|c|}
      \hline
      \text{Fiber} & \text{Model} & \text{Matter Locus} & \text{Matter}
      \cr\hline

      I_5^{s(01||2)} & 
      \begin{array}{c}
        (2,2,2,0,0,0,0,0) \cr
        [\xi_3\xi_4, \xi_2\xi_4 + \xi_3\xi_5, \xi_2\xi_5, \sigma_3\xi_3,
          \sigma_2\xi_3 + \sigma_3\xi_2, \sigma_2\xi_2, -, -]
      \end{array} &
      \begin{array}{c}
        \xi_2 \cr
        \sigma_2 \cr
        \sigma_2\xi_3 - \sigma_3\xi_2 \cr
        \sigma_2\xi_4 - \sigma_3\xi_5 \cr
        \xi_2s_{8,0} - \xi_3s_{9,0} \cr
        \sigma_2s_{8,0} - \sigma_3s_{9,0} \cr
        (\ref{NL12}) \cr
      \end{array} &
      \begin{array}{c}
        {\bf 5}_{3,-1} \oplus {\bf \overline{5}}_{-3,1} \cr
        {\bf 5}_{-2,1} \oplus {\bf \overline{5}}_{2,-1} \cr
        {\bf 10}_{1,0} \oplus {\bf \overline{10}}_{-1,0} \cr
        {\bf 5}_{3,0} \oplus {\bf \overline{5}}_{-3,0} \cr
        {\bf 5}_{3,1} \oplus {\bf \overline{5}}_{-3,-1} \cr
        {\bf 5}_{-2,-1} \oplus {\bf \overline{5}}_{2,1} \cr
        {\bf 5}_{-2,0} \oplus {\bf \overline{5}}_{2,0} \cr
      \end{array} \cr\hline

      I_5^{s(0|1||2)} & 
      \begin{array}{c}
        (2,1,1,1,0,0,0,0) \cr
        [-, \sigma_1\xi_3, \sigma_1\xi_2, -, \sigma_4\xi_3, \sigma_4\xi_2,
          \xi_3\xi_4, \xi_2\xi_4]
      \end{array} &
      \begin{array}{c}
        \xi_2 \cr
        \xi_3 \cr
        \xi_4 \cr
        \sigma_4 \cr
        (\ref{NL13}) \cr
        (\ref{NL14}) \cr
        (\ref{NL15}) \cr
      \end{array} &
      \begin{array}{c}
        {\bf 5}_{2,-6} \oplus {\bf \overline{5}}_{-2,6} \cr
        {\bf 10}_{-1,-2} \oplus {\bf \overline{10}}_{1,2} \cr
        {\bf 5}_{-3,-6} \oplus {\bf \overline{5}}_{3,6} \cr
        {\bf 10}_{-1,3} \oplus {\bf \overline{10}}_{1,-3} \cr
        {\bf 5}_{2,-1} \oplus {\bf \overline{5}}_{-2,1} \cr
        {\bf 5}_{-3,-1} \oplus {\bf \overline{5}}_{3,1} \cr
        {\bf 5}_{2,4} \oplus {\bf \overline{5}}_{-2,-4} \cr
      \end{array} \cr\hline
      
      I_5^{s(01||2)} & 
      \begin{array}{c}
        (2,1,1,1,0,0,0,0) \cr
        [\xi_3\xi_4, \sigma_2\xi_3, \sigma_3\xi_3, \xi_2\xi_4 + \xi_3\xi_5,
          \sigma_2\xi_2, \sigma_3\xi_2, \xi_2\xi_5, -]
      \end{array} &
      \begin{array}{c}
        \xi_2 \cr
        \xi_5 \cr
        \sigma_2 \cr
        \sigma_3 \cr
        \xi_2\xi_5\sigma_2 - \sigma_3s_{9,0} \cr
        (\ref{NL16}) \cr
        (\ref{NL17}) \cr
      \end{array} &
      \begin{array}{c}
        {\bf 10}_{1,1} \oplus {\bf \overline{10}}_{-1,-1} \cr
        {\bf 5}_{3,1} \oplus {\bf \overline{5}}_{-3,-1} \cr
        {\bf 5}_{-2,1} \oplus {\bf \overline{5}}_{2,-1} \cr
        {\bf 10}_{1,0} \oplus {\bf \overline{10}}_{-1,0} \cr
        {\bf 5}_{-2,-1} \oplus {\bf \overline{5}}_{2,1} \cr
        {\bf 5}_{3,0} \oplus {\bf \overline{5}}_{-3,0} \cr
        {\bf 5}_{-2,0} \oplus {\bf \overline{5}}_{2,0} \cr
      \end{array} \cr\hline
      
      I_5^{s(1|0|2)} & 
      \begin{array}{c}
        (2,1,1,1,0,0,0,0) \cr
        [\sigma_3\sigma_4, \sigma_3\xi_1\xi_3, -, \sigma_2\sigma_4 +
          \sigma_3\xi_1\xi_2, \sigma_2\xi_1\xi_3, \xi_3\xi_4,
          \sigma_2\xi_1\xi_2, \xi_2\xi_4]
      \end{array} &
      \begin{array}{c}
        \xi_1 \cr
        \xi_2 \cr
        \xi_3 \cr
        \xi_4 \cr
        \sigma_2 \cr
        \xi_3\xi_4\sigma_3 - \sigma_2s_{3,1} \cr
        (\ref{NL18}) \cr
        (\ref{NL19}) \cr
      \end{array} &
      \begin{array}{c}
        {\bf 10}_{3,2} \oplus {\bf \overline{10}}_{-3,-2} \cr
        {\bf 5}_{4,6} \oplus {\bf \overline{5}}_{-4,-6} \cr
        {\bf 10}_{-2,2} \oplus {\bf \overline{10}}_{2,-2} \cr
        {\bf 5}_{-1,6} \oplus {\bf \overline{5}}_{1,-6} \cr
        {\bf 10}_{-2,-3} \oplus {\bf \overline{10}}_{2,3} \cr
        {\bf 5}_{4,1} \oplus {\bf \overline{5}}_{-4,-1} \cr
        {\bf 5}_{-1,-4} \oplus {\bf \overline{5}}_{1,4} \cr
        {\bf 5}_{-1,1} \oplus {\bf \overline{5}}_{1,-1} \cr
      \end{array} \cr\hline
      
%      I_5^{s(1|02)} & 
%      \begin{array}{c}
%        (4,2,0,2,0,0,0,0) \cr
%        [-, -, \sigma_3\xi_2, -, \sigma_2\xi_2 + \sigma_3\xi_3, \xi_2\xi_4,
%          \sigma_2\xi_3, \xi_3\xi_4]
%      \end{array} &
%      \begin{array}{c}
%        \xi_2 \cr
%        \xi_3 \cr
%        \xi_4 \cr
%        \sigma_2 \cr
%        \sigma_3 \cr
%        \xi_2\sigma_2 - \xi_3\sigma_3 \cr
%        (\ref{NL20}) \cr
%        (\ref{NL21}) \cr
%      \end{array} &
%      \begin{array}{c}
%        {\bf 10}_{0,0} \oplus {\bf 10}_{0,0} \cr
%        ?
%      \end{array} \cr\hline
      
      I_5^{s(0|2||1)} & 
      \begin{array}{c}
        (1,1,1,1,0,0,1,0) \cr
        \begin{aligned}[]
          [\xi_2\xi_3\xi_5\xi_6,\, &\xi_3\xi_6\sigma_4 + \xi_2\xi_5\sigma_3,
          \sigma_3\sigma_4, \cr&\xi_2\xi_6\xi_7 + \xi_3\xi_4\xi_5,
          \xi_1\xi_2\xi_3\xi_6, \xi_1\xi_2\sigma_3, \xi_4\xi_7,
          \xi_1\xi_3\xi_4]
        \end{aligned}
      \end{array} &
      \begin{array}{c}
        \xi_1 \cr
        \xi_3 \cr
        \xi_4 \cr
        \xi_5 \cr
        \xi_6 \cr
        \xi_8 \cr
        \sigma_3 \cr
        (\ref{NL25}) \cr
        (\ref{NL26}) 
      \end{array} &
      \begin{array}{c}
        \text{non-min} \cr
        {\bf 10}_{3,-1} \oplus {\bf \overline{10}}_{-3,1} \cr
        {\bf 5}_{4,2} \oplus {\bf \overline{5}}_{-4,-2} \cr
        {\bf 10}_{3,4} \oplus {\bf \overline{10}}_{-3,-4} \cr
        {\bf 10}_{3,4} \oplus {\bf \overline{10}}_{-3,-4} \cr
        {\bf 5}_{4,7} \oplus {\bf \overline{5}}_{-4,-7} \cr
        {\bf 5}_{4,-3} \oplus {\bf \overline{5}}_{-4,3} \cr
        {\bf 5}_{-1,-3} \oplus {\bf \overline{5}}_{1,3} \cr
        {\bf 5}_{-1,2} \oplus {\bf \overline{5}}_{1,-2} \cr
      \end{array} \cr\hline
      
    \end{array}
  \end{equation*}
  \caption{$U(1)$ charges of the twice non-canonical $I_5$ models from table \ref{TableNonCanonicalord5}.}
  \label{tbl:NC2}
\end{table}
\newpage

\begin{table}[t]
  \centering\footnotesize
  \begin{equation*}
    \begin{array}{|c|c|c|c|}
      \hline
      \text{Fiber} & \text{Model} & \text{Matter Locus} & \text{Matter}
      \cr\hline
      I_5^{s(0|1||2)} & 
      \begin{array}{c}
        (1,1,1,0,0,0,0,0) \cr
        \begin{aligned}[]
        [\xi_3\delta_3\delta_4, 
          \,&\delta_4(\delta_3\xi_2 + \delta_2\xi_3), 
          \xi_2\delta_2\delta_4, 
          \cr&\xi_3\delta_1\delta_4, 
          \delta_1(\delta_2\xi_3 + \delta_3\xi_2),
          \delta_1\delta_2\xi_2, 
          \sigma_1\xi_3, 
          \sigma_1\xi_2]
        \end{aligned}
      \end{array} &
      \begin{array}{c}
        \delta_1 \cr
        \delta_2 \cr
        \xi_2 \cr
        \sigma_1 \cr
        \xi_2\delta_3 - \xi_3\delta_2 \cr
        (\ref{NL22}) \cr
        (\ref{NL23}) \cr
        (\ref{NL24}) \cr
      \end{array} &
      \begin{array}{c}
        {\bf 10}_{1,-3} \oplus {\bf \overline{10}}_{-1,3} \cr
        {\bf 5}_{3,-4} \oplus {\bf \overline{5}}_{-3,4} \cr
        {\bf 5}_{-2,6} \oplus {\bf \overline{5}}_{2,-6} \cr
        {\bf 5}_{3,6} \oplus {\bf \overline{5}}_{-3,-6} \cr
        {\bf 10}_{1,2} \oplus {\bf \overline{10}}_{-1,-2} \cr
        {\bf 5}_{-2,1} \oplus {\bf \overline{5}}_{2,-1} \cr
        {\bf 5}_{3,1} \oplus {\bf \overline{5}}_{-3,-1} \cr
        {\bf 5}_{-2,-4} \oplus {\bf \overline{5}}_{2,4} \cr
      \end{array} \cr\hline
    \end{array}
  \end{equation*}
  \caption{$U(1)$ charges of the single thrice non-canonical $I_5$ model from table \ref{TableNonCanonicalord5}.}
  \label{tbl:NC3}
\end{table}

\section{Exceptional Singular Fibers}\label{sec:Exceptional}

In this section the algorithm is continued up to the exceptional singular fibers. 
In determining the exceptional fibers we recall that the
sections can only intersect the fiber components of multiplicity one, which
means that there is a very restricted number of singular fibers.

\begin{figure}
  \centering
    \includegraphics[height=3cm]{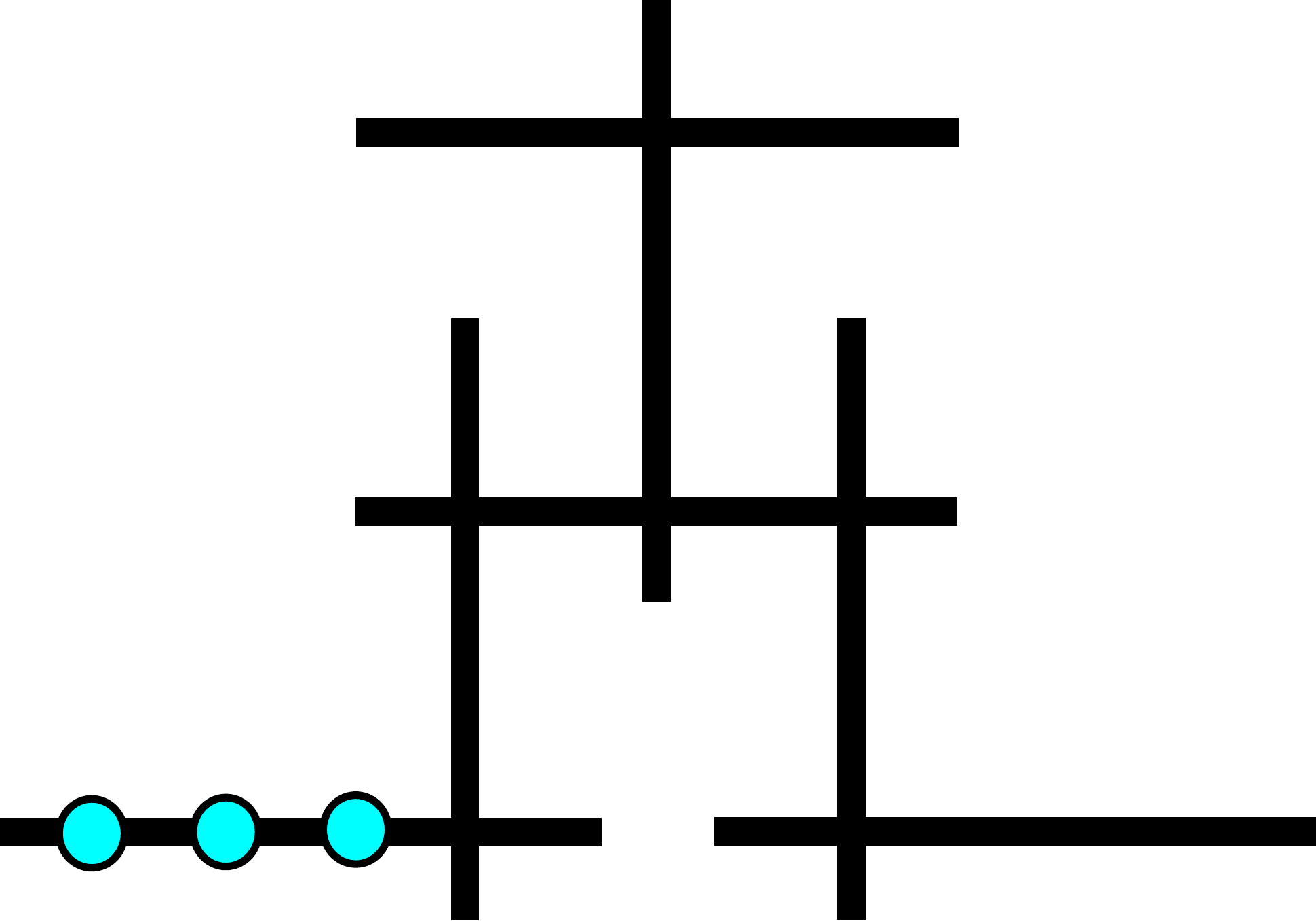} \qquad\qquad
    \includegraphics[height=3cm]{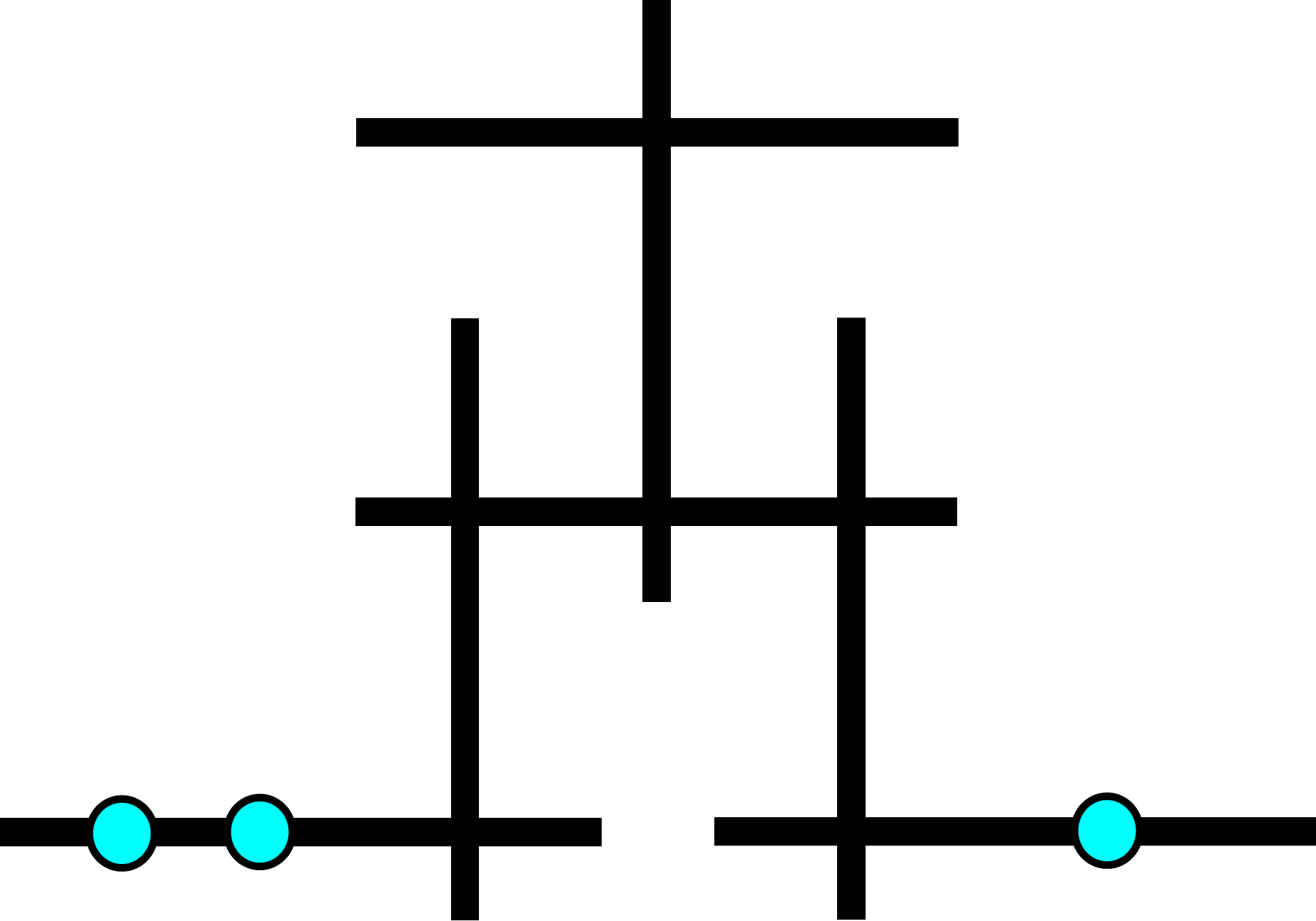} \qquad\qquad
    \includegraphics[height=3cm]{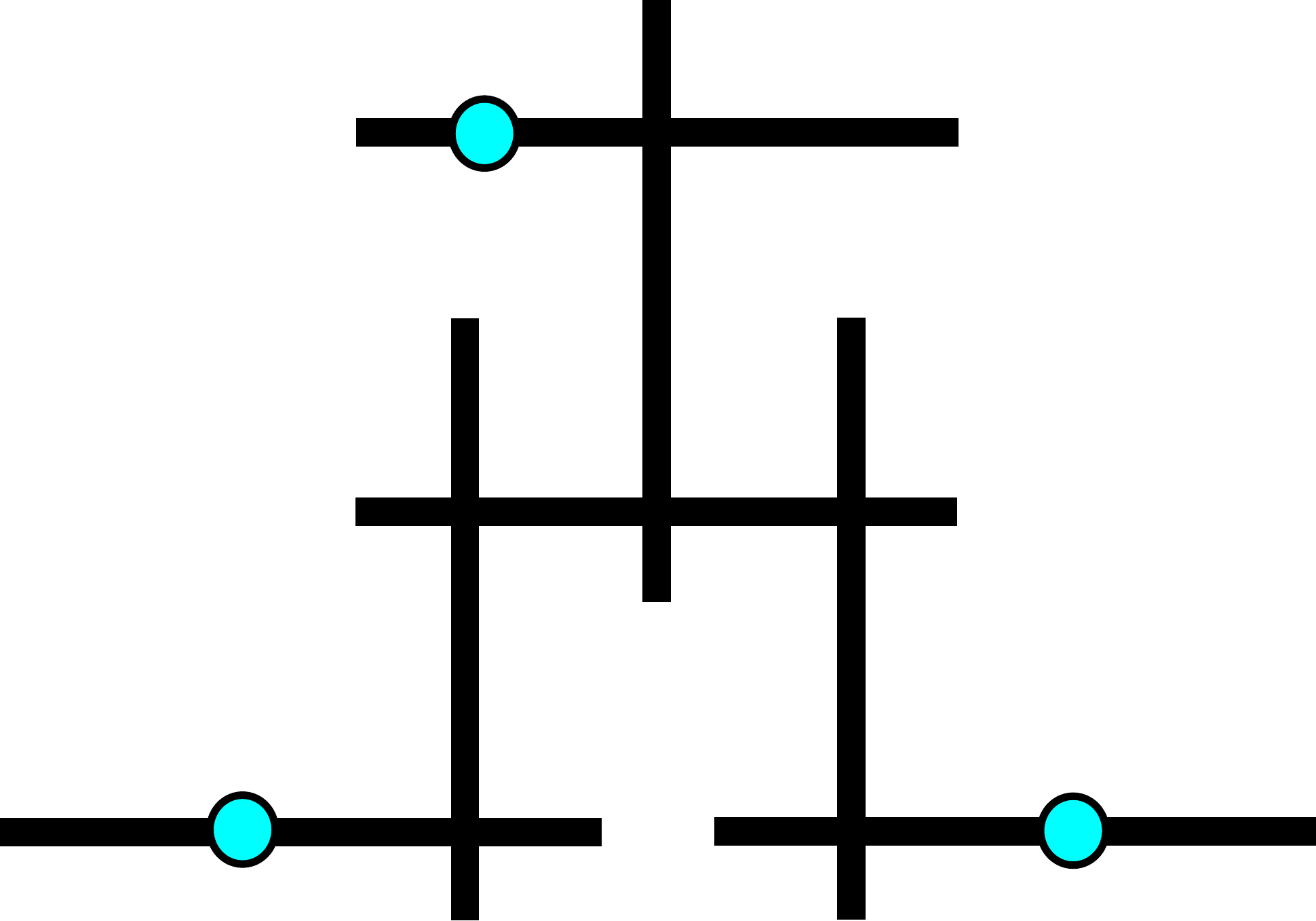}
    \caption{The type $IV^{*s}$ fibers. The sections, which intersect the
    components of the $IV^{*s}$ fiber represented by the blue nodes, are seen
      to intersect only the external, multiplicity one components. Because of
      the $S_3$ symmetry we write these as $IV^{*s(ijk)}$, $IV^{*s(ij|k)}$,
      and $IV^{*s(i|j|k)}$ respectively.}
  \label{IVstarFibs}
\end{figure}

For what concerns the type $IV^*$ singular fiber there are three different
ways in which the sections can intersect the multiplicity one components.
These are the types $IV^{*(012)},IV^{*(01|2)}$ and $IV^{*(0|1|2)}$. As can be
seen from figure \ref{IVstarFibs} the three multiplicity one components of the
$IV^*$ singular fiber appear symmetrically, and so sections separated by a
slash merely indicates that they do not intersect the same multiplicity one
component.

\begin{figure}
  \centering
    \includegraphics[height=2cm]{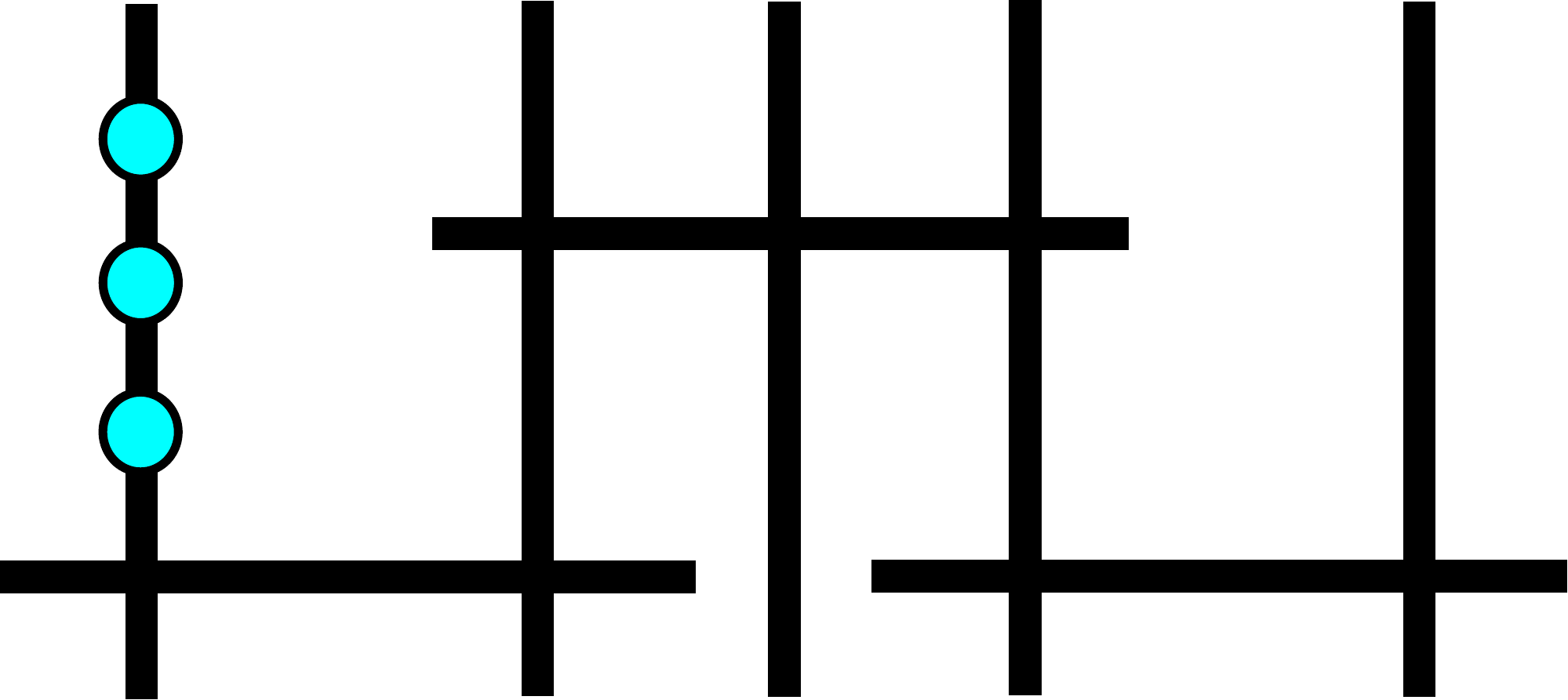} \qquad
    \includegraphics[height=2cm]{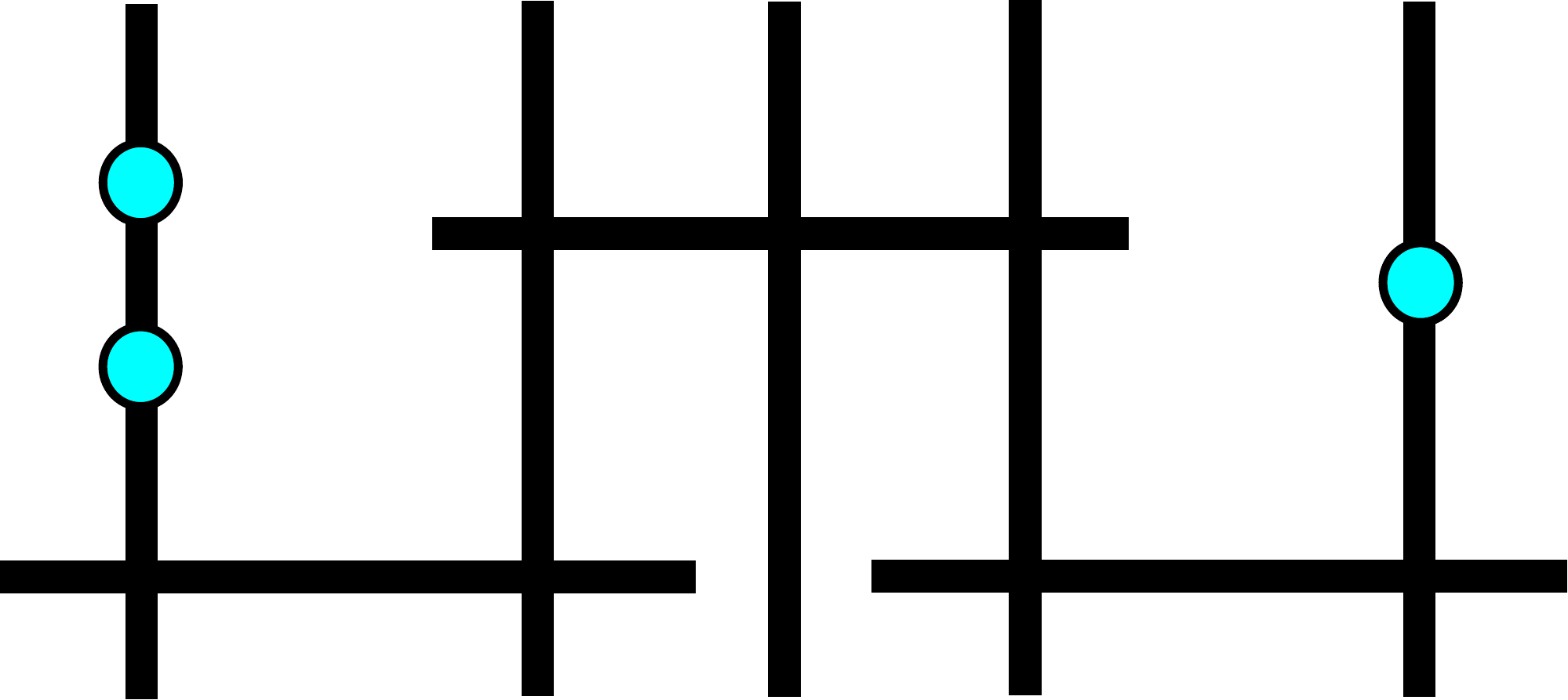} \qquad
    \includegraphics[height=2cm]{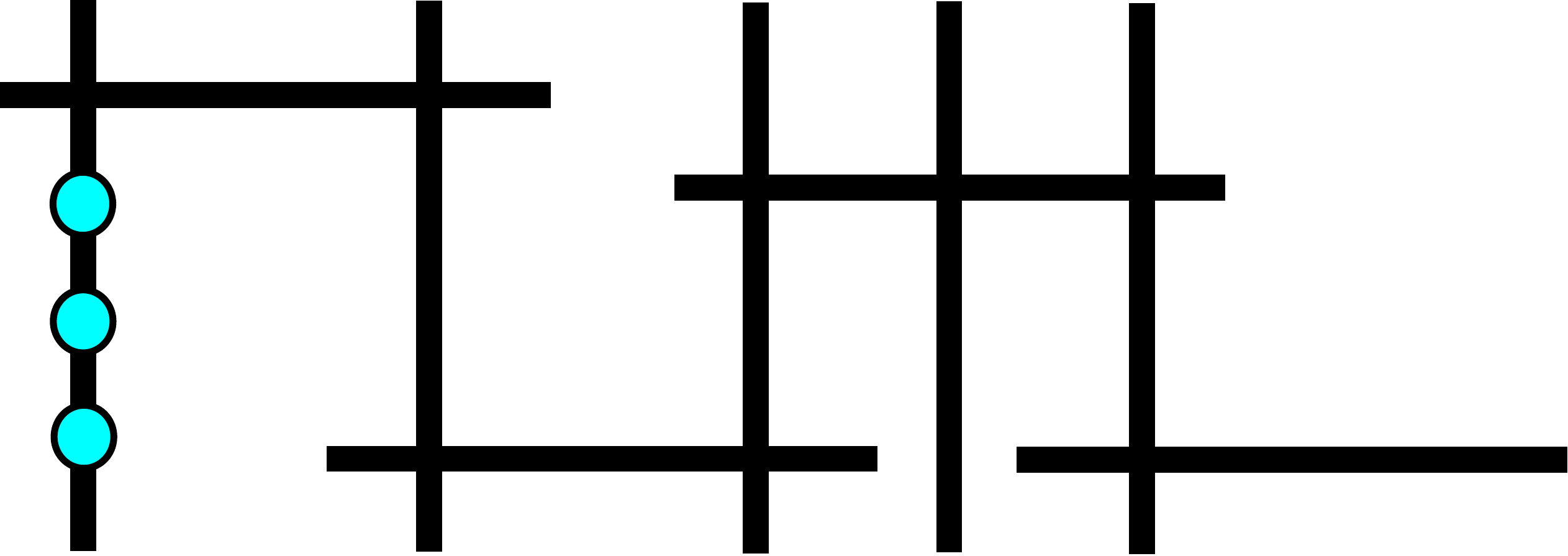}
    \caption{The type $III^*$ and $II^*$ fibers. Shown are the two type $III^{*(ijk)}$ and $III^{*(ij|k)}$ 
    fibers where the sections are distributed over the two multiplicity one
  components, and the single type $II^{*(ijk)}$ fiber, which has all three sections
intersecting the single multiplicity one component.}
  \label{IIIstarFibs}
\end{figure}

Regarding the singular $III^*$ fibers, the possible ways the sections can
intersect the components restrict the range of singular fibers to
$III^{*(012)}$ and $III^{*(01|2)}$. The different singular fibers can be seen
in figure \ref{IIIstarFibs}.

Finally it is clear that the only type $II^*$ fiber one could find (since
there is only one multiplicity one component) is the $II^{*(012)}$. This fiber
is also shown in figure \ref{IIIstarFibs}.

It was also possible to obtain the singular fibers corresponding to gauge
groups $G_2$ and $F_4$ which come from, respectively, the non-split singular
fiber types $I_0^{*ns(012)}$ and $IV^{*ns(012)}$.

Proceeding through these subbranchs of the Tate tree will involve the $I_n^*$
fibers corresponding to Dynkin diagrams of $D$-type in the split case. There
fibers are composed of a chain of multiplicity two nodes with two multiplicity
one nodes connected to each end of the chain. As the rational sections can
only intersect the multiplicity one nodes they are constrained to lie of these
outer legs. The notation of these fibers shall be $(01)$ represents two
sections on the same leg, $(0|1)$ represents two section intersecting two of
the outer legs attached to the same end of the chain, and $(0||1)$ will
represent two sections sitting on multiplicity one component separated by the
length of the chain.

\subsection{Canonical Enhancements to Exceptional Singular Fibers}

The starting point for the enhancements to the possible canonical exceptional singular
fibers is the $I_0^{*ss(01|2)}:(2,2,1,1,1,0,0,0)$. Recall that one of the fiber 
components will split only if the condition $s_{5,1}^2-4 s_{1,2} s_{8,0}= p^2$
is satisfied for some $p$. The discriminant at
sixth order takes the form
\begin{equation}
\Delta = s_{7,0}^2 s_{8,0}^2 (s_{5,1}^2 - 4 s_{1,2} s_{8,0}) ( s_{1,2} s_{7,0}^2 -s_{3,1} s_{5,1} s_{7,0} + 
     s_{3,1}^2 s_{8,0})^2) z^6+\mathcal{O} (z^7) \,.
\end{equation}
First let $z \mid \fks_8$ and the resulting fiber is of type
$I_1^{s(0|2||1)}$. The discriminant at seventh order reads
\begin{equation}
\Delta = s_{5,1}^3 s_{7,0}^3 (s_{3,1} s_{5,1} - s_{1,2} s_{7,0})^2
s_{9,0}^2 z^7+\mathcal{O}(z^8) \,.
\end{equation}
Now let $z^2 \mid \fks_5$ and the first exceptional singular fiber is found;
it is of type $IV^{*(0|1|2)}$
\begin{equation}
  IV^{*(0|1|2)} \, : \quad (2,2,1,2,1,0,1,0) \,.
\end{equation}
This subbranch of the tree does not continue because the discriminant now
takes the form $\Delta = s_{1,2} s_{7,0} s_{9,0} z^8+\mathcal{O}(z^9$)
and the only possible enhancement that remains inside the lop-equivalence
class, the vanishing of $s_{1,2}$, is a non-minimal enhancement.

Looking back at the $I_0^*$ starting point, the discriminant can instead be
enhanced by letting the three-term polynomial vanish, through
the canonical solution $s_{1,2}=s_{3,1}=0$. This gives an $I_1^{*s(01||2)}$ singular fiber. The discriminant at seventh order takes the form
\begin{equation}
\Delta = s_{2,2}^2 s_{5,1}^3 s_{7,0}^5 s_{8,0}^2 z^7+\mathcal{O}(z^8)
\,.
\end{equation}
The discriminant is enhanced further by letting $s_{5,1}=0$. This gives the second exceptional singular fiber, that is a type $IV^{*(01|2)}$
\begin{equation}
  IV^{*(01|2)} \, : \quad (3,2,2,2,1,0,0,0) \,.
\end{equation}
Proceeding in this subbranch, the discriminant now reads
\begin{equation}
\Delta = s_{2,2}^4 s_{7,0}^4 s_{8,0}^4 z^8+\mathcal{O}(z^9) \,.
\end{equation}
The only enhancement which  is possible (as all the others are non-minimal
enhancements) is $s_{2,2}=0$. The canonical excpetional singular fiber that
arises from this enhancement is $III^{*(01|2)}$
\begin{equation}
  III^{*(01|2)} \, : \quad (3,3,2,2,1,0,0,0) \,.
\end{equation}
Every further enhancement in this subbranch is a non-minimal fibration. 

\subsection{Non-canonical Enhancements to Exceptional Singular Fibers}

In this section the remaining exceptional fibers are obtained
through non-canonical enhancements of the discriminant.  The starting point is
the singular $I_3^{ns(012)}$ given by the vanishing orders
$(3,2,0,2,0,0,0,0)$. The discriminant contains $\Delta \supset
(s_{6,0}^2-4 s_{3,0} s_{8,0})$. Following appendix \ref{app:poly} it can be
solved 
non-canonically to find a non-split $I_{0,nc}^{*ns(012)}$ associated to gauge group
$G_2$
\begin{equation}
  I_{0,nc}^{*ns(012)}: \quad \left\{
  \begin{array}{c}
     (3,2,0,2,0,0,0,0) \cr
     [-,-,\mu \sigma_3^2,-,2 \mu
         \sigma_3 \sigma_8,-,\mu \sigma_8^2,-]
\end{array} \right\} \,.
\end{equation}
The next exceptional singular fiber is found through the following series of enhancements
\begin{equation}
\begin{aligned}
I_{0,nc}^{*ns(012)} \quad &\overset{\{s_{1,3},s_{2,2},s_{5,2}=0\}}\longrightarrow \quad I_{1,nc}^{*ns(012)} \quad \overset{P=0}\longrightarrow \quad IV_{nc^2}^{*ns(012)}\\
&P=(\sigma_8^2 s_{3, 1} - \sigma_3 \sigma_8 s_{6, 1} + \sigma_3^2 s_{8,1}) \,.
\end{aligned}
\end{equation}
Where the non-canonical solution to the three-term polynomial was applied to find a singular $IV_{nc^2}^{*ns(012)}$ with gauge group $F_4$
\begin{equation}
  IV_{nc^2}^{*ns(012)}: \quad \left\{
  \begin{array}{c}
     (4,3,0,3,0,0,0,0) \cr
     [-,-,\mu \xi_2^2+\xi_2 \xi_5
         z,-,2\mu \xi_2 \xi_3+(\xi_2 \xi_4+\xi_3 \xi_5)z,-,\mu \xi_3^2+\xi_3
         \xi_4
       z,-]
\end{array} \right\} \,.
\end{equation}
It was also necessary to specialize terms linear in $z$ in the expansion of the
coefficients. From this singular fiber the remaining two
fiber types can be reached through
\begin{equation}
\begin{aligned}
IV_{nc^2}^{*ns(012)}&\overset{s_{1,4}=0}\longrightarrow \quad III_{nc^2}^{*(012)} \overset{Q=0} \longrightarrow\quad II_{nc^3}^{*(012)}\\
&Q=(s_{5,3} \xi_2-s_{2,3} \xi_3) \,.
\end{aligned}
\end{equation}
The singular fibers obtained this way are type $III_{nc^2}^{*(012)}$
\begin{equation}
  III_{nc^2}^{*(012)}: \quad \left\{
  \begin{array}{c}
     (5,3,0,3,0,0,0,0) \cr
      [-,-,\mu \xi_2^2+\xi_2 \xi_5
          z,-,2\mu \xi_2 \xi_3+(\xi_2 \xi_4+\xi_3 \xi_5)z,-,\mu \xi_3^2+\xi_3
          \xi_4
        z,-]
\end{array} \right\} \,.
\end{equation}
and the singular fiber type $II_{nc^3}^{*(012)}$
\begin{equation}
  II_{nc^3}^{*(012)}: \quad \left\{
  \begin{array}{c}
     (5,3,0,3,0,0,0,0) \cr
     [-,\delta_1 \delta_3,\mu
         \delta_3^2+\delta_3 \xi_5 z,\delta_1\delta_2,2\mu \delta_2\delta_3
           +(\delta_3 \xi_4+\delta_2 \xi_5)z,-,\mu \delta_2^2+\delta_2 \xi_4
         z,-]
\end{array} \right\} \,.
\end{equation}

\section*{Acknowledgements}
First and foremost we thank Sakura Schafer-Nameki for numerous
contributions and discussions throughout this work. For frequent comments 
we thank Moritz K\"untzler and Jenny Wong. We would also like to thank Andreas
Braun, Andres Collinucci, and Eran Palti for helpful and interesting discussions. We
acknowledge the support of the STFC grant ST/J0028798/1.

\newpage

\appendix

\addtocontents{toc}{\protect\setcounter{tocdepth}{1}}

\section{Solving Polynomial Equations over UFDs}\label{app:poly}

In this appendix details are included of how to solve polynomial equations in
the sections $\fks_i$ given that they belong to a unique factorization domain \cite{MR0103906}. These
solutions were repeatedly used in the algorithm to enhance the vanishing order
of the discriminant. For convenience a part of this section will be a summary
of the details given in the appendix A of \cite{Kuntzler:2014ila}, however
there are polynomials specific to the case of two additional rational sections
and the derivation of the solution for these is provided here. For more details
on polynomial equations over UFDs that arise in the application of Tate's
algorithm the reader is referred to appendix B of \cite{Katz:2011qp}.

In \cite{Kuntzler:2014ila} solutions were obtained for a three-term polynomial
of the form
\begin{equation}
  s_1^2s_2 - s_1s_3s_4 + s_3^2s_5 = 0 \,.
\end{equation}
Four solutions were found, three of which involve setting pairs of terms to
zero, which are what we refer to as canonical solutions of the polynomials,
and one other solution which we refer to as the non-canonical solution. The
canonical solutions were found to be the pairs
\begin{equation}
  \begin{aligned}
    &s_1=s_3=0\\
    &s_1=s_5=0\\
    &s_2=s_3=0 \,.
  \end{aligned}
\end{equation}
The non-canonical solution is when 
\begin{equation}\label{ufdsolutions}
  \begin{aligned}
    s_1 &= \sigma_1\sigma_2 \\
    s_2 &= \sigma_3\sigma_4 \\
    s_3 &= \sigma_1\sigma_3 \\
    s_4 &= \sigma_2\sigma_4 + \sigma_3\sigma_5 \\
    s_5 &= \sigma_2\sigma_5 \,,
  \end{aligned}  
\end{equation}
where $\sigma_2$ and $\sigma_3$ are coprime over this UFD.

The non-canonical solution of a two-term polynomial was also needed
\begin{equation}\label{twotermsol}
s_1s_2 - s_3s_4 = 0:\quad \begin{cases}
s_1 = \sigma_1\sigma_2 \\
s_2 = \sigma_3\sigma_4 \\
s_3 = \sigma_1\sigma_3 \\
s_4 = \sigma_2\sigma_4 \,.
\end{cases}
\end{equation}
With this solution $\sigma_2$ and $\sigma_3$ are coprime, and so are
$\sigma_1$ and $\sigma_4$.
 
\subsection{Two Term Polynomial}
We now look at the polynomial
\begin{equation}
P=s_1^2 - 4s_2s_3 \,.
\end{equation}
Setting $P=0$ imposes the following conditions:
\begin{itemize}
\item There is an equality between the irreducible components of $s_1^2$ and
  the product of the irreducibles of $s_2$ and $s_3.$
\item Write $\mu$ for the irreducible components common to all the three terms.
\item Write $\sigma_1$ for the irreducible components common to $s_1$ and $s_2$.
\item Write $\sigma_2$ for the irreducible components common to $s_1$ and $s_3$.
\end{itemize}
Note that no conclusion is drawn about irreducibles shared only by $s_2$ and $s_3$. Then the most general solution takes the form
\begin{equation}
s_1^2 - 4s_2s_3=0:\quad\begin{cases}
s_1 = 2\mu\sigma_1\sigma_2 \\
s_2 = \mu\sigma_1^2 \\
s_3 = \mu\sigma_2^2 \,.
\end{cases}
\end{equation}
Since $\mu$ is the greatest common divisor of $s_2$ and $s_3$ we have that
$\sigma_1$ and $\sigma_2$ are coprime.

\subsection{Perfect Square Polynomial}
The first perfect square polynomial is given by
\begin{equation}
   s_1^2 - 4 s_2 s_3 = p^2 \,.
\end{equation}
This can be reformulated as
\begin{equation}
  \left( s_1 + p \right) \left( s_1 - p \right) = 4 s_2 s_3 \,,
\end{equation}
which can be solved in general by applying the solution of the two-term
polynomial (\ref{twotermsol}). In this case, it reads
\begin{equation}
  \begin{aligned}
    s_1 - p &= 2 \sigma_1 \sigma_2 \\
    s_1 + p &= 2 \sigma_3 \sigma_4 \\
    s_2 &= \sigma_1 \sigma_3 \\
    s_3 &= \sigma_2 \sigma_4 \,.
  \end{aligned}
\end{equation}
From the first two of these equations, one finds the generic form of $s_1$
\begin{equation}
  s_1 = \sigma_1 \sigma_2 + \sigma_3 \sigma_4 \,.
\end{equation}
So the general solution to the perfect square condition is
\begin{equation}
 s_1^2 - 4 s_2 s_3 = p^2: \quad \begin{cases}
s_1=\sigma_1 \sigma_2+\sigma_3\sigma_4\\
s_2=\sigma_1 \sigma_3 \\
s_3=\sigma_2 \sigma_4 \,.
\end{cases}
\end{equation}
It follows from the solution of (\ref{twotermsol}) that $\sigma_2$ and
$\sigma_3$ are coprime, as are $\sigma_1$ and $\sigma_4$.

\subsection{Three Term Polynomial}
The three-term polynomial
\begin{equation}
P=s_1^2 s_2 s_3 - s_1 s_4 s_5 + s_5^2 s_6 \,,
\end{equation}
appears in the algorithm. 
By imposing $P=0$ it is seen that $s_1 \mid s_5^2 s_6$, since it divides the
other two terms in the equation. In the same way $s_5 \mid s_1^2 s_2 s_3$. Decompose $s_5=\sigma_1 \sigma_2$ and $s_1=\sigma_1 \sigma_3$, where $\sigma_1=(s_1,s_5)$ is the greatest common divisor of the two terms, so that $\sigma_2$ and $\sigma_3$ have no common irreducibles. Then the equation of the polynomial becomes
\begin{equation}
\sigma_1^2 (s_6 \sigma_2^2 - s_4 \sigma_2 \sigma_3   + 
      s_2 s_3 \sigma_3^2)=0 \,.
\end{equation}
Applying the same reasoning it is now seen that $\sigma_3 \mid s_6 \sigma_2$,
but since $\sigma_2$ and $\sigma_3$ have no common irreducibles one can
conclude that $\sigma_3 \mid s_6$. In the same way it can be deduced that
$\sigma_2 \mid s_2 s_3$. This can be expressed as
\begin{equation}
s_6=\sigma_4 \sigma_3, \qquad s_2 s_3=\kappa \sigma_2 \,,
\end{equation}
where $\kappa$ is some constant of proportionality. The two-term solution
(\ref{twotermsol}) can be applied to the second of these equations to obtain
\begin{equation}
s_2=\sigma_5 \sigma_6, \quad s_3=\sigma_7 \sigma_8,\quad \kappa= \sigma_5
\sigma_7,\quad \sigma_2=\sigma_6 \sigma_8 \,.
\end{equation}
Then the initial polynomial reduces to
\begin{equation}
\sigma_1^2 \sigma_3 \sigma_6 \sigma_8 (  \sigma_3 \sigma_5 \sigma_7 + 
\sigma_4 \sigma_6 \sigma_8-s_4)=0 \,,
\end{equation}
from which can be solved for $s_4$. Then there is a non-canonical solution
\begin{equation}
s_1^2 s_2 s_3 - s_1 s_4 s_5 + s_5^2 s_6=0: \quad \begin{cases}
s_1=\sigma_1 \sigma_3 \\
s_2=\sigma_5 \sigma_6\\
s_3=\sigma_7 \sigma_8\\
s_4=\sigma_3 \sigma_5 \sigma_7+\sigma_4 \sigma_6 \sigma_8\\
s_5=\sigma_1 \sigma_6 \sigma_8\\
s_6=\sigma_3 \sigma_4 \,,
\end{cases}
\end{equation}
where the pairs $(\sigma_5, \sigma_8)$, $(\sigma_6, \sigma_7)$, and
$(\sigma_3,\sigma_6\sigma_8)$ are all coprime. There are also four different canonical solutions
\begin{equation}
\begin{aligned}
\sigma_1=0:& \quad s_1=s_5=0\\
\sigma_3=0:& \quad s_1=s_6=0\\
\sigma_6=0:& \quad s_2=s_5=0\\
\sigma_8=0:& \quad s_3=s_5=0 \,.
\end{aligned}
\end{equation}

\section{Matter Loci of $SU(5)$ Models}

In this appendix we list the matter loci of the $I_5$ fibers whose $U(1)$
charges are studied in section \ref{sec:Spectra}.
\begin{equation}\label{NL1}
  \sigma_3^2s_{1,2} - \sigma_2\sigma_3s_{2,2} + \sigma_2^2s_{3,2}
\end{equation}
\begin{equation}\label{NL2}
  \sigma_4^2s_{1,2} - \sigma_4\sigma_5s_{2,2} + \sigma_5^2s_{3,2}
\end{equation}
\begin{equation}\label{NL3}
  \sigma_1(\sigma_3^2s_{5,1} - \sigma_2\sigma_3s_{6,1} + \sigma_2^2s_{7,1}) -
  (\sigma_2\sigma_4 - \sigma_3\sigma_5)(\sigma_2s_{9,1} - \sigma_3s_{8,1})
\end{equation}
\begin{equation}\label{NL4}
  \sigma_3^2\sigma_4(\sigma_4s_{3,1} - \sigma_1s_{7,0}) +
  \sigma_2^2(\sigma_4^2s_{1,3} - \sigma_1\sigma_4s_{5,2} + \sigma_1^2s_{8,1})
  - \sigma_2\sigma_3(\sigma_4^2s_{2,2} - \sigma_1\sigma_4s_{6,1} +
  \sigma_1^2s_{9,0})
\end{equation}
\begin{equation}\label{NL5}
  \sigma_3^2\sigma_4(\sigma_4s_{1,2} - \sigma_1s_{5,1}) +
  \sigma_2^2\sigma_4(\sigma_4s_{3,2} - \sigma_1s_{7,1}) -
  \sigma_2\sigma_3(\sigma_4^2s_{2,2} - \sigma_1\sigma_4s_{6,1} +
  \sigma_1^2s_{9,0})
\end{equation}
\begin{equation}\label{NL6}
  \sigma_1^2\sigma_2^2s_{1,3} - \sigma_1\sigma_2s_{2,2}s_{5,1} +
  s_{3,1}s_{5,1}^2
\end{equation}
\begin{equation}\label{NL7}
  -\sigma_1\sigma_3^2\sigma_4s_{5,1} - \sigma_2\sigma_3(\sigma_4^2s_{3,1} -
  \sigma_1\sigma_4s_{6,1} + \sigma_1^2s_{8,1}) +
  \sigma_1\sigma_2^2(-\sigma_4s_{7,1} + \sigma_1s_{9,1})
\end{equation}
\begin{equation}\label{NL8}
  \sigma_1^2\sigma_3(\sigma_3^2s_{1,4} - \sigma_2\sigma_3s_{2,3} +
  \sigma_2^2s_{3,2}) + \sigma_1(\sigma_2\sigma_4 -
  \sigma_3\sigma_5)(\sigma_3^2s_{5,2} - \sigma_2\sigma_3s_{6,1} +
  \sigma_2^2s_{7,0}) + \sigma_3(\sigma_2\sigma_4 - \sigma_3\sigma_5)^2s_{8,0}
\end{equation}
\begin{equation}\label{NL9}
  \sigma_1(\sigma_2^2s_{3,1} - \sigma_2\sigma_3s_{6,1} + \sigma_3^2s_{8,1}) -
  (\sigma_2\sigma_4 - \sigma_3\sigma_5)(\sigma_2s_{7,1} - \sigma_3s_{9,1})
\end{equation}
\begin{equation}\label{NL10}
  (\sigma_2\sigma_4 - \sigma_3\sigma_5)^2s_{1,4} + \sigma_2\sigma_5s_{2,2}^2 -
  (\sigma_2\sigma_4 - \sigma_3\sigma_5)s_{2,2}s_{5,2} +
  \sigma_3\sigma_4s_{5,2}^2
\end{equation}
\begin{equation}\label{NL11}
  (\sigma_2\sigma_4 - \sigma_3\sigma_5)s_{1,5} - \sigma_1(\sigma_2\sigma_4 -
  \sigma_3\sigma_5)(\sigma_2s_{5,3} - \sigma_3s_{2,3}) +
  \sigma_1^2(\sigma_3^2s_{3,1} - \sigma_2\sigma_3s_{6,1} + \sigma_2^2s_{8,1})
\end{equation}
\begin{equation}\label{NL12}
  (\xi_3\sigma_2 - \xi_2\sigma_3)(\xi_2^2s_{1,3} - \xi_2\xi_3s_{2,3} +
  \xi_3^2s_{3,3}) + (\xi_2\xi_4 - \xi_3\xi_5)(\xi_2^2s_{5,1} -
  \xi_2\xi_3s_{6,1} + \xi_3^2s_{7,1})
\end{equation}
\begin{equation}\label{NL13}
  \xi_3\xi_4\sigma_1^2 - \sigma_1\sigma_4s_{5,1} + \sigma_4^2s_{1,2}
\end{equation}
\begin{equation}\label{NL14}
  \xi_2^2(\sigma_1s_{5,1} - \sigma_4s_{1,2}) + \xi_2\xi_3(\sigma_4s_{2,2} -
  \sigma_1s_{6,1}) + \xi_3^2(\sigma_1s_{7,1} - \sigma_4s_{3,2}) 
\end{equation}
\begin{equation}\label{NL15}
  \xi_2^2\xi_4s_{5,1} + \xi_2\xi_3(\sigma_4s_{8,1} - \xi_4s_{6,1}) +
  \xi_3^2(\xi_4s_{7,1} - \sigma_4s_{9,1})
\end{equation}
\begin{equation}\label{NL16}
  \xi_2(-\xi_2\sigma_2s_{2,2} + \xi_2\sigma_3s_{3,2} + \xi_3\sigma_2s_{6,1} -
  \xi_3\sigma_3s_{7,1}) - \xi_3^2\sigma_2s_{9,0}
\end{equation}
\begin{equation}\label{NL17}
  \xi_2^3(\sigma_3s_{1,3} - \xi_4s_{2,2}) + \xi_2^2\xi_3(\xi_5s_{2,2} -
  \sigma_3s_{5,2} + \xi_4s_{6,1}) + \xi_3^3\xi_5s_{9,0} -
  \xi_2\xi_3^2(\xi_5s_{6,1} - \sigma_3s_{8,1} + \xi_4s_{9,0})
\end{equation}
\begin{equation}\label{NL18}
  \xi_1\xi_2^2\xi_4s_{3,1} + \xi_2\xi_3(\xi_4^2\sigma_4 - \xi_1\xi_4s_{6,1} +
  \xi_1^2\sigma_2s_{7,1}) + \xi_1\xi_3^2(\xi_4s_{8,1} - \xi_1\sigma_2s_{9,1})
\end{equation}
\begin{equation}\label{NL19}
  (\xi_1\xi_2\sigma_3 - \sigma_2\sigma_4)^2s_{3,1} +
  \xi_3(\xi_1\xi_2\sigma_3 - \sigma_2\sigma_4)(\xi_1\sigma_2s_{2,2} -
  \xi_1\sigma_3s_{6,1} + \xi_4\sigma_3\sigma_4) +
  \xi_1^2\xi_3^2(\sigma_2^2s_{1,3} -
  \sigma_2\sigma_3s_{5,2} + \sigma_3^2s_{8,1})
\end{equation}
%\begin{equation}\label{NL20}
%  \xi_3^2(\xi_4s_{1,1} - \sigma_3s_{7,1}) + \xi_2^2(\xi_4s_{8,1} -
%  \sigma_2s_{9,1}) + \xi_2\xi_3(\sigma_2s_{7,1} + \sigma_3s_{9,1} -
%  \xi_4s_{6,1})
%\end{equation}
%\begin{equation}\label{NL21}
%  s_{1,0}(\xi_2\sigma_2 - \xi_3\sigma_3)^2 + (\xi_3s_{2,0} -
%  \xi_2s_{5,0})(\sigma_2s_{2,0} - \sigma_3s_{5,0})
%\end{equation}}
\begin{equation}\label{NL25}
  \begin{aligned}
    \xi_4\xi_6^3&\xi_8^3\sigma_4 - \xi_3\xi_6^2\xi_8^2(\xi_5\xi_7\sigma_4 -
    s_{1,2} + \xi_4s_{6,1}) \cr &+ \xi_3^2\xi_6\xi_8(\xi_4\xi_7\xi_8\sigma_3 -
    \xi_5\xi_6s_{5,2} + \xi_5\xi_7s_{6,1} + \xi_4\xi_5\xi_6s_{9,1}) - 
    \xi_3^2\xi_5(\xi_7^2\xi_8\sigma_3 - \xi_5\xi_6^2s_{8,2} +
    \xi_5\xi_6\xi_7s_{9,1})
  \end{aligned}
\end{equation}
\begin{equation}\label{NL26}
  \begin{aligned}
    \xi_3^3\xi_4\xi_7\sigma_3^3 - \xi_5\xi_6^3\xi_8\sigma_3\sigma_4^2 &+
    \xi_3^2\xi_6(\xi_5^2\xi_6^2s_{3,2} + \sigma_3^2(s_{1,2} - \xi_4s_{6,1}) -
    \xi_5 \sigma_3(\xi_7\sigma_3\sigma_4 + \xi_6s_{2,2} - \xi_4\xi_6s_{7,1}))
    \cr &+ \xi_3\xi_6^2\sigma_4(\xi_4\xi_8\sigma_3^2 + \xi_5(\sigma_3s_{6,1} -
    \xi_5\xi_6s_{7,1}))
  \end{aligned}
\end{equation}
\begin{equation}\label{NL22}
  \delta_1(\xi_2^2s_{1,2} - \xi_2\xi_3s_{2,2} + \xi_3^2s_{3,2}) -
  \delta_4(\xi_2^2s_{5,1} - \xi_2\xi_3s_{6,1} + \xi_3^2s_{7,1})
\end{equation}
\begin{equation}\label{NL23}
  \xi_2^2(\sigma_1s_{5,1} - \delta_1\sigma_3s_{8,1}) + \xi_3^2(\sigma_2s_{7,1}
  - \delta_1\delta_2s_{9,1}) + \xi_2\xi_3(-\sigma_1s_{6,1} +
  \delta_1\delta_2s_{8,1} + \delta_1\delta_3s_{9,1})
\end{equation}
\begin{equation}\label{NL24}
  \delta_2\delta_4^2\sigma_1(\delta_2\xi_3 - \delta_3\xi_2) +
  \delta_1^2(\delta_2^2s_{1,2} - \delta_2\delta_3s_{2,2} + \delta_3^2s_{3,2})
  - \delta_1\delta_4(\delta_2^2s_{5,1} - \delta_2\delta_3s_{6,1} +
  \delta_3^2s_{7,1})
\end{equation}

\section{Resolution of Generic Singular Fibers}\label{Forms for General Fibers}

In section \ref{sec:Overview} a table (table \ref{TableTateForms}) of canonical forms for many
of the different fiber types as originally denoted by Kodaira was presented.
In this section is is shown by explicitly constructing the resolution that
each of the forms is the stated fiber. 
Given the
set of resolutions and the canonical vanishing orders, the resolved geometry is
uniquely determined and the form of the
resolved geometry will not be written explicitly. For the Cartan divisors the
equations are given after the resolution process and they will intersect
according to the fiber type of
the singularity under consideration.

\subsection{$I_{2k+1}^{s(0|^n1|^m2)}$ \small{($n+m\leq k$)}}
The generic form for the singular $I_{2k+1}^{s(0|^n1|^m2)}$  is $(2k+1-(n+m),
k-n, m, k+1-m, 0, 0, n, 0)$, provided that $(m+n\leq k)$. 
The resolution process involves several steps.
First perform the following blow ups
\begin{equation}
\begin{aligned}
(x,y,z;\zeta_1),\quad (x,y,\zeta_i;\zeta_{i+1}) \qquad & 1\leq
i<\text{min}\{k-n,k+1-m\} \,.
\end{aligned}
\end{equation}
If $n\neq 0$ then the following small resolutions can be applied
\begin{equation}
\begin{aligned}
(x,z;\xi_1),\quad (x,\xi_i;\xi_{i+1}) \qquad &  &1\leq i <n \,.
\end{aligned}
\end{equation}
Similarly if $m \neq 0$ the small resolutions,
\begin{equation}
\begin{aligned}
(y,z;\delta_1),\quad (y,\delta_j;\delta_{j+1}) \qquad &  &1\leq j <m \,,
\end{aligned}
\end{equation}
are possible. If both $n\neq 0$ and $m\neq 0$ we need to use both sets of
resolutions are applied.
The next step depends on the sign of the quantity $m-n-1$. 
We call $\zeta_{max}$ the last exceptional divisor introduced in the initial
blow ups, and from now on the index will be used as
$\text{max}=\text{min}\{k-n,k+1-m\}$
If it is positive
then the resolutions,
\begin{equation}
  (y,\zeta_{max};\chi_1),\quad (y,\chi_r;\chi_{r+1}) \qquad  1\leq r < m-n-1
  \,,
\end{equation}
are used. Whereas if negative then the resolutions are 
\begin{equation}
  (x,\zeta_{max};\Omega_1),\quad (x,\Omega_r;\Omega_{r+1})  \qquad   1\leq r
  <  -(m-n-1) \,.
\end{equation}
If the term is exactly zero then we do neither set.
Finally the process can be completed with the resolutions 
\begin{equation}
\begin{aligned}
(y,\zeta_s;\psi_{s}) \qquad &  &1\leq s < \text{max} \,.
\end{aligned}
\end{equation}
The Cartan divisors are listed, assuming that $n-m-1>0$,
\begin{table}[H]
\centering
\begin{tabular}{c|c}
Exceptional Divisor & Fiber Equation\\ \hline 
$z$& $l_1 l_2 w s_{6, 0} + l_2 s_{9, 0} \zeta_1 \delta_1 + 
 l_1 s_{7, 0} \zeta_1 \xi_1 $\\
$\zeta_{i<max}$ & $s_{6,0} x$
 \\
$\zeta_{max}$ & $  x s_{6, 0} + s_{5,k+1-m} \zeta_{max-1}$\\

$\delta_{j<m}$ & $ l_2 s_{6, 0} + s_{7, 0} \zeta_1$\\
$\delta_{m}$ & $ l_2 y s_{6, 0} + y s_{7, 0} \zeta_1 \psi_1 + 
 s_{3, m} \zeta_1^m \delta_{m-1} \psi_1^{m-1}$\\
$\xi_{i<n}$ & $ l_1 s_{6, 0} + s_{9, 0} \zeta_1$
 \\
$\xi_{n}$ & $ l_1 x s_{6, 0} + \zeta_1 (x s_{9, 0} + s_{8, 0} \zeta_1^{n-1} \xi_{n-1})$
 \\
$\chi_{r<m-n-1}$ & $ x s_{6, 0} + s_{5, k+1-m} \zeta_{max-1}$
 \\
$\chi_{m-n-1}$& $ \begin{aligned} &y (x s_{6, 0} + s_{5, k+1-m} \zeta_{max-1} \psi_{max-1}) + \\
+ \zeta_{\max-1}^{m-n} \psi_{max-1}^{m-n-1}& (x s_{2, k-n} + s_{1, 2k+1-n-m} \zeta_{max-1} \psi_{max-1}) \chi_{m-n-2}\end{aligned}
$
 \\

$\psi_{s< max}$ & $ s_{6,0} y
$
 \\
 
\end{tabular}
\end{table}
Then the ordered set $(z,\xi_1,\cdots,\xi_n,\zeta_1 , \cdots,
\zeta_{max},\chi_1,\cdots, \chi_{m-n-1}, \psi_{max}, \cdots, \psi_1\delta_m,
\cdots, \delta_1)$ gives an $I_{2k+1}^{s(0|^n1|^m2)}$ fiber, where the divisors are
listed in the canonical ordering for the Dynkin diagram. One gets the
analogous result when $n - m - 1 < 0$.

\subsection{$I_{2k+1}^{s(0|^n1|^m2)}$ \small{($k<n+m\leq \left \lfloor{
  \frac{2}{3}(2k+1)}\right \rfloor$)}} 
  
  The generic form for the singular fibers of type $I_{2k+1}^{s(0|^n1|^m2)}$
  with section separation of the form $m+n\leq\left \lfloor{ \frac{2}{3}(2k+1)}\right
  \rfloor$ is given by $(2k+1-(m+n),m,m,n,0,0,n,0)$, where it is assumed that
  $m\geq n$. In order to resolve the geometry the following set of
  resolutions is used

\begin{equation}
\begin{aligned}
(x,z;\xi_1),\qquad (x,\xi_i;\xi_{i+1}) \qquad   & 1\leq i <n \\
(y,z;\delta_1),\qquad (y,\delta_j;\delta_{j+1}) \qquad   &1\leq j <m\\
(x,\delta_r;\chi_{r}) \qquad   &1\leq r \leq m\\
(x,\chi_m;\psi_1),\qquad (x,\psi_s;\psi_{s+1}) \qquad   & 1\leq s < 2k-2m-n
\,.
\end{aligned}
\end{equation}
Notice that the first three set of resolutions (together with $z$) produce
$2m+n+1$ Cartan divisors. The fourth set of resolutions is only necessary if
$2k-2m-n > 0$.
Then the Cartan divisors in the most general case are
\begin{table}[H]
\centering
\begin{tabular}{c|c}
Exceptional Divisor & Fiber Equation\\ \hline 
$z$& $l_1 l_2 w s_{6, 0} + l_2 s_{9, 0} \delta_1 + 
 l_1 s_{7, 0}  \xi_1 $\\
$\delta_1$ & $l_2 s_{6, 0} + 
 s_{7, 0} \xi_1 \xi_2^2 \cdots \xi_n^n\chi_1$\\
$\delta_{j<m}$ & $ l_2 s_{6, 0} + s_{7, 0} \chi_{j-1} \chi_j$\\
$\delta_{m}$ & $ l_2 (y s_{6, 0} + s_{2, m} \delta_{m-1} \chi_{m-1}) + \chi_{m-1} (y s_{7, 0} + s_{3, m} \delta_{m-1} \chi_{m-1}) \chi_m$\\
$\xi_{i<n}$ & $ l_1 s_{6, 0} + s_{9, 0} \delta_1$
 \\
$\xi_{n}$ & $l_1 x s_{6, 0} + x s_{9, 0} \delta_1 \chi_1 + 
 l_1 s_{5, n} \delta_1^n \xi_{n-1} \chi_1^{n-1} + 
 s_{8, n} \delta_1^{n+1} \xi_{n-1} \chi_1^n$
 \\
$\chi_{r<m}$ & $ x s_{6, 0} $
 \\
$\chi_{m}$& $ y s_{6, 0} + s_{2, m} \chi_{m-1}
$
 \\

$\psi_{s< 2k-2m-n}$ & $ y s_{6, 0} + s_{2, m} \chi_{m-1}
$
 \\
$\psi_{2k-2m-n}$ & $x y s_{6, 0} + x s_{2, m} \chi_{m-1} + s_{1, 2k+1-m-n} \psi_{2k-2m-n-1} \chi_{m-1}^{n-1}
$
\\
\end{tabular}
\end{table}
The ordered set $(z,\xi_1,\cdots,\xi_n,\chi_1 , \cdots,
\chi_{m-1},\psi_{2k-2m-n} , \cdots, \psi_{1}, \chi_{m},\delta_m,\cdots,
\delta_1)$ gives an $I_{2k+1}^{s(0|^n1|^m2)}$ singular fiber.

\subsection{$I_{2k}^{s(0|^n1|^m2)}$ \small{($n+m\leq k,\quad m<k$)}}

The generic form for the singular fiber of type $I_{2k}^{s(0|^n1|^m2)}$,
where $m+n\leq k$, is given by $(2k-(n+m),k-n,m,k-m,0,0,n,0)$. 
The analysis follows closely that carried out for $I_{2k+1}^{s(0|^n1|^m2)}$  where more details can
be found.  In order to resolve the geometry perform the 
resolutions
\begin{equation}
\begin{aligned}
(x,y,z;\zeta_1),\quad (x,y,\zeta_i;\zeta_{i+1}) \qquad & 1\leq i<\text{min}\{k-n,k-m\} \\
(x,z;\xi_1),\qquad (x,\xi_i;\xi_{i+1}) \qquad   & 1\leq i <n \\
(y,z;\delta_1),\qquad (y,\delta_j;\delta_{j+1}) \qquad   &1\leq j <m \,.
\end{aligned}
\end{equation}
Then the sign of the quantity $m-n$ and then use the according set
of small resolutions, where the index in $\zeta_{max}$ again means the last
exceptional divisor introduced in the blow ups, that is, $\text{max}=\text{min} \{k-n,k-m\}$
\begin{equation}
\begin{aligned}
(y,\zeta_{max};\chi_1),\quad (y,\chi_r;\chi_{r+1}) \qquad   &1\leq r < m-n\\
(x,\zeta_{max};\Omega_1),\quad (x,\Omega_r;\Omega_{r+1})  \qquad   &1\leq r <
-(m-n) \,.
\end{aligned}
\end{equation}
Finally the resolution process is completed with
\begin{equation}
\begin{aligned}
  (y,\zeta_s;\psi_{s}) \qquad &  &1\leq s < \text{max} \,.
\end{aligned}
\end{equation}
The Cartan divisors are, assuming $m-n>0$,
\begin{table}[H]
\centering
\begin{tabular}{c|c}
Exceptional Divisor & Fiber Equation\\ \hline 
$z$& $l_1 l_2 w s_{6, 0} + l_2 s_{9, 0} \zeta_1 \delta_1 + 
 l_1 s_{7, 0} \zeta_1 \xi_1 $\\
$\zeta_{i<max}$ & $s_{6,0} x$
 \\
$\zeta_{max}$ & $  x s_{6, 0} + s_{5,k-m} \zeta_{max-1}$\\

$\delta_{j<m}$ & $ l_2 s_{6, 0} + s_{7, 0} \zeta_1$\\
$\delta_{m}$ & $ l_2 y s_{6, 0} + y s_{7, 0} \zeta_1 \psi_1 + 
 s_{3, 0} \zeta_1^m \delta_{m-1} \psi_1^{m-1}$\\
$\xi_{i<n}$ & $ l_1 s_{6, 0} + s_{9, 0} \zeta_1$
 \\
$\xi_{n}$ & $ l_1 x s_{6, 0} + \zeta_1 (x s_{9, 0} + s_{8, n} \zeta_1^{n-1} \xi_{n-1})$
 \\
$\chi_{r<m-n}$ & $ x s_{6, 0} + s_{5, k-m} \zeta_{max-1}$
 \\
$\chi_{m-n}$& $\begin{aligned} &y (x s_{6, 0} + s_{5, k-m} \zeta_{max-1} \psi_{max-1}) +\\+ 
 \zeta_{\max-1}^{m-n+1} \psi_{max-1}^{m-n} (x &s_{2, k-n} + s_{1, 2k-m-n} \zeta_{max-1} \psi_{max-1}) \chi_{m-n-1} \end{aligned}
$
 \\

$\psi_{s< max}$ & $ s_{6,0} y
$
 \\
 
\end{tabular}
\end{table}

Then the ordered set $(z,\xi_1,\cdots,\xi_n,\zeta_1 ,\cdots,
\zeta_{max},\chi_1, \cdots, \chi_{m-n-1}, \psi_{max} ,\cdots, \psi_1,\delta_m,
\cdots, \delta_1)$ gives an $I_{2k+1}^{s(0|^n1|^m2)}$, and again analogously
for $m - n < 0$. 
Notice that if $m=k$ and $n=0$ the vanishing orders $(k,k,k,0,0,0,0,0)$
specify the singular fibers $I_{2k}^{ns(01|^n2)}$ as listed in table
\ref{TableTateForms}. The $k$ small resolutions that resolve the singularity
are
\begin{equation}
(y,z;\delta_1),\qquad (y,\delta_j;\delta_{j+1}) \qquad   1\leq j <k \,.
\end{equation}
The resolved geometry has $k+1$ Cartan divisors, $k-1$ of which will split if $s_{6,0}^2-4 s_{5,0} s_{7,0}$ is a perfect, non-zero, square.

\subsection{$I_{2k}^{s(0|^n1|^m2)}$ \small{($n+m\leq \left\lfloor{\frac{4}{3}k}\right \rfloor$)}}

The generic form for the singular fibers of type $I_{2k}^{s(0|^n1|^m2)}$ with
section separation such that $m+n\leq\left\lfloor{\frac{4}{3}k}\right \rfloor$ is given
by $(2k-(m+n),m,m,n,0,0,n,0)$, where it is assumed that $m\geq n$. In order to
resolve the geometry the following set of resolutions is used
\begin{equation}
\begin{aligned}
(x,z;\xi_1),\qquad (x,\xi_i;\xi_{i+1}) \qquad   & 1\leq i <n \\
(y,z;\delta_1),\qquad (y,\delta_j;\delta_{j+1}) \qquad   &1\leq j <m\\
(x,\delta_r;\chi_{r}) \qquad   &1\leq r \leq m\\
(x,\chi_m;\psi_1),\qquad (x,\psi_s;\psi_{s+1}) \qquad   & 1\leq s < 2k-2m-n-1
\,.
\end{aligned}
\end{equation}
Notice that the first three sets of resolutions produce $2m+n+1$ Cartan divisors. The fourth set of resolutions is then necessary if $2k-2m-n-1\neq 0$. The Cartan divisors in the most general case are
\begin{table}[H]
\centering
\begin{tabular}{c|c}
Exceptional Divisor & Fiber Equation\\ \hline 
$z$& $l_1 l_2 w s_{6, 0} + l_2 s_{9, 0} \delta_1 + 
 l_1 s_{7, 0}  \xi_1 $\\
$\delta_1$ & $l_2 s_{6, 0} + 
 s_{7, 0} \xi_1 \xi_2^2 \cdots \xi_n^n\chi_1$\\
$\delta_{j<m}$ & $ l_2 s_{6, 0} + s_{7, 0} \chi_{j-1} \chi_j$\\
$\delta_{m}$ & $ l_2 (y s_{6, 0} + s_{2, m} \delta_{m-1} \chi_{m-1}) + \chi_{m-1} (y s_{7, 0} + s_{3, m} \delta_{m-1} \chi_{m-1}) \chi_m$\\
$\xi_{i<n}$ & $ l_1 s_{6, 0} + s_{9, 0} \delta_1$
 \\
$\xi_{n}$ & $l_1 x s_{6, 0} + x s_{9, 0} \delta_1 \chi_1 + 
 l_1 s_{5, n} \delta_1^n \xi_{n-1} \chi_1^{n-1} + 
 s_{8, n} \delta_1^{n+1} \xi_{n-1} \chi_1^n$
 \\
$\chi_{r<m}$ & $ x s_{6, 0} $
 \\
$\chi_{m}$& $ y s_{6, 0} + s_{2, m} \chi_{m-1}
$
 \\

$\psi_{s< 2k-2m-n-1}$ & $ y s_{6, 0} + s_{2, m} \chi_{m-1}
$
 \\
$\psi_{2k-2m-n-1}$ & $x y s_{6, 0} + x s_{2, m} \chi_{m-1} + s_{1, 2k-m-n} \psi_{2k-2m-n-2} \chi_{m-1}^{n-2}
$
\\
\end{tabular}
\end{table}
The ordered set $(z,\xi_1,\cdots,\xi_n,\chi_1 ,\cdots,
\chi_{m-1},\psi_{2k-2m-n-1},\cdots, \psi_{1}, \chi_{m}, \delta_m,  \cdots,
\delta_1)$ gives an $I_{2k}^{s(0|^n1|^m2)}$ type singular fiber.

\subsection{$I_{2k+1}^{ns(012)}$}
The generic form for $I_{2k+1}^{ns(012)}$ is $(2k+1, k+1, 0, k+1, 0, 0, 0,
0)$. The geometry is singular at $x=y=z=0$ and it can be resolved by performing
a blow up $(x,y,z;\zeta_1)$. This process can be repeated $k$ times, with the
$i^{\text{th}}$ resolution being $(x,y,\zeta_{i-1};\zeta_{i})$. 
The Cartan divisors are then
\begin{table}[H]
\centering
\begin{tabular}{c|c}
Exceptional Divisor & Fiber Equation\\ \hline 
$z$& $l_1 w x (l_1 x s_{3, 0} + l_2 y s_{6, 0}) + l_2^2 w y^2 s_{8, 0} + 
  x y (l_1 x s_{7, 0} + l_2 y s_{9, 0}) \zeta_1 $\\
$\zeta_{i\leq k}$ & $x^2 s_{3, 0} + x y s_{6, 0} + y^2 s_{8, 0}
$ 
\end{tabular}
\end{table}

It is easily seen by considering the projective relations introduced by the
resolutions the ordered set $(z,\zeta_1,\cdots, \zeta_k)$ of Cartan divisors
intersects in an $I_{2k+1}^{ns(012)}$. Notice that if $s_{6,0}^2-4 s_{3,0} s_{8,0}$ is a perfect square, each of the fiber components along $\{\zeta_i=0\}$ splits into two, thus giving the split version $I_{2k+1}^{s(012)}$.

\subsection{$I_{2k}^{ns(012)}$}
The generic form for $I_{2k}^{ns(012)}$ is $(2k, k, 0, k, 0, 0, 0, 0)$. 
The singular geometry can be blown up  $k$ times with the $i^{\text{th}}$ resolution being
$(x,y,\zeta_{i-1};\zeta_{i})$. 
The Cartan divisors are
\begin{table}[H]
\centering
\begin{tabular}{c|c}
Exceptional Divisor & Fiber Equation\\ \hline 
$z$& $l_1 w x (l_1 x s_{3, 0} + l_2 y s_{6, 0}) + l_2^2 w y^2 s_{8, 0} + 
  x y (l_1 x s_{7, 0} + l_2 y s_{9, 0}) \zeta_1 $\\
$\zeta_{i<k}$ & $x^2 s_{3, 0} + x y s_{6, 0} + y^2 s_{8, 0}
$ \\
$\zeta_k$ & $x^2 s_{3, 0} + x y s_{6, 0} + y^2 s_{8, 0} + 
  \zeta_{k-1}  x s_{2, k} + \zeta_{k-1} y s_{5, k} s_{1, 2k} \zeta_{k-1}^2$\\ 
\end{tabular}
\end{table}
The ordered set of $(k+1)$ Cartan divisors $(z,\zeta_1,\cdots,\zeta_k)$ gives
an $I_{2k}^{ns(012)}$. If, in addition, $s_{6,0}^2-4 s_{3,0} s_{8,0}$ is a
perfect square the $(k-1)$ Cartan divisors along $\zeta_i$ split into two,
giving an $I_{2k}^{s(012)}$ fiber.

\subsection{$I_{2k+1}^{*s(0|1||2)}$}
The generic forms for the singular fibers of type $I_{2k+1}^{*s(0|1||2)}$ are
characterized by the vanishing orders $(k+2,k+2,k+1,1,1,0,1,0)$. In order to
resolve the geometry perform the resolutions
\begin{equation}
\begin{aligned}
(x,y,z;\zeta_1),\quad (z,\zeta_1;\zeta_2),\quad &(x,z;\zeta_3),\quad (y,z;\zeta_4),\quad (y,\zeta_2;\delta_1)  \\ 
&(y,\delta_{i};\delta_{i+1})  \qquad 1 \leq i \leq 2k \,.
\end{aligned}
\end{equation}
The Cartan divisors are
\begin{table}[H]
\centering
\begin{tabular}{c|c}
Exceptional Divisor & Fiber Equation\\ \hline 
$z$ & $l_1 s_{7, 0} \zeta_3 + l_2 s_{9, 0} \zeta_4$

   \\ 
$\zeta_1$ &  $ y s_{5, 1} $\\
$\zeta_2$ &$s_{5, 1} z \zeta_4 + x \zeta_1 (x s_{7, 0} \zeta_3 + s_{9, 0} \zeta_4 \delta_1)$
 
 \\
$\zeta_3$ & $x s_{9, 0} + z (l_1 s_{5, 1} + s_{8, 1} \zeta_2)$ 

 \\
$\zeta_4$ & $ y s_{7, 0} $ \\
$\delta_{i\leq 2k}$ & $s_{7, 0} \zeta_1  + s_{5, 1}  \zeta_4$\\
$\delta_{2k+1}$ &$ y (s_{7, 0} \zeta_1 + s_{5, 1} \zeta_4) + 
 \zeta_1^k \zeta_4^k (s_{3, k+1} \zeta_1 + s_{1, k+2} \zeta_4) \delta_{2k}$  \\
\end{tabular}
\end{table}
The ordered set of divisors $(z,\zeta_3,\zeta_2,\delta_1,\cdots,\delta_{2k+1},\zeta_1,
\zeta_4)$ specifies an $I_{2k+1}^{*s(0|1||2)}$ fiber in the canonical ordering.

\subsection{$I_{2k}^{*s(0|1||2)}$}
The generic forms for the singular fibers of type $I_{2k+1}^{*s(0|1||2)}$ are
given by the vanishing orders $(k+2,k+1,k+1,1,1,0,1,0)$. In order to resolve
the geometry the following resolutions are used
\begin{equation}
\begin{aligned}
(x,y,z;\zeta_1),\quad (z,\zeta_1;\zeta_2),\quad &(x,z;\zeta_3),\quad (y,z;\zeta_4),\quad (y,\zeta_2;\delta_1)  \\ 
&(y,\delta_{i};\delta_{i+1})  \qquad 1 \leq i \leq 2k-1 \,.
\end{aligned}
\end{equation}
The Cartan divisors are listed, where, as always, all coordinates
that are constrained to be non-zero by the projective relations have been
scaled to one,
\begin{table}[H]
\centering
\begin{tabular}{c|c}
Exceptional Divisor & Fiber Equation\\ \hline 
$z$ & $l_1 s_{7, 0} \zeta_3 + l_2 s_{9, 0} \zeta_4$
 %$l_1 x s_{7, 0} \zeta_3 + 
   %l_2 y s_{9,0} \zeta_4 \delta_1 \delta_2^2 \cdots \delta_{2k}^{2k}$ 
   \\ 
$\zeta_1$ &  $ y s_{5, 1} $\\
$\zeta_2$ &$s_{5, 1} z \zeta_4 + x \zeta_1 (x s_{7, 0} \zeta_3 + s_{9, 0} \zeta_4 \delta_1)$
 %
%$l_1 x^2 s_{7, 0} \zeta_1 \zeta_3 + l_1 l_2^2 w^2 s_{5, 0} z \zeta_4 + 
 %l_2 x y s_{9, 0} \zeta_1 \zeta_4 \delta_1 \delta_2^2 \cdots \delta_{2k}^{2k}$ 
 \\
$\zeta_3$ & $x s_{9, 0} + z (l_1 s_{5, 1} + s_{8, 1} \zeta_2)$ 
%$l_1 l_2 w^2 s_{5, 0} z + 
% y \zeta_1 \delta_1 \delta_2^2 \cdots \delta_{2k}^{2k} (x s_{9, 0} + 
 %   l_2 w s_{8, 0} z \zeta_2 \zeta_4 \delta_1 \delta_2 \cdots \delta_{2k})$ 
 \\
$\zeta_4$ & $ y s_{7, 0} $ \\
$\delta_{i<2k}$ & $s_{7, 0} \zeta_1  + s_{5, 1}  \zeta_4$\\
$\delta_{2k}$ &$ y s_{7, 0} \zeta_1  + y s_{5, 1}  \zeta_4 + 
 s_{2, k+1}  \zeta_1^k \zeta_4^k  \delta_{2k-1}$  \\
\end{tabular}
\end{table}
Then the ordered set $(z,\zeta_3,\zeta_2,\delta_1,\cdots,\delta_{2k},\zeta_1,
\zeta_4)$ is an $I_{2k}^{*s(0|1||2)}$ fiber.

\subsection{$I_{2k+1}^{*s(01||2)}$}
The standard forms for the $I_{2k+1}^{*s(01||2)}$ type of singular fibers 
are given through the vanishing orders $(k+3,k+2,k+2,1,1,0,0,0)$. 
In order to resolve the geometry use the resolutions 
\begin{equation}
\begin{aligned}
(x,y,z;\zeta_1),\quad (y,z;\zeta_2),\quad &(\zeta_1,\zeta_2;\zeta_3),\quad (y,\zeta_1;\zeta_4),\quad (y,\zeta_3;\delta_1)  \\ 
&(y,\delta_{i};\delta_{i+1})  \qquad 1 \leq i \leq 2k \,.
\end{aligned}
\end{equation}
The Cartan divisors are
\begin{table}[H]
\centering
\begin{tabular}{c|c}
Exceptional Divisor & Fiber Equation\\ \hline 
$z$ & $l_1 x^2 s_{7, 0} \zeta_1 + l_2 \zeta_2 (l_2 w s_{8, 0} + x s_{9, 0} \zeta_1 \zeta_3)$
   \\ 
$\zeta_1$ &  $s_{5, 1} z + s_{8, 0} \zeta_4$
%$
\\
$\zeta_2$ & $ y s_{7, 0} $ \\
$\zeta_3$ &$s_{5, 1} z \zeta_2 + \zeta_4 (s_{7, 0} \zeta_1 + s_{8, 0} \zeta_2 \delta_1)$
 
  \\
$\zeta_4$ & $ y s_{5, 1} $ \\
$\delta_{i\leq2k}$ & $ s_{7, 0} \zeta_4 +  s_{5, 1}  \zeta_2$\\
$\delta_{2k+1}$ &$y s_{5, 1} \zeta_2 + y s_{7, 0} \zeta_4 + s_{2, k+2} \zeta_2^{k+1} \zeta_4^k \delta_{2k}$

  \\

\end{tabular}
\end{table}
Then the ordered set
$(z,\zeta_1,\zeta_3,\delta_1,\cdots,\delta_{2k+1},\zeta_2, \zeta_4)$
intersects in  an $I_{2k+1}^{*s(01||2)}$ type fiber.

\subsection{$I_{2k}^{*s(01||2)}$}
The generic forms for singular fibers of type $I_{2k}^{*s(01||2)}$ are given
by the vanishing orders $(k+2,k+2,k+1,1,1,0,0,0)$. The geometry is
non-singular after the resolutions
\begin{equation}
\begin{aligned}
(x,y,z;\zeta_1),\quad (y,z;\zeta_2),\quad &(\zeta_1,\zeta_2;\zeta_3),\quad (y,\zeta_1;\zeta_4),\quad (y,\zeta_3;\delta_1)  \\ 
&(y,\delta_{i};\delta_{i+1})  \qquad 1 \leq i \leq 2k-1 \,.
\end{aligned}
\end{equation}
The Cartan divisors after these resolutions take the form
\begin{table}[H]
\centering
\begin{tabular}{c|c}
Exceptional Divisor & Fiber Equation\\ \hline 
$z$ & $l_1 x^2 s_{7, 0} \zeta_1 + l_2 \zeta_2 (l_2 w s_{8, 0} + x s_{9, 0} \zeta_1 \zeta_3)$
%$l_1 x^2 s_{7, 0} \zeta_1 + 
   %l_2 y  \zeta_2 \delta_1 \delta_2^2 \cdots \delta_{2k}^{2k}(l_2 w s_{8, 0} + 
   % x s_{9, 0} \zeta_1 \zeta_3 \zeta_4 \delta_1 \delta_2 \cdots \delta_{2k})$ 
   \\ 
$\zeta_1$ &  $s_{5, 1} z + s_{8, 0} \zeta_4$
%$l_1 w s_{5, 0} z + 
% y s_{8, 0} \zeta_4 \delta_1 \delta_2^2 \cdots \delta_{2k}^{2k}
%$
\\
$\zeta_2$ & $ y s_{7, 0} $ \\
$\zeta_3$ &$s_{5, 1} z \zeta_2 + \zeta_4 (s_{7, 0} \zeta_1 + s_{8, 0} \zeta_2 \delta_1)$

 %$l_1 l_2^2 w^2 s_{5, 0} z \zeta_2 + l_1 x^2 s_{7, 0} \zeta_1 \zeta_4 + 
 %l_2^2 w y s_{8, 0} \zeta_2 \zeta_4 \delta_1 \delta_2^2 \cdots \delta_{2k}^{2k}$
 
  \\
$\zeta_4$ & $ y s_{5, 1} $ \\
$\delta_{i<2k}$ & $ s_{7, 0} \zeta_4 +  s_{5, 1}  \zeta_2$\\
$\delta_{2k}$ &$y (s_{5, 1} \zeta_2 + s_{7, 0} \zeta_4) + 
 \zeta_2^k \zeta_4^{k-1} (s_{1, k+2} \zeta_2 + s_{3, k+1} \zeta_4) \delta_{2k-1}$

%$ l_2^2 w^2 y s_{5, 0} z \zeta_2 + x^2 y s_{7, 0} \zeta_1 \zeta_4 + 
%   l_1 l_2 w^2 x s_{2, 0} z^{k+1} \zeta_1^{k} \zeta_2^{k} \zeta_3^{2k} %\zeta_4^{k-1} \delta_1^{2k-1} \delta_2^{2(k-1)} \cdots \delta_{2k-1}
%$

  \\
\end{tabular}
\end{table}
The set of divisors $(z,\zeta_1,\zeta_3,\delta_1,\cdots,\delta_{2k},\zeta_2,
\zeta_4)$ then has the intersection structure of an $I_{2k}^{*s(01||2)}$ fiber.

\subsection{$I_{2k+1}^{*ns(01|2)}$}
The generic forms for the singular fibers of type $I_{2k+1}^{*ns(01|2)}$ are
given by the vanishing orders $(2k+3,k+2,1,k+2,1,0,0,0)$. In order to resolve
the geometry perform the resolutions
\begin{equation}
\begin{aligned}
(x,y,z;\zeta_1),\quad (x,y,\zeta_i;\zeta_{i+1}) \qquad &1\leq i\leq k \\
(y,z;\delta_1), \qquad (y,\zeta_i;\delta_{i+1})\qquad &1\leq i \leq k+1 \\
 (\zeta_j,\delta_j;\xi_j)  \qquad &1\leq j\leq k+1 \\
 (\zeta_{k+1},\delta_{k+2};\chi) \,.
\end{aligned}
\end{equation}
The Cartan divisors are
\begin{table}[H]
\centering
\begin{tabular}{c|c}
Exceptional Divisor & Fiber Equation\\ \hline 
$z$ &$l_1 x^2 s_{7, 0} \zeta_1 + 
 l_2 \delta_1 (l_2 w s_{8, 0} + x s_{9, 0} \zeta_1 \xi_1)$

 %\begin{array} 
  %l_1 x^2 s_{7, 0} \zeta_1 \zeta_2^2\cdots \zeta_{k+1}^{k+1} \delta_3 \delta_4^2 \cdots\delta_{k+2}^k \xi_2^2 \xi_3^4 \cdots \xi_{k+1}^{2k} \chi_1 \chi_2^3 \cdots \chi_{k+1}^{2k+1} + \\
%+ l_2 y \delta_1 (l_2 w s_{8, 0} + 
 %   x s_{9, 0} \zeta_1 \zeta_2^2 \cdots \zeta_{k+1}^{k+1} \delta_2 \delta_3^2\cdots \delta_{k+2}^{k+1} \xi_1 \xi_2^3 \cdots \xi_{k+1}^{2k+1} \chi_1^2 \chi_2^4 \cdots \chi_{k+1}^{2(k+1)})\end{array}$ 
 
 \\ 

$\zeta_{i\leq k}$ & $\delta_{i+1}$ 
\\ 
$\delta_1$ & $s_{3, 1} z + y s_{7, 0} \delta_2$
 
  %l_1 w s_{3, 0} z + 
% y s_{7, 0} \zeta_2 \zeta_3^2 \cdots \zeta_{k+1}^k \delta_2 \delta_3^2\cdots \delta_{k+2}^{k+1} \xi_2^2 \xi_3^4\cdots \xi_{k+1}^{2k} \chi_1 \chi_2^3 \cdots\chi_{k+1}^{2k+1}$ 
\\
$\delta_{k+2}$ & $x^2 s_{3, 1} + x s_{2, k+2} \delta_{k+1} \xi_{k+1} + 
 s_{1, 2k+3} \delta_{k+1}^2 \xi_{k+1}^2$

%$\begin{aligned}
% x^2 s_{3, 0} + 
 %l_2 w x s_{2, 0} z^{k+1} \zeta_1^k \cdots \zeta_k \delta_1^{k+1} \delta_2^k \cdots \delta_{k+1} \xi_1^{2k+1} \xi_2^{2k-1} \cdots \xi_{k+1} \chi_1^{2k} \chi_2^{2(k-1)} \cdots \chi_{k}^2 + \\
%+ %l_2^2 w^2 s_{1, 0} z^{2(k+1)} \zeta_1^{2k} \zeta_2^{2(k-1)} \cdots %\zeta_k^2 \delta_1^{2k} \cdots \delta_{k+1}^2 \xi_1^{2(2k+1)} \xi_2^{2(2k-1)} \cdots \xi_{k+1}^2 \chi_1^{4k} \chi_2^{4(k-1)} \cdots \chi_k^4\end{aligned}$ 

\\
$\xi_1$ & $s_{3, 1} z \zeta_1 + (s_{7, 0} \zeta_1 + s_{8, 0} \delta_1) \delta_2$

%\begin{aligned}l_1^2 w x^2 s_{3, 0} z \zeta_1 \zeta_2\cdots \zeta_{k+1} + 
 %l_2^2 w y^2 s_{8, 0} \delta_1 \delta_2\cdots \delta_{k+2} + \\+
 %l_1 x^2 y s_{7, 0} \zeta_1 \zeta_2^2\cdots \zeta_{k+1}^{k+1} \delta_2 \delta_3^2 \cdots \delta_{k+2}^{k+1} \xi_2^2 \xi_3^4 \cdots \xi_{k+1}^{2k} 
 %\chi_1 \chi_2^3\cdots \chi_{k+1}^{2k+1}\end{aligned}$ 
 
 \\
$\xi_{j\leq k+1}$ & $s_{3, 1} \zeta_{j-1} \zeta_j + s_{8, 0} \delta_j \delta_{j+1}$

%l_1^2 x^2 s_{3, 0} z \zeta_1 \zeta_2\cdots \zeta_{k+1} + 
% l_2^2 y^2 s_{8, 0} \delta_1 \delta_2\cdots \delta_{k+2}$
\\
$\chi$ & $s_{8, 0} \delta_{k+2} + 
 \zeta_{k+1} (x^2 s_{3, 1} + x s_{2, k+2} \xi_{k+1} + s_{1, 2k+3} \xi_{k+1}^2)$

%l_1^2 x^2 s_{3, 0} z \zeta_1 \zeta_2\cdots \zeta_{k+1} + 
 %l_2^2 y^2 s_{8, 0} \delta_1 \delta_2\cdots \delta_{k+2}$
 
 \\
%
%\begin{aligned}
%l_2^2 y^2 s_{8, 0} \delta_1 \delta_2\cdots \delta_{k+2} + 
 %l_1^2 z \zeta_1 \zeta_2 \cdots \zeta_{k+1} (x^2 s_{3, 0} +
  %  l_2 w x s_{2, 0} z^{k+1} \zeta_1^k \cdots\zeta_k\\ \delta_1^{k+1} \delta_2^k\cdots\delta_{k+1} \xi_1^{2k+1} \xi_2^{2k-1} \cdots \xi_{k+1} \chi_1^{2k} \chi_2^{2(k-1)} \cdots\chi_k^2 + 
  %  l_2^2 w^2 s_{1, 0} z^{2(k+1)} \zeta_1^{2k} \cdots\zeta_k^2\\ \delta_1^{2(k+1)} \delta_2^{2k} \cdots \delta_{k+1}^2 \xi_1^{2(2k+1)} \xi_2^{2(2k-1)} \xi_{k+1}^2 \chi_1^{4k} \chi_2^{4(k-1)} \cdots \chi_k^4)
%\end{aligned}

\end{tabular}
\end{table}
Then the set $(z,\delta_1,\xi_1,\zeta_1,\xi_2, \zeta_2 ,\cdots,
\zeta_{k},\xi_{k+1},\chi,\delta_{k+2})$ is an $I_{2k+1}^{*ns(01|2)}$ fiber.
Notice that if $s_{2,k+2}^2-4s_{1,2k+3} s_{3,1}$ is a perfect, non zero square
then the Cartan divisor $\delta_{k+2}$ splits into two and the fiber is an 
$I_{2k+1}^{*s(01|2)}$.

\subsection{$I_{2k}^{*ns(01|2)}$}
The standard forms for the singular fibers of type $I_{2k}^{*ns(01|2)}$ are
expressed through the vanishing orders $(2k+2,k+2,1,k+1,1,0,0,0)$. The space
is resolved by the following sequence of resolutions
\begin{equation}
\begin{aligned}
(x,y,z;\zeta_1),\quad (x,y,\zeta_i;\zeta_{i+1}) \qquad &1\leq i\leq k \\
(y,z;\delta_1), \qquad (y,\zeta_i;\delta_{i+1})\qquad &1\leq i \leq k \\
 (\zeta_j,\delta_j;\xi_j)  \qquad &1\leq j\leq k+1 \,.
\end{aligned}
\end{equation}
The Cartan divisors in the resolved geometry are then
\begin{table}[H]
\centering
\begin{tabular}{c|c}
Exceptional Divisor & Fiber Equation\\ \hline 
$z$ & %\begin{array} 
  %l_1 x^2 s_{7, 0} \zeta_1 \zeta_2^2\cdots \zeta_{k+1}^{k+1} \delta_3 %\delta_4^2 \cdots\delta_{k+1}^{k-1} \xi_2^2 \xi_3^4 \cdots \xi_{k+1}^{2k} %\chi_1 \chi_2^3 \cdots \chi_{k}^{2k-1} + \\
%+ l_2 y \delta_1 (l_2 w s_{8, 0} + 
 %   x s_{9, 0} \zeta_1 \zeta_2^2 \cdots \zeta_{k+1}^{k+1} \delta_2 \delta_3^2\cdots \delta_{k+1}^{k} \xi_1 \xi_2^3 \cdots \xi_{k+1}^{2k+1} \chi_1^2 \chi_2^4 \cdots \chi_{k}^{2k})\end{array}
 
 $l_1 x^2 s_{7, 0} \zeta_1 + 
 l_2 \delta_1 (l_2 w s_{8, 0} + x s_{9, 0} \zeta_1 \xi_1)$ \\ 

$\zeta_{i\leq k}$ & $\delta_{i+1}$  
\\ 
$\zeta_{k+1}$&$y^2 s_{8, 0} + y s_{5, k+1} \zeta_k + s_{1, 2k+2} \zeta_k^2$\\
$\delta_1 $ & %l_1 w s_{3, 0} z + 
% y s_{7, 0} \zeta_2 \zeta_3^2 \cdots \zeta_{k+1}^k \delta_2 \delta_3^2\cdots \delta_{k+1}^{k} \xi_2^2 \xi_3^4\cdots \xi_{k+1}^{2k} \chi_1 \chi_2^3 \cdots\chi_{k}^{2k-1}

$s_{3, 1} z + y s_{7, 0} \delta_2$ \\

$\xi_1$ & %\begin{aligned}l_1^2 w x^2 s_{3, 0} z \zeta_1 \zeta_2\cdots \zeta_{k+1} + 
 %l_2^2 w y^2 s_{8, 0} \delta_1 \delta_2\cdots \delta_{k+1} + \\+
 %l_1 x^2 y s_{7, 0} \zeta_1 \zeta_2^2\cdots \zeta_{k+1}^{k+1} \delta_2 \delta_3^2 \cdots \delta_{k+1}^{k} \xi_2^2 \xi_3^4 \cdots \xi_{k+1}^{2k} \chi_1 \chi_2^3\cdots \chi_{k}^{2k-1}\end{aligned}
 
 $s_{3, 1} z \zeta_1 + \delta_2 (s_{8, 0} \delta_1 + s_{7, 0} \zeta_1 )$ \\
$\xi_{j\leq k}$&%l_1^2 x^2 s_{3, 0} z \zeta_1 \zeta_2\cdots \zeta_{k+1} + 
 %l_2^2 y^2 s_{8, 0} \delta_1 \delta_2\cdots \delta_{k+1}$\\
$s_{3, 1} \zeta_{j-1} \zeta_j + s_{8, 0} \delta_j \delta_{j+1}$ \\
$\xi_{k+1}$ & 
$s_{3, 1} \zeta_k \zeta_{k+1} + (y^2 s_{8, 0} + y s_{5, k+1} \zeta_k + s_{1, 2k+2} \zeta_k^2) \delta_{k+1}$  \\

\end{tabular}
\end{table}
The ordered set $(z,\delta_1,\xi_1,\zeta_1,\xi_2, \zeta_2 ,\cdots,
\zeta_{k},\xi_{k+1},\zeta_{k+1})$ represents an $I_{2k}^{*ns(01|2)}$ fiber. We note 
that if $s_{5,k+2}^2-4 s_{1,2k+2}s_{8,0}$ is a perfect, non-zero square then
the Cartan divisor $\zeta_{k+1}$ splits into two and the fiber is an
$I_{2k}^{*s(01|2)}$ fiber.

\section{Determination of the Cubic Equation}\label{app:RR}

In this appendix a non-singular elliptic curve with three marked points is
constructed following
\cite{MR0387292,MR2514094} and it is embedded into the projective space
$\mathbb{P}^2$. This non-singular elliptic curve is then fibered over some
arbitrary base, $B_3$, to create a non-singular elliptic fibration.

\begin{floatingtable}[r]{
    \begin{tabular}{c|c|c|c}
      Function & \multicolumn{3}{c}{Order} \cr
       & P & Q & R \cr\hline
      $1$ & 0 & 0 & 0 \cr
      $x$ & 1 & 1 & 0 \cr
      $y$ & 1 & 0 & 1 \cr\hline
      $xy$ & 2 & 1 & 1 \cr
      $x^2$ & 2 & 2 & 0 \cr
      $y^2$ & 2 & 0 & 2 \cr\hline
      $x^2y$ & 3 & 2 & 1 \cr
      $xy^2$ & 3 & 1 & 2 \cr
      $x^3$ & 3 & 3 & 0 \cr
      $y^3$ & 3 & 0 & 3 \cr
    \end{tabular}
  }
  \label{tbl:p2poles}
\end{floatingtable}

Begin by considering a genus one algebraic curve, $X$, with three marked
divisors $P$, $Q$, and $R$. The line bundle $\mathcal{O}(P+Q+R)$ is
identified with the vector space of meromorphic functions
on $X$, with poles of at worst order one at the points $P$, $Q$, and $R$, and
regular elsewhere. The Riemann-Roch theorem for algebraic curves fixes the
dimension of such vector spaces. Any divisor in an algebraic curve $X$ can be
written as a formal sum over the points of $X$: $D = \sum_{P \in X} n_P P$,
where $n_P = 0$ for all by finitely many $P$.
The Riemann-Roch theorem then states that for any such divisor
\begin{equation} 
  \text{dim }\mathcal{O}(D) = \text{deg}(D) + 1 - g \,,
\end{equation} 
where $\text{deg}(D)$ is the sum over the $n_P$ associated to $D$. Thus
it follows that the vector space $\mathcal{O}(P+Q+R)$ has dimension 3. Let the
three generators of this space be denoted by the functions $1$, $x$, and $y$.
We can determine the pole structure of these functions. Consider first the
vector space $\mathcal{O}(P)$, which has dimension 1 for any $P \in X$, and
which must contain the one dimensional space of constant functions. As it has
dimension 1 it can only contain these holomorphic functions, and therefore
there are no functions with a pole of order one at any single point of $X$.
The pole structure of $1$, $x$, and $y$ can then be determined to be as given
in table \ref{tbl:p2poles}, up to linear combinations.

Similarly one can consider the vector space $\mathcal{O}(2(P+Q+R))$ which has
degree, and thus dimension, 6. Clearly $1$, $x$, and $y$ are generators of
half this space, and the other three generators can be written as $x^2$, $y^2$
and $xy$, which have the pole structures given in table \ref{tbl:p2poles}.
Finally consider $\mathcal{O}(3(P+Q+R))$ which has dimension nine. Out of
the six generators for $\mathcal{O}(2(P+Q+R))$ one can construct ten
meromorphic functions inside $\mathcal{O}(3(P+Q+R))$, which must be linearly
dependent for the space to be of dimension nine. We write this relation as 
\begin{equation}
  A_1 + A_2x + A_3y + A_4xy +
  A_5x^2 + A_6y^2 + A_7x^2y + A_8xy^2 + A_9x^3 + A_{10}y^3 = 0 \,.
\end{equation}

The right-hand side of this equation is the zero function, which does not have
poles anywhere. It must then be the case that the left-hand side must not have
any poles for such a relation to hold. There are two terms with poles of order
three at the points $Q$, $R$, which are the $x^3$ and $y^3$ terms
respectively. There is no other term which contributes a pole of these orders
and so could be tuned to cancel it off, therefore the only solution is to set
the coefficients, $A_9$ and $A_{10}$, to zero.

This leaves exactly two terms with a pole of order three at $P$ and, by the
same argument as above, if either of these coefficients vanish then the other
must also vanish. Let us follow this line of argument and demonstrate that it
leads to a contradiction. If $A_7 = A_8 = 0$ then it is clear that both $A_5 =
0$ and $A_6 = 0$ as these are the only terms remaining with a pole of order
two in $Q$, $R$. Further if these terms are vanishing the arguments above lead us to
conclude that $A_4 = A_3 = A_2 = A_1 = 0$. If this is the case then this is
not a non-trivial relation among these ten meromorphic functions, and so 
the relation cannot have either of $A_7$ or $A_8$ vanishing.

After the embedding of the elliptic curve into projective space the relation 
defines the curve by a hypersurface equation which we write as
\begin{equation}\label{eqn:cubic}
  \fks_1 w^3 + \fks_2 w^2 x + \fks_3 w x^2 + \fks_5 w^2 y + \fks_6 w x y + 
  \fks_7 x^2 y + \fks_8 w y^2 + \fks_9 x y ^2 = 0 \,,
\end{equation}
where $[x : y : w]$ are the coordinates of a $\mathbb{P}^2$ and $\fks_i$ lie
in some base coordinate ring $R$. This will be taken as the defining equation of our
elliptic fibration. 

The cubic equation (\ref{eqn:cubic}) can always be mapped into the form of a
Weierstrass model using Nagell's algorithm \cite{MR1555271, MR1144763}. 
For the convenience of the reader we write here only the $f$ and $g$ of the corresponding 
Weierstrass model. The complete derivation of the Weierstrass model from the
cubic (\ref{eqn:cubic}) is given in \cite{Borchmann:2013jwa, Cvetic:2013nia,
Borchmann:2013hta, Cvetic:2013jta}
and we do not repeat it here.
The Weierstrass equation is 
\begin{equation}
  \widetilde{y}^2 = \widetilde{x}^3 + f\widetilde{x} + g \,,
\end{equation}
where $f$ and $g$ are given in terms of the coefficients of
(\ref{eqn:P111cubic}) as
\begin{align}
  f = &\frac{1}{48}(
    - \fks_6^4 
    + 8\fks_6^2(\fks_5\fks_7 + \fks_3\fks_8 + \fks_2\fks_0) 
    - 24\fks_6(\fks_2\fks_7\fks_8 + \fks_3\fks_5\fks_9 + \fks_1\fks_7\fks_9) \cr 
    &+ 16(
      - \fks_5^2\fks_7^2
      + 3\fks_1\fks_7^2\fks_8
      - \fks_3^2\fks_8^2 
      + \fks_2\fks_3\fks_8\fks_9 
      - \fks_2^2\fks_9^2
      + 3\fks_1\fks_3\fks_9^2
      + \fks_5\fks_7(\fks_3\fks_8 + \fks_2\fks_9)
      )
    ) \, 
\end{align}
\begin{align}
  g = &\frac{1}{864}(
    \fks_6^6
    - 12\fks_6^4(\fks_5\fks_7 + \fks_3\fks_8 + \fks_2\fks_9)
    + 36\fks_6^3(\fks_2\fks_7\fks_8 + \fks_3\fks_5\fks_9 + \fks_1\fks_7\fks_9)
    \cr
    &+ 24\fks_6^2(
      2\fks_5^2\fks_7^2
      + 2\fks_3^2\fks_8^2
      + \fks_2\fks_3\fks_8\fks_9
      + 2\fks_2^2\fks_9^2
      + \fks_5\fks_7(\fks_3\fks_8 + \fks_2\fks_9) 
      - 3\fks_1(\fks_7^2\fks_8 + \fks_3\fks_9^2)
      )
    \cr
    &+ 8(
      - 8\fks_5^3\fks_7^3 
      - 72\fks_1\fks_3\fks_7^2\fks_8^2
      - 8\fks_3^3\fks_8^3
      + 27\fks_1^2\fks_7^2\fks_9^2
      - 72\fks_1\fks_3^2\fks_8\fks_9^2
      - 8\fks_2^3\fks_9^3 
      \cr
      &+ 3\fks_2^2\fks_8(9\fks_7^2\fks_8 + 4\fks_3\fks_9^2) 
      + 6\fks_5\fks_7(
        6\fks_1\fks_7^2\fks_8 
        + 2\fks_3^2\fks_8^2
        + \fks_2\fks_3\fks_8\fks_9
        + 2\fks_2^2\fks_9^2
        - 3\fks_1\fks_3\fks_9^2
        )
      \cr
      &+ 6\fks_2\fks_9(
        - 3\fks_1\fks_7^2\fks_8 
        + 2\fks_3^2\fks_8^2
        + 6\fks_1\fks_3\fks_9^2
        )
      + 3\fks_5^2(
        4\fks_3\fks_7^2\fks_8
        + 4\fks_2\fks_7^2\fks_9
        + 9\fks_3^2\fks_9^2
        )
      )
    \cr
    &- 144\fks_6(
      \fks_2^2\fks_7\fks_8\fks_9
      + \fks_9(
        \fks_1\fks_5\fks_7^2 
        + \fks_3^2\fks_5\fks_8 
        + \fks_3\fks_8(\fks_5^2 - 5\fks_1\fks_8)
        )
      \cr
      &+ \fks_2(
        \fks_5\fks_7^2\fks_8 
        + \fks_3\fks_7\fks_8^2 
        + \fks_1\fks_7\fks_9^2
        )
      )
    ) \,.
\end{align}

\newpage
%\bibliography{Quarticbib}{}

\begin{thebibliography}{10}

\bibitem{Vafa:1996xn}
C.~Vafa, {\it {Evidence for F-Theory}},  {\em Nucl. Phys.} {\bf B469} (1996)
  403--418, [\href{http://xxx.lanl.gov/abs/hep-th/9602022}{{\tt
  hep-th/9602022}}].

\bibitem{Morrison:1996pp}
D.~R. Morrison and C.~Vafa, {\it {Compactifications of F-Theory on Calabi--Yau
  Threefolds -- II}},  {\em Nucl. Phys.} {\bf B476} (1996) 437--469,
  [\href{http://xxx.lanl.gov/abs/hep-th/9603161}{{\tt hep-th/9603161}}].

\bibitem{Morrison:1996na}
D.~R. Morrison and C.~Vafa, {\it {Compactifications of F-Theory on Calabi--Yau
  Threefolds -- I}},  {\em Nucl. Phys.} {\bf B473} (1996) 74--92,
  [\href{http://xxx.lanl.gov/abs/hep-th/9602114}{{\tt hep-th/9602114}}].

\bibitem{Donagi:2008ca}
R.~Donagi and M.~Wijnholt, {\it {Model Building with F-Theory}},  {\em
  Adv.Theor.Math.Phys.} {\bf 15} (2011) 1237--1318,
  [\href{http://xxx.lanl.gov/abs/0802.2969}{{\tt 0802.2969}}].

\bibitem{Beasley:2008dc}
C.~Beasley, J.~J. Heckman, and C.~Vafa, {\it {GUTs and Exceptional Branes in
  F-theory - I}},  {\em JHEP} {\bf 01} (2009) 058,
  [\href{http://xxx.lanl.gov/abs/0802.3391}{{\tt 0802.3391}}].

\bibitem{Beasley:2008kw}
C.~Beasley, J.~J. Heckman, and C.~Vafa, {\it {GUTs and Exceptional Branes in
  F-theory - II: Experimental Predictions}},  {\em JHEP} {\bf 01} (2009) 059,
  [\href{http://xxx.lanl.gov/abs/0806.0102}{{\tt 0806.0102}}].

\bibitem{Weigand:2010wm}
T.~Weigand, {\it {Lectures on F-theory compactifications and model building}},
  {\em Class.Quant.Grav.} {\bf 27} (2010) 214004,
  [\href{http://xxx.lanl.gov/abs/1009.3497}{{\tt 1009.3497}}].

\bibitem{Heckman:2010bq}
J.~J. Heckman, {\it {Particle Physics Implications of F-theory}},  {\em
  Ann.Rev.Nucl.Part.Sci.} {\bf 60} (2010) 237--265,
  [\href{http://xxx.lanl.gov/abs/1001.0577}{{\tt 1001.0577}}].

\bibitem{Maharana:2012tu}
A.~Maharana and E.~Palti, {\it {Models of Particle Physics from Type IIB String
  Theory and F-theory: A Review}},  {\em Int.J.Mod.Phys.} {\bf A28} (2013)
  1330005, [\href{http://xxx.lanl.gov/abs/1212.0555}{{\tt 1212.0555}}].

\bibitem{BerasaluceGonzalez:2011wy}
M.~Berasaluce-Gonzalez, L.~E. Ibanez, P.~Soler, and A.~M. Uranga, {\it
  {Discrete gauge symmetries in D-brane models}},  {\em JHEP} {\bf 1112} (2011)
  113, [\href{http://xxx.lanl.gov/abs/1106.4169}{{\tt 1106.4169}}].

\bibitem{BerasaluceGonzalez:2012zn}
M.~Berasaluce-Gonzalez, P.~Camara, F.~Marchesano, and A.~Uranga, {\it {Zp
  charged branes in flux compactifications}},  {\em JHEP} {\bf 1304} (2013)
  138, [\href{http://xxx.lanl.gov/abs/1211.5317}{{\tt 1211.5317}}].

\bibitem{Mayrhofer:2014haa}
C.~Mayrhofer, E.~Palti, O.~Till, and T.~Weigand, {\it {Discrete Gauge
  Symmetries by Higgsing in four-dimensional F-Theory Compactifications}},
  \href{http://xxx.lanl.gov/abs/1408.6831}{{\tt 1408.6831}}.

\bibitem{Garcia-Etxebarria:2014qua}
I.~García-Etxebarria, T.~W. Grimm, and J.~Keitel, {\it {Yukawas and discrete
  symmetries in F-theory compactifications without section}},  {\em JHEP} {\bf
  1411} (2014) 125, [\href{http://xxx.lanl.gov/abs/1408.6448}{{\tt
  1408.6448}}].

\bibitem{Mayrhofer:2014laa}
C.~Mayrhofer, E.~Palti, O.~Till, and T.~Weigand, {\it {On Discrete Symmetries
  and Torsion Homology in F-Theory}},
  \href{http://xxx.lanl.gov/abs/1410.7814}{{\tt 1410.7814}}.

\bibitem{Hayashi:2008ba}
H.~Hayashi, R.~Tatar, Y.~Toda, T.~Watari, and M.~Yamazaki, {\it {New Aspects of
  Heterotic--F Theory Duality}},  {\em Nucl.Phys.} {\bf B806} (2009) 224--299,
  [\href{http://xxx.lanl.gov/abs/0805.1057}{{\tt 0805.1057}}].

\bibitem{Donagi:2009ra}
R.~Donagi and M.~Wijnholt, {\it {Higgs Bundles and UV Completion in F-Theory}},
   {\em Commun.Math.Phys.} {\bf 326} (2014) 287--327,
  [\href{http://xxx.lanl.gov/abs/0904.1218}{{\tt 0904.1218}}].

\bibitem{Marsano:2009gv}
J.~Marsano, N.~Saulina, and S.~Schafer-Nameki, {\it {Monodromies, Fluxes, and
  Compact Three-Generation F-theory GUTs}},  {\em JHEP} {\bf 08} (2009) 046,
  [\href{http://xxx.lanl.gov/abs/0906.4672}{{\tt 0906.4672}}].

\bibitem{Marsano:2009wr}
J.~Marsano, N.~Saulina, and S.~Schafer-Nameki, {\it {Compact F-theory GUTs with
  $U(1)_{PQ}$}},  {\em JHEP} {\bf 04} (2010) 095,
  [\href{http://xxx.lanl.gov/abs/0912.0272}{{\tt 0912.0272}}].

\bibitem{Dudas:2009hu}
E.~Dudas and E.~Palti, {\it {Froggatt-Nielsen models from E(8) in F-theory
  GUTs}},  {\em JHEP} {\bf 1001} (2010) 127,
  [\href{http://xxx.lanl.gov/abs/0912.0853}{{\tt 0912.0853}}].

\bibitem{Hayashi:2010zp}
H.~Hayashi, T.~Kawano, Y.~Tsuchiya, and T.~Watari, {\it {More on Dimension-4
  Proton Decay Problem in F-theory -- Spectral Surface, Discriminant Locus and
  Monodromy}},  {\em Nucl.Phys.} {\bf B840} (2010) 304--348,
  [\href{http://xxx.lanl.gov/abs/1004.3870}{{\tt 1004.3870}}].

\bibitem{Dudas:2010zb}
E.~Dudas and E.~Palti, {\it {On hypercharge flux and exotics in F-theory
  GUTs}},  {\em JHEP} {\bf 09} (2010) 013,
  [\href{http://xxx.lanl.gov/abs/1007.1297}{{\tt 1007.1297}}].

\bibitem{Dolan:2011iu}
M.~J. Dolan, J.~Marsano, N.~Saulina, and S.~Schafer-Nameki, {\it {F-theory GUTs
  with U(1) Symmetries: Generalities and Survey}},  {\em Phys.Rev.} {\bf D84}
  (2011) 066008, [\href{http://xxx.lanl.gov/abs/1102.0290}{{\tt 1102.0290}}].

\bibitem{Grimm:2010ez}
T.~W. Grimm and T.~Weigand, {\it {On Abelian Gauge Symmetries and Proton Decay
  in Global F-theory GUTs}},  {\em Phys.Rev.} {\bf D82} (2010) 086009,
  [\href{http://xxx.lanl.gov/abs/1006.0226}{{\tt 1006.0226}}].

\bibitem{Grimm:2011tb}
T.~W. Grimm, M.~Kerstan, E.~Palti, and T.~Weigand, {\it {Massive Abelian Gauge
  Symmetries and Fluxes in F-theory}},  {\em JHEP} {\bf 1112} (2011) 004,
  [\href{http://xxx.lanl.gov/abs/1107.3842}{{\tt 1107.3842}}].

\bibitem{Braun:2011zm}
A.~P. Braun, A.~Collinucci, and R.~Valandro, {\it {G-flux in F-theory and
  algebraic cycles}},  {\em Nucl.Phys.} {\bf B856} (2012) 129--179,
  [\href{http://xxx.lanl.gov/abs/1107.5337}{{\tt 1107.5337}}]. 55 pages, 1
  figure/ added refs, corrected typos.

\bibitem{Grimm:2011fx}
T.~W. Grimm and H.~Hayashi, {\it {F-theory fluxes, Chirality and Chern-Simons
  theories}},  {\em JHEP} {\bf 1203} (2012) 027,
  [\href{http://xxx.lanl.gov/abs/1111.1232}{{\tt 1111.1232}}]. 53 pages, 5
  figures/ v2: typos corrected, minor improvements.

\bibitem{Krause:2012yh}
S.~Krause, C.~Mayrhofer, and T.~Weigand, {\it {Gauge Fluxes in F-theory and
  Type IIB Orientifolds}},  {\em JHEP} {\bf 1208} (2012) 119,
  [\href{http://xxx.lanl.gov/abs/1202.3138}{{\tt 1202.3138}}].

\bibitem{Morrison:2012ei}
D.~R. Morrison and D.~S. Park, {\it {F-Theory and the Mordell-Weil Group of
  Elliptically-Fibered Calabi-Yau Threefolds}},  {\em JHEP} {\bf 1210} (2012)
  128, [\href{http://xxx.lanl.gov/abs/1208.2695}{{\tt 1208.2695}}].

\bibitem{Cvetic:2012xn}
M.~Cvetic, T.~W. Grimm, and D.~Klevers, {\it {Anomaly Cancellation And Abelian
  Gauge Symmetries In F-theory}},  {\em JHEP} {\bf 1302} (2013) 101,
  [\href{http://xxx.lanl.gov/abs/1210.6034}{{\tt 1210.6034}}].

\bibitem{Mayrhofer:2012zy}
C.~Mayrhofer, E.~Palti, and T.~Weigand, {\it {U(1) symmetries in F-theory GUTs
  with multiple sections}},  {\em JHEP} {\bf 1303} (2013) 098,
  [\href{http://xxx.lanl.gov/abs/1211.6742}{{\tt 1211.6742}}].

\bibitem{Braun:2013yti}
V.~Braun, T.~W. Grimm, and J.~Keitel, {\it {New Global F-theory GUTs with U(1)
  symmetries}},  {\em JHEP} {\bf 1309} (2013) 154,
  [\href{http://xxx.lanl.gov/abs/1302.1854}{{\tt 1302.1854}}].

\bibitem{Kuntzler:2014ila}
M.~Kuntzler and S.~Schafer-Nameki, {\it {Tate Trees for Elliptic Fibrations
  with Rank one Mordell-Weil group}},
  \href{http://xxx.lanl.gov/abs/1406.5174}{{\tt 1406.5174}}.

\bibitem{Krippendorf:2014xba}
S.~Krippendorf, D.~K. Mayorga~Pena, P.-K. Oehlmann, and F.~Ruehle, {\it
  {Rational F-Theory GUTs without exotics}},  {\em JHEP} {\bf 1407} (2014) 013,
  [\href{http://xxx.lanl.gov/abs/1401.5084}{{\tt 1401.5084}}].

\bibitem{Borchmann:2013jwa}
J.~Borchmann, C.~Mayrhofer, E.~Palti, and T.~Weigand, {\it {Elliptic fibrations
  for $SU(5)\times U(1)\times U(1)$ F-theory vacua}},  {\em Phys.Rev.} {\bf
  D88} (2013), no.~4 046005, [\href{http://xxx.lanl.gov/abs/1303.5054}{{\tt
  1303.5054}}].

\bibitem{Cvetic:2013nia}
M.~Cvetic, D.~Klevers, and H.~Piragua, {\it {F-Theory Compactifications with
  Multiple U(1)-Factors: Constructing Elliptic Fibrations with Rational
  Sections}},  {\em JHEP} {\bf 1306} (2013) 067,
  [\href{http://xxx.lanl.gov/abs/1303.6970}{{\tt 1303.6970}}].

\bibitem{Braun:2013nqa}
V.~Braun, T.~W. Grimm, and J.~Keitel, {\it {Geometric Engineering in Toric
  F-Theory and GUTs with U(1) Gauge Factors}},  {\em JHEP} {\bf 1312} (2013)
  069, [\href{http://xxx.lanl.gov/abs/1306.0577}{{\tt 1306.0577}}].

\bibitem{Cvetic:2013uta}
M.~Cvetic, A.~Grassi, D.~Klevers, and H.~Piragua, {\it {Chiral Four-Dimensional
  F-Theory Compactifications With SU(5) and Multiple U(1)-Factors}},  {\em
  JHEP} {\bf 1404} (2014) 010, [\href{http://xxx.lanl.gov/abs/1306.3987}{{\tt
  1306.3987}}].

\bibitem{Borchmann:2013hta}
J.~Borchmann, C.~Mayrhofer, E.~Palti, and T.~Weigand, {\it {SU(5) Tops with
  Multiple U(1)s in F-theory}},  {\em Nucl.Phys.} {\bf B882} (2014) 1--69,
  [\href{http://xxx.lanl.gov/abs/1307.2902}{{\tt 1307.2902}}].

\bibitem{Cvetic:2013jta}
M.~Cvetic, D.~Klevers, and H.~Piragua, {\it {F-Theory Compactifications with
  Multiple U(1)-Factors: Addendum}},  {\em JHEP} {\bf 1312} (2013) 056,
  [\href{http://xxx.lanl.gov/abs/1307.6425}{{\tt 1307.6425}}].

\bibitem{Cvetic:2013qsa}
M.~Cvetic, D.~Klevers, H.~Piragua, and P.~Song, {\it {Elliptic fibrations with
  rank three Mordell-Weil group: F-theory with U(1) x U(1) x U(1) gauge
  symmetry}},  {\em JHEP} {\bf 1403} (2014) 021,
  [\href{http://xxx.lanl.gov/abs/1310.0463}{{\tt 1310.0463}}].

\bibitem{Klevers:2014bqa}
D.~Klevers, D.~K. Mayorga~Pena, P.-K. Oehlmann, H.~Piragua, and J.~Reuter, {\it
  {F-Theory on all Toric Hypersurface Fibrations and its Higgs Branches}},
  \href{http://xxx.lanl.gov/abs/1408.4808}{{\tt 1408.4808}}.

\bibitem{Braun:2014qka}
V.~Braun, T.~W. Grimm, and J.~Keitel, {\it {Complete Intersection Fibers in
  F-Theory}},  \href{http://xxx.lanl.gov/abs/1411.2615}{{\tt 1411.2615}}.

\bibitem{Candelas:1996su}
P.~Candelas and A.~Font, {\it {Duality between the webs of heterotic and type
  II vacua}},  {\em Nucl.Phys.} {\bf B511} (1998) 295--325,
  [\href{http://xxx.lanl.gov/abs/hep-th/9603170}{{\tt hep-th/9603170}}].

\bibitem{MR0393039}
J.~Tate, {\it Algorithm for determining the type of a singular fiber in an
  elliptic pencil},  in {\em Modular functions of one variable, {IV} ({P}roc.
  {I}nternat. {S}ummer {S}chool, {U}niv. {A}ntwerp, {A}ntwerp, 1972)},
  pp.~33--52. Lecture Notes in Math., Vol. 476.
\newblock Springer, Berlin, 1975.

\bibitem{Bershadsky:1996nh}
M.~Bershadsky, K.~A. Intriligator, S.~Kachru, D.~R. Morrison, V.~Sadov, {\em
  et.~al.}, {\it {Geometric singularities and enhanced gauge symmetries}},
  {\em Nucl.Phys.} {\bf B481} (1996) 215--252,
  [\href{http://xxx.lanl.gov/abs/hep-th/9605200}{{\tt hep-th/9605200}}].

\bibitem{Katz:2011qp}
S.~Katz, D.~R. Morrison, S.~Schafer-Nameki, and J.~Sully, {\it {Tate's
  algorithm and F-theory}},  {\em JHEP} {\bf 1108} (2011) 094,
  [\href{http://xxx.lanl.gov/abs/1106.3854}{{\tt 1106.3854}}].

\bibitem{MR0132556}
K.~Kodaira, {\it On compact complex analytic surfaces. {I}},  {\em Ann. of
  Math. (2)} {\bf 71} (1960) 111--152.

\bibitem{MR0184257}
K.~Kodaira, {\it On compact analytic surfaces. {II}, {III}},  {\em Ann. of
  Math. (2) 77 (1963), 563--626; ibid.} {\bf 78} (1963) 1--40.

\bibitem{MR0179172}
A.~N{\'e}ron, {\it Mod\`eles minimaux des vari\'et\'es ab\'eliennes sur les
  corps locaux et globaux},  {\em Inst. Hautes \'Etudes Sci. Publ.Math. No.}
  {\bf 21} (1964) 128.

\bibitem{Lin:2014qga}
L.~Lin and T.~Weigand, {\it {Towards the Standard Model in F-theory}},
  \href{http://xxx.lanl.gov/abs/1406.6071}{{\tt 1406.6071}}.

\bibitem{Grassi:2014zxa}
A.~Grassi, J.~Halverson, J.~Shaneson, and W.~Taylor, {\it {Non-Higgsable QCD
  and the Standard Model Spectrum in F-theory}},
  \href{http://xxx.lanl.gov/abs/1409.8295}{{\tt 1409.8295}}.

\bibitem{Hayashi:2013lra}
H.~Hayashi, C.~Lawrie, and S.~Schafer-Nameki, {\it {Phases, Flops and F-theory:
  SU(5) Gauge Theories}},  {\em JHEP} {\bf 1310} (2013) 046,
  [\href{http://xxx.lanl.gov/abs/1304.1678}{{\tt 1304.1678}}].

\bibitem{Hayashi:2014kca}
H.~Hayashi, C.~Lawrie, D.~R. Morrison, and S.~Schafer-Nameki, {\it {Box Graphs
  and Singular Fibers}},  {\em JHEP} {\bf 1405} (2014) 048,
  [\href{http://xxx.lanl.gov/abs/1402.2653}{{\tt 1402.2653}}].

\bibitem{Esole:2014bka}
M.~Esole, S.-H. Shao, and S.-T. Yau, {\it {Singularities and Gauge Theory
  Phases}},  \href{http://xxx.lanl.gov/abs/1402.6331}{{\tt 1402.6331}}.

\bibitem{Braun:2014kla}
A.~P. Braun and S.~Schafer-Nameki, {\it {Box Graphs and Resolutions I}},
  \href{http://xxx.lanl.gov/abs/1407.3520}{{\tt 1407.3520}}.

\bibitem{Esole:2014hya}
M.~Esole, S.-H. Shao, and S.-T. Yau, {\it {Singularities and Gauge Theory
  Phases II}},  \href{http://xxx.lanl.gov/abs/1407.1867}{{\tt 1407.1867}}.

\bibitem{Morrison:2014era}
D.~R. Morrison and W.~Taylor, {\it {Sections, multisections, and U(1) fields in
  F-theory}},  \href{http://xxx.lanl.gov/abs/1404.1527}{{\tt 1404.1527}}.

\bibitem{Braun:2014oya}
V.~Braun and D.~R. Morrison, {\it {F-theory on Genus-One Fibrations}},  {\em
  JHEP} {\bf 1408} (2014) 132, [\href{http://xxx.lanl.gov/abs/1401.7844}{{\tt
  1401.7844}}].

\bibitem{Anderson:2014yva}
L.~B. Anderson, I.~García-Etxebarria, T.~W. Grimm, and J.~Keitel, {\it
  {Physics of F-theory compactifications without section}},
  \href{http://xxx.lanl.gov/abs/1406.5180}{{\tt 1406.5180}}.

\bibitem{MR0103906}
M.~Auslander and D.~A. Buchsbaum, {\it Unique factorization in regular local
  rings},  {\em Proc. Nat. Acad. Sci. U.S.A.} {\bf 45} (1959) 733--734.

\bibitem{Lawrie:2012gg}
C.~Lawrie and S.~Schafer-Nameki, {\it {The Tate Form on Steroids: Resolution
  and Higher Codimension Fibers}},  {\em JHEP} {\bf 1304} (2013) 061,
  [\href{http://xxx.lanl.gov/abs/1212.2949}{{\tt 1212.2949}}].

\bibitem{Grimm:2009yu}
T.~W. Grimm, S.~Krause, and T.~Weigand, {\it {F-Theory GUT Vacua on Compact
  Calabi-Yau Fourfolds}},  {\em JHEP} {\bf 1007} (2010) 037,
  [\href{http://xxx.lanl.gov/abs/0912.3524}{{\tt 0912.3524}}].

\bibitem{Krause:2011xj}
S.~Krause, C.~Mayrhofer, and T.~Weigand, {\it {$G_4$ flux, chiral matter and
  singularity resolution in F-theory compactifications}},  {\em Nucl.Phys.}
  {\bf B858} (2012) 1--47, [\href{http://xxx.lanl.gov/abs/1109.3454}{{\tt
  1109.3454}}]. 53 pages, 2 figures.

\bibitem{Esole:2011sm}
M.~Esole and S.-T. Yau, {\it {Small resolutions of SU(5)-models in F-theory}},
  {\em Adv.Theor.Math.Phys.} {\bf 17} (2013) 1195--1253,
  [\href{http://xxx.lanl.gov/abs/1107.0733}{{\tt 1107.0733}}].

\bibitem{MS}
J.~Marsano and S.~Schafer-Nameki, {\it {Yukawas, G-flux, and Spectral Covers
  from Resolved Calabi-Yau's}},  {\em JHEP} {\bf 1111} (2011) 098,
  [\href{http://xxx.lanl.gov/abs/1108.1794}{{\tt 1108.1794}}].

\bibitem{Braun:2013cb}
A.~P. Braun and T.~Watari, {\it {On Singular Fibres in F-Theory}},  {\em JHEP}
  {\bf 1307} (2013) 031, [\href{http://xxx.lanl.gov/abs/1301.5814}{{\tt
  1301.5814}}].

\bibitem{MR1078016}
R.~Miranda, {\em The basic theory of elliptic surfaces}.
\newblock Dottorato di Ricerca in Matematica. [Doctorate in Mathematical
  Research]. ETS Editrice, Pisa, 1989.

\bibitem{Smooth}
M.~Kuntzler and C.~Lawrie, {\it {Smooth: A Mathematica package for studying
  resolutions of singular fibrations, Version 0.4}}, .

\bibitem{MR1312368}
J.~H. Silverman, {\em Advanced topics in the arithmetic of elliptic curves},
  vol.~151 of {\em Graduate Texts in Mathematics}.
\newblock Springer-Verlag, New York, 1994.

\bibitem{Mayrhofer:2014opa}
C.~Mayrhofer, D.~R. Morrison, O.~Till, and T.~Weigand, {\it {Mordell-Weil
  Torsion and the Global Structure of Gauge Groups in F-theory}},  {\em JHEP}
  {\bf 1410} (2014) 16, [\href{http://xxx.lanl.gov/abs/1405.3656}{{\tt
  1405.3656}}].

\bibitem{Aspinwall:1998xj}
P.~S. Aspinwall and D.~R. Morrison, {\it {Nonsimply connected gauge groups and
  rational points on elliptic curves}},  {\em JHEP} {\bf 9807} (1998) 012,
  [\href{http://xxx.lanl.gov/abs/hep-th/9805206}{{\tt hep-th/9805206}}].

\bibitem{Park:2011ji}
D.~S. Park, {\it {Anomaly Equations and Intersection Theory}},  {\em JHEP} {\bf
  1201} (2012) 093, [\href{http://xxx.lanl.gov/abs/1111.2351}{{\tt
  1111.2351}}]. 29 pages + appendices, 8 figures/ v2: minor corrections,
  references added.

\bibitem{MR1030197}
T.~Shioda, {\it Mordell-{W}eil lattices and {G}alois representation. {I}},
  {\em Proc. Japan Acad. Ser. A Math. Sci.} {\bf 65} (1989), no.~7 268--271.

\bibitem{MR0387292}
P.~Deligne, {\it Courbes elliptiques: formulaire d'apr\`es {J}. {T}ate},  in
  {\em Modular functions of one variable, {IV} ({P}roc. {I}nternat. {S}ummer
  {S}chool, {U}niv. {A}ntwerp, {A}ntwerp, 1972)}, pp.~53--73. Lecture Notes in
  Math., Vol. 476.
\newblock Springer, Berlin, 1975.

\bibitem{MR2514094}
J.~H. Silverman, {\em The arithmetic of elliptic curves}, vol.~106 of {\em
  Graduate Texts in Mathematics}.
\newblock Springer, Dordrecht, second~ed., 2009.

\bibitem{MR1555271}
T.~Nagell, {\it Sur les propri\'et\'es arithm\'etiques des cubiques planes du
  premier genre},  {\em Acta Math.} {\bf 52} (1929), no.~1 93--126.

\bibitem{MR1144763}
J.~W.~S. Cassels, {\em Lectures on elliptic curves}, vol.~24 of {\em London
  Mathematical Society Student Texts}.
\newblock Cambridge University Press, Cambridge, 1991.

\end{thebibliography}
\bibliographystyle{JHEP}

\providecommand{\href}[2]{#2}\begingroup\raggedright\endgroup

\end{document}